\documentclass[apj,twocolumn]{emulateapj}
\usepackage{longtable,graphics}
\shorttitle{Evolution of Cluster AGN from $z$$\approx$1.2 to $z$$=$0.2}
\shortauthors{Hart et al.}
\slugcomment{{\sc Accepted for publication in the Astrophysical Journal:} July 8, 2011} 

\def\Tinytwo{\fontsize{6pt}{6pt}\selectfont}

\def\xraylimit{$10^{42}$ erg s$^{-1}$}

\def\lum{erg s$^{-1}$}
\def\radiounits{W Hz$^{-1}$}
\def\rlim{P$_{1.4GHz}$$\geq$3$\times$10$^{23}$ W Hz$^{-1}$}
\def\plim{P$_{1.4GHz}$$\geq$5$\times$10$^{23}$ W Hz$^{-1}$}
\def\rlimb{5$\times$10$^{23}$ W Hz$^{-1}$}
\def\xlim{10$^{42}$ erg s$^{-1}$}
\def\xlimb{5$\times$10$^{42}$ erg s$^{-1}$}
\def\ergs{erg s$^{-1}$}
\def\flux{erg cm$^{-2}$ s$^{-1}$}
\def\fluxunits{erg cm$^{-2}$ s$^{-1}$}
\def\chandra{{\it Chandra}}
\def\hst{{\it HST}}
\def\gminusr{({\it g--r})}
\def\rminusi{({\it r--i})}
\def\iminusz{({\it i--z})}

\def\arcsec{$^{\prime\prime}$}
\def\arcmin{$^{\prime}$}

\def\oii{[O II] $\lambda$3727}

\def\cahk{Ca HK $\lambda$4000}
\def\cahkbreak{Ca HK break}

\def\oiii{[O III] $\lambda$5007}

\def\rtc{``Road to Coma"}
\def\allz{0.2$<$$z$$<$1.1}
\def\lowz{0.2$<$$z$$<$0.4}
\def\midz{0.4$<$$z$$<$0.8}
\def\highz{0.8$<$$z$$<$1.2}
\def\highzsmall{0.8$<$$z$$<$1.1}

\def\comboz{0.4$<$$z$$<$1.2}
\def\combozb{0.4$<$$z$$<$1.1}
\def\lxbb{L$_{0.3-8.0keV}$}

\def\kms{km s$^{-1}$}
\def\logns{Log N - Log S}
\def\logp{log(P$_{1.4GHz}$)}

\def\arcsec{$^{\prime\prime}$}

\def\arcmin{$^{\prime}$}

\begin{document}
\title{The Evolution of Radio Galaxies and X-ray Point Sources in Coma Cluster Progenitors Since $z$$\sim$1.2}
\author{Quyen N. Hart\altaffilmark{1}, John T. Stocke\altaffilmark{1}, August E. Evrard\altaffilmark{2}, Erica E. Ellingson\altaffilmark{1}, and Wayne A. Barkhouse\altaffilmark{3}}
\altaffiltext{1}{Center for Astrophysics and Space Astronomy, Department of Astrophysical and Planetary Sciences,
        UCB-389, University of Colorado, Boulder, CO 80309, USA}
\altaffiltext{2}{Departments of Physics and Astronomy and Michigan Center for Theoretical Physics,
        University of Michigan, Ann Arbor, MI 48109-1040, USA}
\altaffiltext{3}{(For the ChaMP collaboration) Department of Physics and Astrophysics, University of North Dakota, Grand Forks, ND 58202, USA}
\keywords{galaxies: active -- galaxies: clusters: general -- radio continuum: galaxies -- X-rays: galaxies: clusters -- X-rays: galaxies}

\begin{abstract}
Using \chandra\ imaging spectroscopy and Very Large Array (VLA) L-band radio maps, we have identified
radio sources at P$_{1.4GHz}$$\geq$5$\times$10$^{23}$ \radiounits\ and X-ray point sources (XPSs) at 
L$_{0.3-8keV}$$\geq$5$\times$10$^{42}$ \lum\ in L$>$L$^*$ galaxies in 12 high-redshift (0.4$<$$z$$<$1.2) clusters of galaxies.  
The radio galaxies and XPSs in this cluster sample, chosen to be consistent with Coma Cluster progenitors 
at these redshifts, are compared to those found at low-$z$ analyzed in \citet{hart09}.
Within a projected radius of 1 Mpc of the cluster cores,
we find 17 cluster radio galaxies (11 with secure redshifts, including one 
luminous FR~II radio source at $z$$=$0.826, and 6 more with host galaxy colors similar to cluster ellipticals).
The radio luminosity function (RLF) of the cluster radio galaxies 
as a fraction of the cluster red sequence (CRS) galaxies reveals significant evolution of this 
population from high-$z$ to low-$z$, with higher power radio galaxies situated in lower temperature clusters at earlier epochs.  
Additionally, there is some evidence that cluster radio galaxies become more centrally concentrated
than CRS galaxies with cosmic time.
Within this same projected radius, we identify 7 
spectroscopically-confirmed cluster XPSs, all with CRS host galaxy colors.  
Consistent with the results from \citet{2009ApJ...701...66M}, we estimate a minimum X-ray active fraction of 1.4$\pm$0.8\%
for CRS galaxies in high-$z$ clusters, corresponding to an approximate 10-fold increase from 0.15$\pm$0.15\% at low-$z$.
Although complete redshift information is lacking for several XPSs in $z$$>$0.4 cluster fields, the increased numbers and luminosities of the 
CRS radio galaxies and XPSs suggest a substantial (9-10 fold) increase in the heat injected into high redshift clusters by AGN
compared to the present epoch.
\end{abstract}
\maketitle

\section{Introduction}
\label{sec:p2_intro}

Clusters of galaxies are special environments in which to study the evolution of galaxies and active
galactic nuclei (AGN) because these massive structures contain a hot, diffuse intracluster medium (ICM) that
is detectable in the X-ray.  The disturbed appearance of extended radio emission from individual cluster galaxies
\citep[e.g., NGC 1265 in the Perseus cluster;][]{1986ApJ...301..841O} was the first indication that an ICM existed 
\citep[e.g.,][]{1972Natur.237..269M} and early X-ray observations detected it at T$=$10$^{7-8}$ K 
\citep[][and references therein]{1986RvMP...58....1S}.  With the high-resolution
of the \chandra\ Advanced CCD Imaging Spectrometer (ACIS), it was found that the radio-lobes of nearby AGN in the brightest cluster galaxy (BCG)
appear to be coincident with cavities in X-ray surface brightness \citep[e.g., Perseus cluster;][]{2002MNRAS.331..369F}.
The physical connection suggested by this spatial coincidence allows an indirect measurement of the
kinetic power produced by these AGN.

The connection between AGN and galaxy evolution is not well-understood, but theoretical  
results suggest that the absence of extremely massive galaxies is due to the truncation
of star formation \citep[e.g.,][]{2004ApJ...608..752B}.  AGN have been implicated as the mechanism to ``shut-off" star-formation
by heating the surrounding gas and/or ejecting the gas reservoirs necessary for galaxy growth during outburst episodes 
\citep[e.g.,][]{2006MNRAS.370..645B}.
Heating by AGN has also been suggested to account for anomalies in  
the scaling relationship between cluster X-ray luminosity and temperature \citep[][and references therein]{2004MNRAS.348.1078B}.
Particularly in low-mass clusters (i.e., clusters with low X-ray temperatures), an ``entropy" floor is observed; \citet{1999Natur.397..135P}
suggest that preheating of the ICM is required to explain this feature.  This entropy floor may be due to 
the heat injection by AGN in these clusters over time.

In general the X-ray surface brightness of a cluster ICM is well-fit by a $\beta$-model, which
is similar to a King profile.  However, some clusters display a strong peak
of X-ray emission at their centers.  Assuming that thermal bremsstrahlung is the dominant cooling
process in cluster cores, the cooling time for the X-ray emitting gas is shorter than
the Hubble time in these ``cool-core clusters".  Therefore, prodigious amounts of star formation ($\geq$100~M$_{\sun}$/year) are 
expected; however, spectroscopic X-ray observations fail to detect line emission from 
this gas at kT$<$1~keV \citep{2003ApJ...590..207P}, nor do optical observations reveal significant amounts of star formation
\citep[e.g.,][]{1989AJ.....98.2018M,1993AJ....105..417M}.  The lack of strong evidence for wide-spread star formation in cluster cores suggests
that a source of heat is preventing the ICM from cooling in cluster cores.  AGN are likely
contributors to ``feedback" processes that are required to prevent this large-scale cooling.

The connection between AGN and their surrounding environments, both small and large, proves to be a difficult problem to simulate.  
Many cosmological simulations are large volume boxes in which computational efficiency is balanced by the need to resolve small
structures, such as galaxies, and to track cluster scale-properties that are several orders of
magnitudes larger than galaxies.  Thus, in practice, some cosmological simulations inject a pre-determined 
amount of ``heat" at the centers of evolving clusters to simulate pre-heating of the cluster
cores.  Hydrodynamical simulations of the plasma-filled lobes created by the AGN in cluster BCGs can recreate the 
observed X-ray structures \citep[e.g.,][]{2004ApJ...611..158R} and 
halt the formation of strong cool-core clusters;
however, these simulations in general cannot efficiently distribute the heat throughout the cluster core
\citep[e.g.,][]{2007ApJ...671..171V}. One potential solution to this difficulty is that more than just the one AGN
in the BCG contribute to heating the ICM.  

Previous studies have found statistical excesses of radio galaxies \citep[e.g.,][]{2007ApJS..170...71L} and
X-ray point sources, or XPSs, in the vicinity of galaxy clusters \citep[e.g.,][]{2005ApJ...623L..81R, 2005AA...430...39C}.
The higher surface density of AGN in and around
clusters brings up some interesting questions: (1) Is the enhanced surface density simply a
direct result of a higher density of galaxies? (2) Are cluster AGN triggered into
activity as they pass through the dense cluster cores? (3) Is the cluster environment especially
suited in some specific way to create AGN activity?  Radio galaxies and XPSs are clearly associated with AGN
activity at the core of their host galaxies.  In order to
consistently study cluster AGN, observations obtained in at least three wavelength regimes are required
to detect these objects and to understand the energy feedback processes between them and the
surrounding ICM.

In \citet{hart09}, hereafter Paper I, we examined the cluster radio galaxy and XPS population in 
eleven \lowz\ clusters.  Within 1 Mpc from the cluster center, we detected 20 radio galaxies at \rlim\
and 8 XPSs at \lxbb$\geq$\xlim, all spectroscopically-confirmed as cluster members with host galaxies at L$\geq$L$^*$.
These cluster sources are 
hosted by cluster ``red sequence" (CRS) ellipticals, but are more centrally concentrated than the CRS galaxies as a whole.
The cluster XPSs in our low-$z$ cluster sample show little evidence for obscuration based both on their optical colors, which are consistent with
the CRS, and also on the absence of significant obscuration in the X-ray.  Based on the
optical and X-ray properties of these cluster XPSs, we suggested that these sources
are most similar to low-luminosity, high-energy peaked BL Lacertae (BL Lac) objects.
Based on the extrapolations of the radio luminosity function (RLF) of radio galaxies and the
X-ray luminosity function (XLF) of BL Lacs to lower radio and X-ray luminosities, we estimate that almost all
CRS galaxies are radio-loud at P$_{1.4GHz}$$\geq$10$^{21.4}$ \radiounits\ and X-ray-loud at \lxbb$\geq$10$^{40}$ \lum.  
Therefore, most CRS galaxies possess AGN jets which can deposit heat into the ICM as they move through the inner cluster
regions.  

Paper I concluded that the radio-loud CRS population between 10$^{21.4}$$<$P$_{1.4GHz}$$<$10$^{23.5}$ \radiounits\ can inject $\sim$55\%
of the total energy to the ICM if we assume a shallow scaling law of AGN jet power to radio luminosity, namely P$_{jet} 
\propto$ L$_r^{0.5}$ from \citet{2008ApJ...686..859B}.  More recently \citet{2010ApJ...720.1066C} extended the
\citet{2008ApJ...686..859B} sample to lower-power radio galaxies and measured a steeper slope for this
scaling relationship (P$_{jet} \propto$ L$_r^{0.7}$), which decreases the estimated heat input by low-luminosity AGN
to 30\%.  In Paper I the number of confirmed cluster radio galaxies is much larger than cluster XPSs and
so cluster radio galaxies will dominate the ICM heating regardless of the interpretation of XPSs. 

In this paper, we continue our study of cluster AGN begun in Paper I with the goal of
studying the evolution of radio galaxies and XPSs in galaxy clusters. 
We examine twelve \comboz\ clusters to detect radio galaxies at P$_{1.4GHz}$$\geq$5$\times$10$^{23}$ \radiounits\
and XPSs at L$_{0.3-8.0keV}$$\geq$5$\times$10$^{42}$ \ergs\ within a 1 Mpc projected radius of the cluster cores to determine
the distribution of AGN luminosities, their locations relative to the ICM, and their numbers and luminosities as a function of redshift. 
In \S\ref{sec:p2_rtc} we describe the cosmological simulations that guide the selection of our cluster sample,
uniquely designed to permit an accurate evolutionary study.  In \S\ref{sec:p2_obs}, we detail the X-ray, radio, and optical
datasets used to measure the ICM X-ray temperatures, to identify potential cluster XPSs and radio sources, and to 
obtain optical colors of typical cluster galaxies and AGN host galaxies.  In \S\ref{sec:p2_obs_results}
we present the sample of potential cluster radio galaxies and XPSs, summarize their optical properties 
through composite color-magnitude diagrams, and describe the typical colors of
cluster radio galaxies and XPSs relative to the CRS.  In \S\ref{sec:p2_rg}--\ref{sec:p2_rlf} we
present the RLF of cluster radio galaxies out to $z$$=$1.1 and compare it to
other cluster RLFs.  In \S\ref{sec:p2_xps} we examine the XPS population in our cluster fields out to $z$$=$1.2 to determine if
any statistical XPS excess above the expected background is observed and estimate the X-ray active fraction of CRS galaxies.
In \S\ref{sec:p2_implications} we discuss our results and in \S\ref{sec:p2_conclusions} we summarize our conclusions from this
study.  X-ray luminosities are defined for the rest-frame (0.3--8.0 keV) bandpass, unless otherwise noted.  
Radio powers are defined for a 1.4 GHz rest-frame frequency.  Throughout this paper, 
we use H$_0$ = 70~km~s$^{-1}$~Mpc$^{-1}$, $\Omega_{\Lambda}=0.70$, and $\Omega_{M}=0.3$.

\section{The ``Road to Coma'' Cluster Sample}
\label{sec:p2_rtc}

Hierarchical structure formation predicts that the highest density fluctuations in the
primordial universe grow and assemble into the large galaxy clusters in the present
epoch \citep{1999coph.book.....P}.  As clusters accrete gas and galaxies, the cluster dark matter halos
increase in mass.   The evolution of such massive structures has been
simulated (e.g., \citealt{2002ApJ...573....7E}; Millennium Simulation, 
\citealt{2005Natur.435..629S}; ENZO simulations, \citealt{2007ApJ...671...27H}; \citealt{2007ApJ...668....1N})
to follow the physics of the intracluster gas, to predict cluster observables, and 
to suggest the need for AGN feedback.

In particular, AGN feedback is difficult to simulate in large-scale cosmological
simulations because of the unknown range of AGN powers and duty cycles, energy transport 
mechanisms through the ICM, and feedback processes within the host galaxies themselves 
and their surrounding environment \citep[e.g.,][and references therein]{2006MNRAS.370..645B}.  Recent simulations of AGN feedback in galaxy cluster
environments include the evolution of over-pressurized ``bubbles" in the ICM, presumably
inflated by the radio-emitting lobes associated with the radio-loud AGN in the BCG 
\citep[e.g.,][]{2005ApJ...630..740B, 2009MNRAS.395.2210B}.  These simulations successfully recreate
X-ray features observed in nearby clusters like the Perseus cluster \citep{2002MNRAS.331..369F}.  Other
numerical approaches include detailed AGN jet simulations of radio sources both at the cluster 
center and moving through the ICM \citep[e.g.,][]{2006MNRAS.373L..65H,2010ApJ...710..180O}.
Observational details of the radio galaxy population is crucial to compile 
AGN-related input parameters that can be utilized in these large-scale cosmological simulations.

Observationally, X-ray surveys of galaxy clusters (e.g., {\it Einstein} Extended Medium Sensitivity Survey (EMSS), 
\citealt{1994ApJS...94..583G}; Wide Angle ROSAT Pointed Survey (WARPS), \citealt{2002ApJS..140..265P};
Massive Cluster Survey (MACS), \citealt{2001ApJ...553..668E}; ROSAT Deep Cluster Survey (RDCS), \citealt{1998ApJ...492L..21R})
are flux-limited surveys that preferentially select X-ray luminous (and thus high mass)
galaxy clusters at high-$z$.  Additionally, depending on the selection method used to identify
the extended emission of the ICM, these surveys also
tend to identify clusters with centrally peaked X-ray emission typically associated
with cool-core clusters \citep[specifically the EMSS;][]{1990MNRAS.244...58P,2002ApJ...566..744L}.
Therefore, comparisons using flux-limited samples may significantly bias cluster evolutionary studies.

Simulations of cluster evolution indicate that, while clusters experience a
variety of formation histories, the mean progenitor mass increases with time
in a regular manner \citep{2002ApJ...568...52W}.  Evolutionary studies in
galaxy clusters can be improved by choosing clusters that are 
statistically likely to evolve into a particular system at low redshift. This approach
reduces the inherent bias in choosing luminous X-ray-selected clusters
from flux-limited surveys at high-$z$ and using their galaxy and AGN content to
compare to clusters at low-$z$.
Thus, to minimize the potential biases present in many sample selections, we utilize the results of cosmological simulations
from \citet{2001ApJ...555..597B} to select galaxy clusters at different redshifts, which permit a consistent study of the 
evolution of galaxies or AGN within these dense environments.

The details of the numerical nature of the cosmological simulations can be found 
in \citet{2001ApJ...555..597B}, but we summarize the major points here. 
Smooth particle hydrodynamics (SPH) and N-body codes are incorporated into a
Lagrangian code P3MSPH \citep{1988MNRAS.235..911E} to track the gaseous components,
as well as the collisionless particles (i.e., dark matter) in these simulations.
These simulations use a standard $\Lambda$CDM cosmology with $\Omega_{M}$=0.3, $\Omega_{\Lambda}$=0.7, $\sigma_8$=1.0 and
$h$=0.8, where the Hubble constant, H, is defined as 100 $h$~\kms\ Mpc$^{-1}$ and $\sigma_8$
is the normalization of the matter power spectrum on 8~$h^{-1}$~Mpc scales.
The simulations were run in a box 300 $h^{-1}$ Mpc on a side and
the most massive cluster dark matter halos (M$>$10$^{15}$M$_\sun$) were 
tracked throughout the simulation.  To match the observed scaling laws of
cluster X-ray temperatures and luminosities, which deviate from the
expected scaling due to pure gravitational collapse models, these gaseous components are
``preheated" by specifying an initial hot temperature for the baryons
at earlier epochs \citep{2001ApJ...555..597B}.

\begin{figure*}
        \begin{center}
        \includegraphics[scale=0.7]{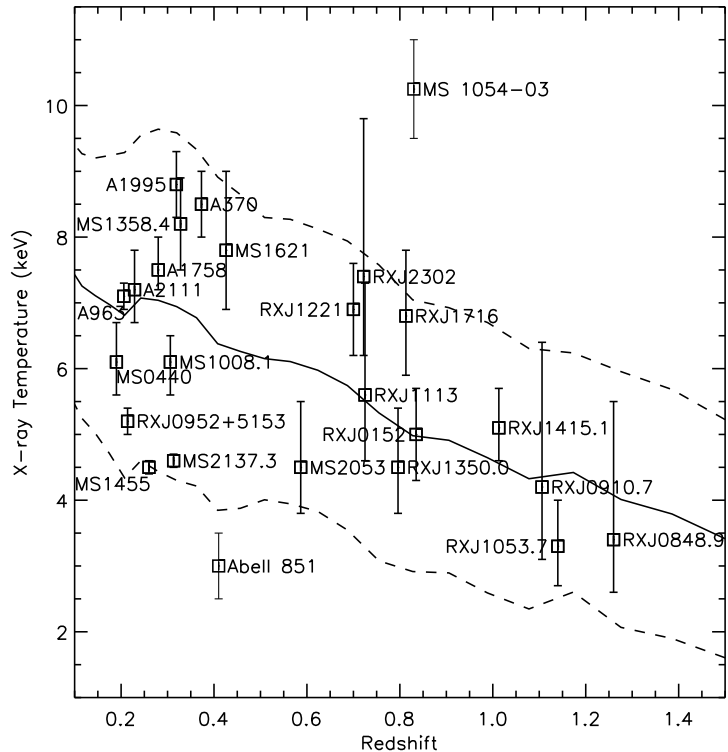}
        \caption{
        Predicted ICM temperature versus redshift of Coma Cluster progenitors. The simulated ICM temperatures of
        massive cluster halos are tracked in cosmological simulations.  For each cluster halo, the ratio of
        T$_z$/T$_0$ is normalized to the current ICM temperature of the Coma Cluster.  The mean value of
        T$_z$ is displayed as the solid line, while the 25th and 75th percentiles of that
        mean are displayed as dashed-lines.  There are eleven clusters between \lowz\ that comprise the low-$z$ cluster sample
        of Paper I.  The remaining twelve clusters between \comboz\ are the high-$z$ sample discussed in this current study.  At the
        present epoch the Coma Cluster X-ray temperature is 8.2 keV \citep{2008ApJ...687..968L}.  Note that the x-axis
        begins at $z$$=$0.1 and not $z$$=$0.
        }
        \label{fig:cluster_sim}
        \end{center}
\end{figure*}

Sixty-eight massive cluster halos chosen from these large N-body simulations were re-simulated with gas dynamics on a 
higher resolution grid. Emission-weighted temperatures of their ICM were extracted at various epochs (T$_z$) and then
compared to their ICM temperatures at the current epoch (T$_{0}$) in the simulation.  
We identify Coma Cluster progenitors between 0.2$<$$z$$<$1.2 in the following manner.  
Using the results of the \citet{2001ApJ...555..597B} simulations above,
the ratio of T$_{z}$/T$_0$ for the fifteen most massive (M$>$10$^{15}$M$_\sun$) cluster halos
as a function of redshift is normalized to the bulk ICM temperature of the present-day Coma Cluster
\citep[kT=8.2$\pm$0.2;][]{2008ApJ...687..968L}.  
Figure~\ref{fig:cluster_sim} displays the median temperature of these simulated progenitors
as a function of redshift (solid line) and the 25th and 75th percentiles about this distribution (dashed lines).
At $z$$=$0.3, the median ICM temperature of a Coma Cluster progenitor is predicted to be 7.0$\pm$2.6 keV.  The
temperature spread reflects the variation in the formation histories of Coma Cluster mass-scale objects in these
simulations.  At $z$$=$1.0, the permitted temperature range for our sample slowly decreases to 4.0$\pm$2.0 keV.

We identify potential ``Road to Coma" clusters as having bulk ICM X-ray temperatures that
fall within the temperature distribution displayed in Figure~\ref{fig:cluster_sim}.  
Observationally, nearly all cool-core clusters \citep{1990AJ.....99...14B} host luminous radio sources in their BCGs.  
Therefore, their exclusion in any cluster sample would decrease the numbers of luminous radio galaxies.
In the future this selection method could be improved by using only those simulated clusters which do {\it not} 
develop cool cores in the current epoch to better match the Coma Cluster.  

The Coma Cluster (or Abell 1656) is a prototypical massive cluster in the present-epoch 
at $z$$=$0.023, with two central dominant galaxies (NGC 4874 and NGC 4889)
and a galaxy population dominated by passive ellipticals and S0s.
XMM-Newton observations of the Coma Cluster out to the virial radius do not reveal any strong XPSs 
with L$_{0.5-2.0keV}$$>$10$^{41}$~\ergs~\citep{2004A&A...419...47F}.  For an off-center
region 1 Mpc from the cluster core, \citet{2006ApJ...643..144H} did not detect sources
with L$_{0.5-8.0keV}$$>$10$^{41}$~\ergs\ either and report that the number of detected XPSs is consistent
with the background estimates.  NGC 4874, the western-most central dominant galaxy is detected
at P$_{1.4GHz}$=2.4$\times$10$^{23}$~\radiounits\ and NGC 4869, part of an in-falling group to the
southwest of NGC 4874, is also detected at P$_{1.4GHz}$=4.8$\times$10$^{23}$~\radiounits~\citep{2009AJ....137.4450M}.
Thus, even for this nearby, massive prototypical cluster, the Coma Cluster does not
host powerful radio sources at P$_{1.4GHz}$$>$10$^{24}$~\radiounits, nor XPSs at \lxbb$>$10$^{42}$~\ergs.
In Paper I, we detected 20 radio galaxies at P$_{1.4GHz}$$\geq$3$\times$10$^{23}$~\radiounits\ and 8 XPSs at \lxbb$\geq$\xraylimit,
yielding 1.8$\pm$0.5 radio galaxies and 0.7$\pm$0.4 XPSs per cluster.  Therefore, detections of
two radio galaxies and no XPSs in the Coma Cluster at these power levels are consistent with the expectations 
based on our previous low-$z$ work in Paper I.

Table~\ref{tab:p2_rtc_sample} lists the full \rtc\ cluster sample between 0.2$<$$z$$<$1.2 that is 
displayed in Figure~\ref{fig:cluster_sim}. We broadly define three redshift bins that are used throughout 
this paper (summarized in Columns 1--2 of Table~\ref{tab:p2_rtc_sample}).  The low-$z$ redshift bin
includes eleven clusters between \lowz, and has been presented and detailed in Paper I.
The mid-$z$ redshift bin includes five clusters between
\midz, while the high-$z$ redshift bin includes seven clusters between \highz.
Columns~3--5 of Table~\ref{tab:p2_rtc_sample} list the clusters, redshifts, and ICM bulk X-ray 
temperatures plotted in Figure~\ref{fig:cluster_sim}, respectively.  The
details of the X-ray temperature determinations, measured so as to match most closely the
temperatures determined in the simulations, are described in \S\ref{subsec:p2_obs_xray}.
Hereafter, we shorten the names of clusters to the first several letters and first four numbers of the
full cluster name when referencing them throughout the text.

Note that Abell 851, a well-studied, optically-selected galaxy cluster at $z$$=$0.41, is well below the ``Road to Coma"
\citep[kT$=$2.9$^{+1.3}_{-0.8}$ keV;][]{1996AA...313..113S} and is predicted to evolve into a cluster of lower mass compared to the Coma Cluster.  
Abell 851 has a large blue, star-forming galaxy fraction \citep{1994ApJ...435L..23D}. A rapid evolution ($\sim$ few Gyr)
of the blue galaxy population in Abell 851 must take place to be consistent with the
galaxy population of Coma-like clusters in the current epoch.  
However, when the star-forming galaxy population of Abell 851 is compared with 
the Virgo Cluster, very little evolution in the galaxy population is required.
On the other-hand, MS 1054-03 ($z$$=$0.83) lies well above 
the ``Road to Coma" and it is predicted to evolve into a cluster much more massive than Coma.  Therefore, evolutionary
studies that utilize MS1054 as a proto-typical cluster at high-$z$ can be skewed in the opposite direction; e.g., MS1054
has an early-type galaxy fraction similar to the Coma Cluster, suggesting little evolution in galaxy types
\citep{2006ApJ...642L.123H} between $z$$=$0.8 and the present epoch.  It is clear that the picture of cluster
galaxy evolution looks very different if one chooses Abell 851 or MS1054 to compare against the Coma Cluster.
These two examples provide anecdotal evidence supporting our selection procedure.

\section{Multiwavelength Observations}
\label{sec:p2_obs}

\subsection{\chandra\ Imaging Spectroscopy}
\label{subsec:p2_obs_xray}

Archival \chandra\ observations of the twelve 0.4$<$$z$$<$1.2 clusters 
in our sample were re-processed using CIAO version 4.1.2 and CALDB 4.1.3.  These
clusters have published bulk ICM temperatures that fall along the ``Road to Coma" in Figure~\ref{fig:cluster_sim}.
Between $z$$=$0.4 and $z$$=$1.2, a circular region within a 1 Mpc projected radius of the cluster center 
covers between 40\% and 20\% of one ACIS chip, respectively.
For this redshift range, the extended emission of the ICM falls on the ACIS-I3 chip, except for RXJ0849 and RXJ1053, which fall on the
ACIS-I2 and ACIS-S3 chips, respectively.
We followed the typical pipeline processing of event-1 files by removing bad pixels, afterglow pixels, and streaking
patterns.  The most recent calibration files for charge transfer inefficiency and
time-dependent gain were also applied.  The final event file was filtered on event
status and grade.  Time periods with background count rates $>$3$\sigma$
above the observed mean were flagged and removed from further analysis.

Low-resolution ICM spectra were extracted from regions within 1 Mpc (projected) radius from
the cluster X-ray emission centroid, except for six $z$$>$0.8 clusters where the ICM
spectra were extracted from within 500 kpc. (The BCG is identified as the brightest galaxy
located closest to the X-ray emission centroid.)
These extraction radii were chosen to be as large as possible based on photon statistics in order
to compare them with the temperatures derived from the simulations.
Background spectra from within an equivalent extraction radius were obtained from the same 
chip as the cluster detection or on one of the adjacent ACIS chips.  The ICM spectra were binned to
have a minimum of 20 counts per bin, then modeled using XSPEC \citep{2007A&A...461...71P} with 
the MEKAL model, foreground extinction held fixed at the Galactic hydrogen column at the
cluster location \citep{1990ARA&A..28..215D}, and a constant 0.3 solar metal abundance.
Column 4 of Table~\ref{tab:p2_rtc_sample} lists the modeled temperatures with 1$\sigma$
error bars.  To identify the cluster X-ray emission centroid, the ICM emission is modeled with a 
2-dimensional Gaussian profile.  This centroid is identified as the cluster center and a region 
within 1 Mpc of this center is our analysis region.  

We used the CIAO routine {\it wavdetect}\footnote{http://cxc.harvard.edu/ciao/ahelp/wavdetect.html}, a
detection routine that correlates each image with a Mexican-hat wavelet at
different pixel scales (1 to 16) to identify the XPSs in the \chandra\ observations.
In our cluster fields at $z$$>$0.4, the maximum off-axis angle location for a
{\it wavdetect} source is 4--5\arcmin\ with a minimum source count of 5--9. 
\citet{2007ApJS..169..401K} performed extensive simulations on the detection
probability of {\it wavdetect} selection of point sources in 
\chandra\ ACIS observations.  Because the PSF
increases in size with increasing off-axis angle from the aimpoint of the
\chandra\ observations, these authors simulated sources with counts between 3 and 3000, placed
them 0--8\arcmin\ from the aimpoint, and ran {\it wavdetect} with the {\it
sigthresh} parameter set to 1$\times$10$^{-6}$. In 2\arcmin\ square
images containing simulated sources that incorporated the different parameter
space above, \citet{2007ApJS..169..401K} report a false detection rate of
$\sim$1\% overall. Referring to Figure 13 in \citet{2007ApJS..169..401K},
the false detection rate is $<$0.8\% for sources $<$5\arcmin\ from the
aimpoint and $<$1.7\% for a minimum source count of 5.  Therefore, we expect
only a small fraction of the {\it wavdetect} sources found by us to be spurious, 
or ``false", detections.

Each {\it wavdetect} source identification was inspected to check source
authenticity.  Total broadband counts (0.3--8.0 keV) were extracted within a 95\% encircled energy radius
estimated for a monochromatic energy source of 1.5 keV at the observed off-axis 
angle\footnote{http://cxc.harvard.edu/cal/Hrma/psf/ECF/hrmaD1996-12-20hrci\_ecf\_N0002.fits}.
Background counts were extracted within 2-3 times the source extraction radius.
XSPEC MEKAL modeling of a typical power-law spectrum
with $\Gamma$=1.7 (N$_{E}\propto E^{-\Gamma}$) was employed to
convert source net counts to unabsorbed X-ray flux.  Table~\ref{tab:p2_rtc_xray} lists
the X-ray observations for our $z$$>$0.4 clusters, including coordinates of the
ICM emission centroid (i.e., the ``cluster center"), the \chandra\ observation ID, the ACIS chip with the ICM detection, and
the cleansed exposure time.
Column 7 of Table~\ref{tab:p2_rtc_xray} lists the estimated flux limit for 
each observation, determined as the flux of an XPS located 1 Mpc from the 
cluster center with a count rate 3$\sigma$ above the background noise.  Column 8 of
Table~\ref{tab:p2_rtc_xray} is the K-corrected X-ray luminosity limit of an XPS 
at the cluster redshift (K-correction assumes a power-law spectrum with $\Gamma$=1.7).
Note that we include RXJ0152 here for completeness, although this cluster is not included in 
the XPS statistics later in this paper because the XPS limiting luminosity of
the available \chandra\ images is $\ge$5.0$\times$10$^{42}$ \ergs.

The limiting X-ray luminosities detected in our low-$z$ cluster sample from Paper I is
L$_{0.3-8.0kev}$ $\geq$ 10$^{42}$ \ergs. Due to the prohibitively long exposure
times required to detect equally luminous XPSs at high-$z$, there are few
deep archival \chandra\ cluster observations at $z$$>$0.8 that have published
temperatures identifying them as Coma Cluster progenitors.  As a result
the archival \chandra\ images do not uniformly reach down to the same
limiting luminosity of $\sim$10$^{42}$ \ergs. The limiting X-ray
luminosity limit for our mid-$z$ clusters is
2.2$\times$10$^{42}$ \ergs\ and 5.0$\times$10$^{42}$
\ergs\ for our \highz\ cluster sample, or roughly a factor of 2 increase
from one redshift bin to the next. In order to compare our high-$z$
cluster XPSs to the mid-$z$ and low-$z$ cluster XPSs, the
X-ray luminosity limit across the entire sample is set at
5.0$\times$10$^{42}$ \ergs. We note that after we apply a
luminosity cut at this value, the minimum source
counts range from 9--30 counts in eleven of twelve \highz\ cluster fields listed in
Table~\ref{tab:p2_rtc_xray} (excludes RXJ0152).  This corresponds to a false detection
probability between 0.5--1.5\%.  We detect 86 XPSs in these 11 fields, of
which $<$2 sources are expected to be false detections.  
Given that only a few percent of the $<$1 Mpc field-of-view (FOV) of our analysis falls within 5\arcsec\
of a $\geq$L$^*$ galaxy (at the cluster redshift), we expect no ``false positives" in our final
XPS sample.

\subsection{Radio Imaging}
\label{subsec:p2_obs_radio}

In September/December 2006 we obtained Very Large Array (VLA) continuum observations at 1.4 GHz (50 MHz bandwidth) for five clusters 
(RXJ1221, RXJ1113, RXJ1350, RXJ1716, and RXJ0910 under VLA Program AS873).
RXJ1221, RXJ1350 and RXJ1716 were observed for 2 hours each in B-array, RXJ1113 for 3.25 hours in BnA-array, and RXJ0910 was 
observed for 8 hours in B-array.  The duration of a typical target scan was 15 minutes, bracketed by 1 minute scans of a nearby
phase calibrator.  A flux calibrator (e.g., 3C286) was observed for 3 minutes at the beginning and the end of each observing session.
We used the National Radio Astronomy Observatory (NRAO)\footnote{The National Radio Astronomy                                       
Observatory is a facility of the National Science Foundation operated under cooperative agreement by Associated Universities, Inc.}
Astronomical Imaging Processing System (AIPS), version 31DEC09, in the usual manner
to flag, calibrate, transform, and clean the images, so that radio sources could be detected manually. 
Source flux densities are estimated with the AIPS tasks TVWIN and IMEAN.
The typical 3$\sigma$ limit of these observations is 0.1--0.2 mJy. 

The 1.4~GHz radio maps and source identifications for two EMSS clusters, MS1621 and MS2053, 
are published in \citet{1999AJ....117.1967S}. Radio maps from the Faint Images of the Radio Sky at 20cm \citep[FIRST; ][]{1995ApJ...450..559B} 
are also available for MS1621.  Eric Perlman kindly 
provided the 1.4~GHz radio maps and source identifications for two WARPS clusters (RXJ2302 and RXJ1415).
The 1.4~GHz radio map of RXJ1053 was provided by Chris Carilli at NRAO.  This observation was obtained 
in the spectral-line mode in the A-configuration under the VLA Program AC587 and it is the deepest radio image 
in our sample with a 3$\sigma$ rms of 0.03 mJy.
For the remaining clusters, RXJ0152 and RXJ0849, we utilized 1.4~GHz maps retrieved from
the VLA archives.
Higher resolution A-array maps for two clusters (MS1621 and RXJ0152) were reduced in a similar manner.  
For RXJ0152, a lower-resolution B-array map does not go deep enough for our purposes, but the A-array map does.  Therefore,
we use the A-array map to identify radio sources, but use the B-array map to measure a source flux, so there is less missing
flux for the extended sources.

Table~\ref{tab:p2_rtc_radio} lists the VLA 1.4~GHz radio observations for our \comboz\ cluster sample, 
including the VLA program ID and array configuration, the flux density limit, and the
K-corrected radio power and luminosity limit for a source located at the cluster redshift.  
The flux density limit is the 3$\sigma$ rms in mJy/Beam at the edge of our survey region 
(i.e., 1 Mpc from the cluster center).  The radio power limits for our mid-$z$ clusters are 
similar to our low-$z$ cluster sample presented in Paper I (P$_{1.4GHz}$$\geq$3$\times$10$^{23}$ \radiounits). 
However, for our high-$z$ cluster sample, the radio power
limits are slightly higher at P$_{1.4GHz}$$\geq$5$\times$10$^{23}$ \radiounits, and thus we
set the latter value as the radio power limit for our survey in this paper.  RXJ0910 and RXJ0849 are excluded from
the subsequent radio-related analyses, as the 1.4~GHz map 3$\sigma$ rms falls 2--3 times above our selected radio limit. 
Also, the redshift limit of our cluster radio galaxies drops from $z$$=$1.26 to $z$$=$1.13 when we exclude RXJ0849.
The X-ray observations of these two clusters are within our X-ray luminosity limits (refer to Column 8 of Table~\ref{tab:p2_rtc_xray})
and hence are included in later X-ray-related statistics. Thus, our radio galaxy sample extends to $z$$=$1.1, while our XPS sample extends to $z$$=$1.2.

\subsection{Optical Imaging}
\label{subsec:p2_obs_optical}

The color-magnitude diagrams of clusters generally display a tight color sequence for the
cluster ellipticals, i.e., the CRS.  Two-color images of our cluster fields provide a simple classification 
scheme for cluster galaxies, particularly if the chosen filters span the \cahk\ break at the cluster 
redshift to give the maximum contrast to the CRS.
Table~\ref{tab:p2_rtc_optical} lists the optical colors utilized for each $z$$>$0.4 cluster,
the observed mean color of CRS galaxies and the 3$\sigma$ spread (with the analysis described in 
\S\ref{subsec:p2_obs_results_crs}), the expected color of a passive elliptical, as estimated from 
Tables~6--8 in \citet{fukugita95}, the 3$\sigma$ limiting magnitude of the optical images, and 
the source of the survey images.  The optical imaging data are summarized below.

Archival Hubble Space Telescope (\hst) Advanced Camera for Surveys (ACS) images
of four clusters in Table~\ref{tab:p2_rtc_optical} (RXJ0152, RXJ1415, RXJ0910, and RXJ0849) were obtained from the 
Multimission Archive at Space Telescope Science Institute (MAST).  These clusters were observed with the F775W 
(Sloan {\it i}) and F850LP (Sloan {\it z}) filters, and RXJ0152 was also observed with the F625W (Sloan {\it r}) filter.
The initial science goals from these imaging campaigns include the identification of the CRS in 
RXJ0849 \citep[Program 9919,][]{2006ApJ...644..759M}, the fraction of early-type galaxies in the cluster cores of RXJ0152
\citep[Program 9290,][]{2006ApJ...642L.123H}, and the detailed morphological classification of
galaxies in RXJ0152, RXJ0910, and RXJ0849 \citep{2005ApJ...623..721P}.
These two-color images, which cover the inner 1 Mpc radius from the cluster center, were reprocessed with
Multidrizzle\footnote{http://stsdas.stsci.edu/multidrizzle/} to correct image distortions, to identify and
remove cosmic rays, and to combine dithered images from multiple visits.  
For the remaining 8 clusters listed in Table~\ref{tab:p2_rtc_optical}, HST archival imaging does not fully survey
1 Mpc from the cluster center, or the images do not survey down to the required limiting magnitude, and/or the imaging does
not exist in the requisite filters for this study.

For RXJ1053, two-color optical images (Sloan {\it i} and {\it z}), obtained with Suprime-Cam on the Subaru 8.2m telescope at 
Mauna Kea, were provided by Yasuhiro Hashimoto.  
For MS1621, two-color optical images (Sloan {\it i} and {\it z}) were obtained from the Sloan Digital Sky Survey Data Release 6 
\citep[SDSS DR6;][]{2008ApJS..175..297A}.
In Spring 2006 we obtained Sloan ({\it r,i}) images of RXJ1221, RXJ1350, and RXJ1716 with SPIcam on the 
Astrophysical Research Consortium (ARC) 3.5m Telescope
at the Apache Point Observatory (APO).  The FOV of SPIcam is 4.8$\times$4.8 arcmin, which adequately covers the entire 1 Mpc radius region for
these three clusters at 0.7$<$$z$$<$0.8.  We used the Image Reduction and Analysis Facility (IRAF), v2.14.1, from the
National Optical Astronomy Observatory (NOAO) in the usual manner to bias-subtract and flat-field individual images.  Images were aligned and subsequently stacked
with the IRAF task IMCOMBINE to produce one final image.  United States Naval Observatory (USNO) B catalog stars
\citep{2003AJ....125..984M} were used to correct the image astrometry.
In-field SDSS stars were used to determine the photometric
zero-points and color terms for images of RXJ1221 and RXJ1350, while off-field SDSS stars were used to calibrate RXJ1716.
Using the tabulated values of E(B-V) retrieved from the NASA/IPAC Extragalactic Database (NED) and based on the infrared dust maps of \citet{1998ApJ...500..525S}, 
source magnitudes are corrected for the foreground Galactic extinction in each filter, by using the empirical relationships
presented by \citet{2002AJ....123..567S}.  For these clusters, E(B-V)$\leq$0.08.

For the APO, Subaru, and \hst\ images, we used the Picture Processing Program \citep[PPP; ][]{1991PASP..103..396Y} to detect and classify
objects as stars or galaxies.  PPP is a robust detection algorithm, especially in crowded fields, such as
galaxy clusters.  Additionally, total source magnitudes are calculated using optimal extraction radii that are based
on photometric growth curves.  These PPP magnitudes have been shown to be highly accurate measurements 
\citep[$\pm$0.03 magnitudes,][]{1996ApJS..102..269Y} of total galaxy magnitudes down to the limiting magnitude of the image.
For the \hst\ images, the photometric
zeropoints were obtained from the \hst\ ACS website\footnote{http://www.stsci.edu/hst/acs/analysis/zeropoints}.
For the APO and Subaru images, the photometric zeropoints and color terms were determined using SDSS DR6 objects 
in the image's FOV.

Two-color optical images (Sloan {\it r} and {\it i}) and photometry of MS2053, RXJ2302, and RXJ1113 were obtained 
from the Chandra Multiwavelength Project (ChaMP) \citep{2004ApJS..150...43G}.
ChaMP is a wide-area ($\sim$ 14 deg$^2$) serendipitous survey of 100 \chandra\ observations compiled 
to detect XPSs at F$_{0.3-8.0keV}\geq$10$^{-14}$-10$^{-15}$ \flux\ that are used to study the evolution of luminous 
AGN \citep[e.g.,][]{2004ApJ...600...59K,2009ApJ...705.1336C,2010ApJ...723.1447H}. 
Wide-field optical images of ChaMP clusters were obtained with the Mosaic 
CCD cameras \citep{1998SPIE.3355..577M} on the NOAO 4-meter telescopes.
Details of the image reductions and photometry with SExtractor \citep{1996A&AS..117..393B} can be found in \citet{2004ApJS..150...43G}.

\section{Observational Results}
\label{sec:p2_obs_results}

\subsection{Cluster Red Sequence and Composite Color-Magnitude Diagrams}
\label{subsec:p2_obs_results_crs}

The most massive, and hence most luminous, early-type cluster galaxies probably formed
at $z$$>$2.0 \citep[e.g.,][]{1998ApJ...492..461S,2006ApJ...644...30B}.  With similar colors down to 
M$^*$+3 \citep{1997ApJ...483..582E}, these objects form a CRS in color-magnitude diagrams.
Several optically-selected cluster surveys 
(e.g., Red-sequence Cluster Survey (RCS), \citealt{2005ApJS..157....1G};
MaxBCG method, \citealt{2007ApJ...660..239K}) utilize this ubiquitous feature 
to photometrically identify clusters in optical surveys.  Color comparisons between
objects with or without CRS colors provide simple galaxy classifications, i.e.,
objects bluer than CRS galaxies could be late-type cluster galaxies or foreground
objects along the line-of-sight.  Objects with redder colors than CRS galaxies
could be background objects or highly-reddened cluster galaxies.

In Paper I, we examined radio galaxies and XPSs in M$_r$$<$-20.8 galaxies ($\approx$ L$^*$ galaxies).
Here, we assume pure luminosity evolution (PLE) of the CRS galaxies to limit our study to
AGN hosts with M$_r$$\leq$-21.2 for \midz\ clusters or 
M$_r$$\leq$-21.6 for \highz\ clusters.  These magnitude limits correspond approximately to L$^*$ once the PLE corrections of $\Delta$0.1 magnitudes 
per $\Delta$0.1 redshift \citep{2007ApJ...655...30V} are applied.
Foreground and background galaxies are statistically subtracted from each cluster by using the 
Sloan {\it r}-band background counts from \citet{2001AJ....122.1104Y} for clusters with a limiting m$_r$$<$21.0 or 
using the COSMOS {\it i}-band background counts from \citet{2007ApJS..172...99C} with a limiting m$_i$$<$26.0.
We determine the mean CRS color by fitting a bimodal Gaussian distribution
to both the red and blue galaxy populations as a function of color (similar to Paper I).
After identifying galaxies within 3$\sigma$ of the mean CRS color, we define the CRS more precisely by fitting 
a line to the observed color versus magnitude of these sources and thus verifying the well-known ``tilt" of the 
CRS in cluster color-magnitude diagrams \citep[e.g.,][]{1977ApJ...216..214V}.  

For the four highest redshift clusters in our sample (RXJ1415, RXJ0910, RXJ1053, and RXJ0848) no statistical subtraction of background galaxies was applied
because the estimated number of background objects in a given magnitude bin in a small region of sky exceeds the observed galaxy 
counts in our optical images.  In a weak-lensing study of RXJ0849, \citet{2006ApJ...642..720J} found no significant
excess in the background galaxy counts due to cluster galaxies at 24.0$<$$m_{z}$$<$27.0.  Similarly, \citet{2006ApJ...639...81M}
estimated that the background contamination is small for early-type galaxies (i.e., CRS galaxies) in RXJ0910.  Additionally, the optical magnitudes of the 
BCGs in RXJ0910 and RXJ0849 are {\it i}=22.9 and {\it i}=23.5, respectively, and potential CRS galaxies are expected to be fainter than this magnitude.
The CRS colors of RXJ0910 and RXJ0849 are $\approx$0.5 magnitudes redder than the typical SDSS field galaxies, which have
({\it i}-{\it z}) colors between 0.0 and 0.5 \citep{1999ApJS..123..377N}.  
In the four highest redshift clusters listed above we assume that any galaxy with CRS colors is a likely cluster member. 
Therefore, we fit a single Gaussian component to the expected CRS galaxy population in the color-magnitude diagrams of these
four clusters to determine the mean color of typical CRS galaxies.

For our 0.4$<$$z$$<$1.0 clusters, we do not expect over-subtraction of background galaxies to be a potential problem because of the
field-to-field differences in the background galaxy counts due to cosmic variance.  In their study of the global environments around 
bright quasars, \citet{1986ApJS...62..681Y} estimated the uncertainty in the background galaxy counts using several control fields and
find it to be 1.3 times larger than Poisson errors alone.  These authors attribute the increased variance to intrinsic clustering of galaxies.
Therefore, we expect the error on the cumulative CRS galaxy number for our 0.4$<$$z$$<$1.0 clusters to be larger than pure counting statistics alone. 
For simplicity we assume Poisson errors on our estimates of CRS galaxy counts.

\begin{figure*}
\begin{center}
\includegraphics[scale=0.7]{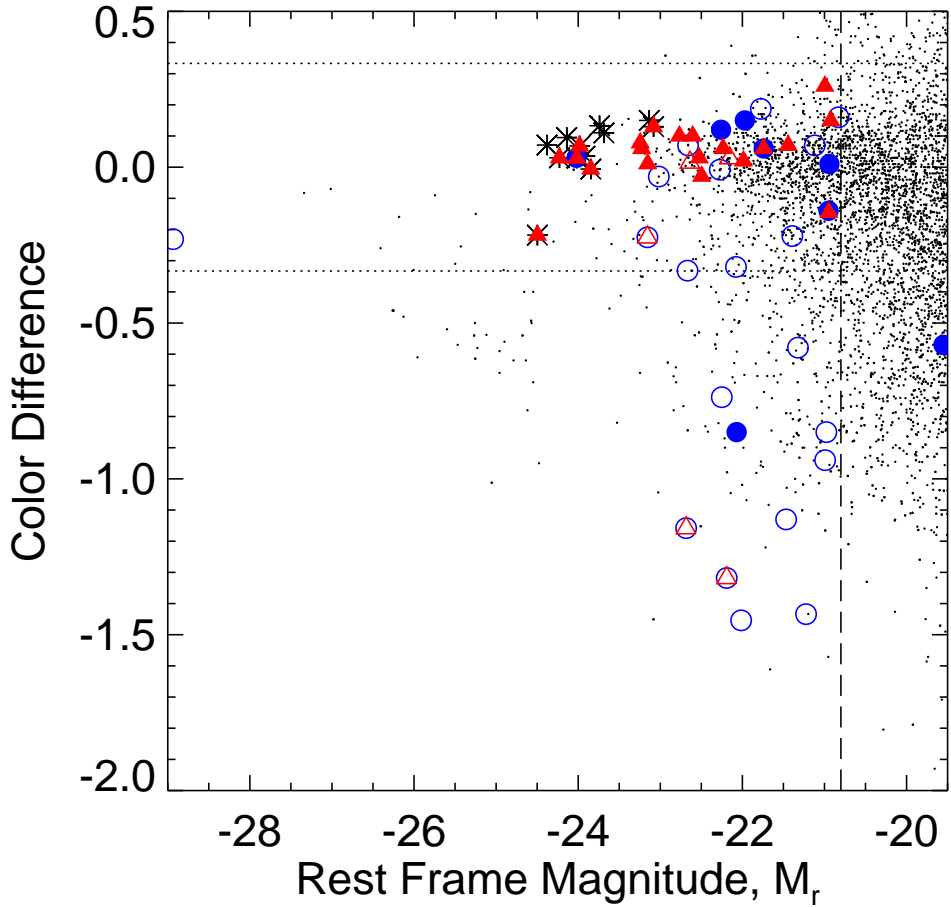}
\caption{
Composite color-magnitude diagram for eleven \lowz\ clusters.
The color difference is the magnitude difference between the
object's observed \gminusr\ color and the mean CRS \gminusr\ color of the cluster, with the
exception of MS2137, A1995, MS1358, and A370 for which the color difference is \rminusi.
Small black dots are optical sources located within a 1 Mpc projected
radius of the cluster X-ray emission centroid. Filled red triangles (blue circles) are host galaxies of cluster radio galaxies (XPSs).
Open red triangles (blue circles) are host galaxies of non-cluster radio sources (XPSs).
Brightest cluster galaxies are identified with an asterisk.  The horizontal short-dashed lines mark $\pm$0.33 mags from the
mean CRS color of each cluster, which is the maximum width of the CRS population in these clusters (see Table~\ref{tab:p2_rtc_optical}).
The vertical long-dashed line marks L$^*$ galaxies with M$^{*}_{r}$$\approx$-20.8.  All AGN at L$\geq$L$^*$
have been spectroscopically confirmed. This is a modified version of Figure~2 in Paper I.
}
\label{fig:p2_cmd_low_z}
\end{center}
\end{figure*}

\begin{figure*}
\begin{center}
\includegraphics[scale=0.7]{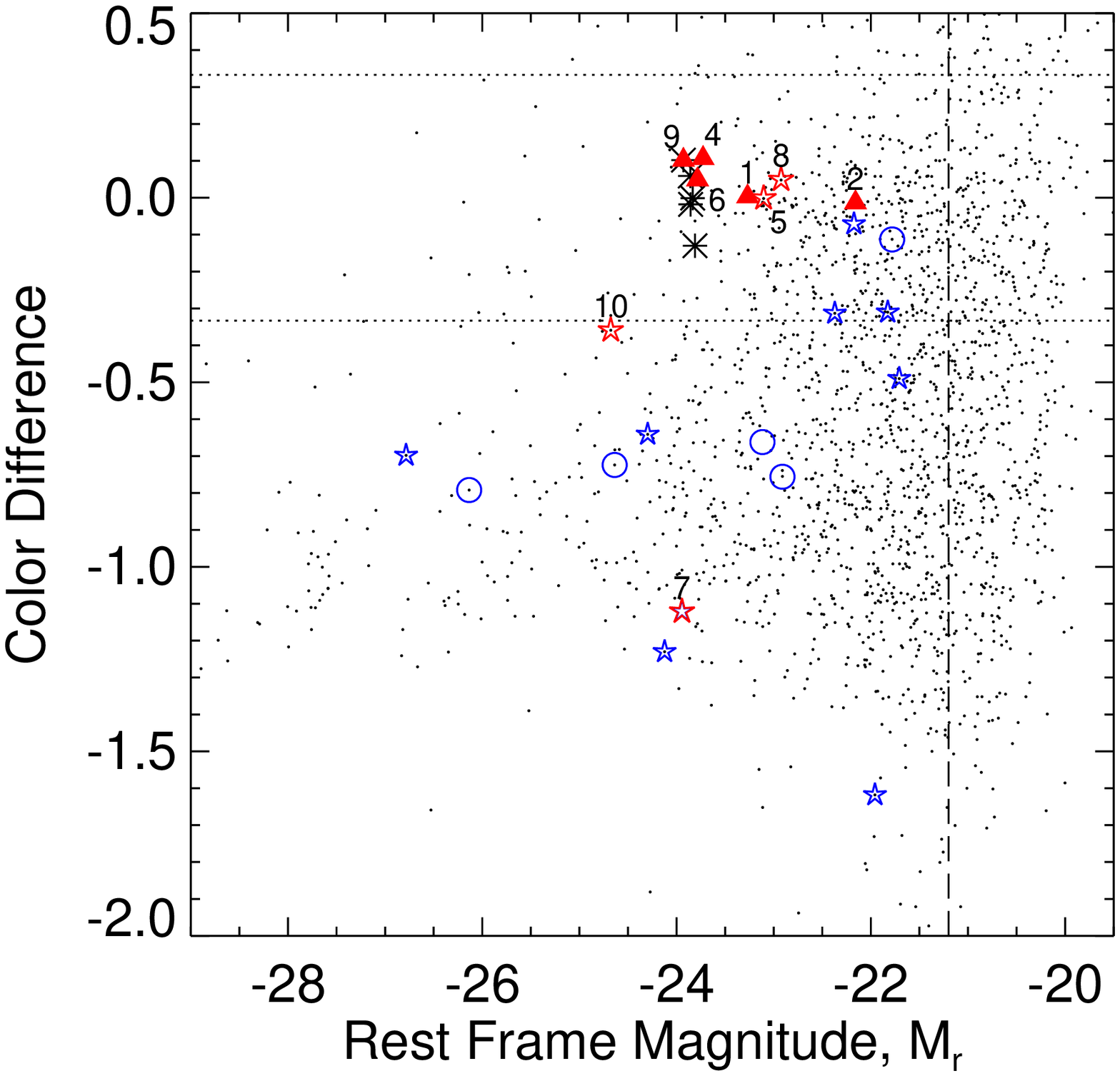}
\caption{
Composite color-magnitude diagram for five \midz\ clusters.
The ``color difference" is the magnitude difference between the
object's observed \rminusi\ color and the mean CRS \rminusi\ color of the cluster as listed in Table~\ref{tab:p2_rtc_optical}.
Symbols and dashed lines are similar to Figure~\ref{fig:p2_cmd_low_z}.  Open red (blue) stars are the
host galaxies of radio galaxies (XPSs) with unknown redshifts.  The vertical long-dashed line marks M$^{*}_{r}$$\approx$-21.2
at $z$$=$0.6, corrected for 0.4 magnitudes of pure luminosity evolution (see \S\ref{subsec:p2_obs_results_crs} for details).
Host galaxies of the radio sources with unknown redshifts and located on or near the CRS
(i.e., between the short-dashed horizontal lines) are assumed to be cluster members (see \S\ref{sec:p2_rlf} for details).
Number identifiers correspond to the radio sources listed in Table~\ref{tab:p2_radio_gal}.
}
\label{fig:p2_cmd_mid_z}
\end{center}
\end{figure*}

\begin{figure*}
\begin{center}
\includegraphics[scale=0.7]{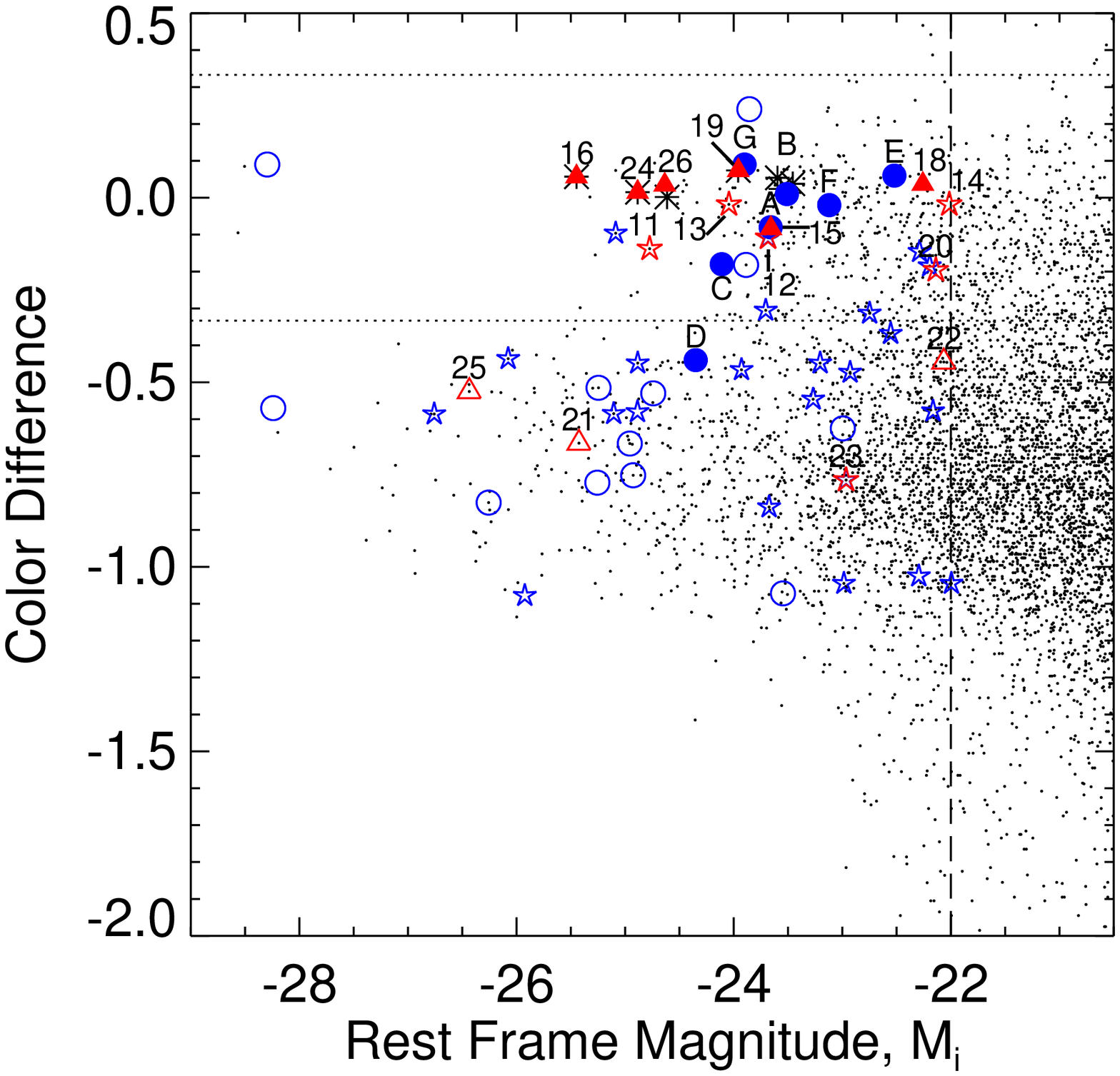}
\caption{
Composite color-magnitude diagram for seven \highz\ clusters.
The ``color difference" is the magnitude difference between the
object's observed \iminusz\ color and the mean CRS \iminusz\ color of the cluster, as listed in Table~\ref{tab:p2_rtc_optical}, with
the exception of RXJ1350 and RXJ1716 for which the color difference is  \rminusi.
Symbols and horizontal dashed lines are similar to Figure~\ref{fig:p2_cmd_mid_z}.
The vertical long-dashed line marks M$^{*}_{i}\approx$-22.0 at $z$$=$1.0, corrected
for 0.8 magnitudes of pure luminosity evolution. Number identifiers correspond to the radio sources listed in Table~\ref{tab:p2_radio_gal},
while letter identifiers correspond to the seven XPSs listed in Column 1 of Table~\ref{tab:p2_xps_high}.
}
\label{fig:p2_cmd_high_z}
\end{center}
\end{figure*}

Figures~\ref{fig:p2_cmd_low_z}--\ref{fig:p2_cmd_high_z} display the composite color-magnitude
diagrams of our low-$z$, mid-$z$, and high-$z$ cluster samples, respectively, as a function of the object's
rest-frame magnitude at the cluster redshift.   
Rest-frame magnitudes are calculated using the k-corrections
for E0 galaxies from \citet{fukugita95}. These k-corrections slightly overestimate the rest-frame magnitudes of bluer cluster galaxies.
The ``color difference" in Figures~\ref{fig:p2_cmd_low_z}--\ref{fig:p2_cmd_high_z} is the magnitude difference
between the object's observed color and the mean CRS color at the given rest-frame magnitude (Columns 3-4 of Table~\ref{tab:p2_rtc_optical}).

The color difference is \gminusr\ in Figure~\ref{fig:p2_cmd_low_z}, with the exception of
MS2137, A1995, MS1358, and A370 where the color difference is \rminusi.  In Figure~\ref{fig:p2_cmd_mid_z} the 
color difference is \rminusi, while in Figure~\ref{fig:p2_cmd_high_z} it is \iminusz\ with the
exception of RXJ1350 and RXJ1716 where the color difference is \rminusi.
All optical sources within a 1 Mpc projected radius of the cluster center are displayed 
(small black dots in Figures~\ref{fig:p2_cmd_low_z}--\ref{fig:p2_cmd_high_z}).
Optical sources that are 2-4 magnitudes brighter than the BCGs are most likely foreground objects. 
The host galaxies of radio galaxies and XPSs, described in the next section,
at L$\geq$L$^*$ and corrected for PLE are also identified in Figures~\ref{fig:p2_cmd_low_z}--\ref{fig:p2_cmd_high_z}
(radio sources in red; XPSs in blue).
Figure~\ref{fig:p2_cmd_low_z} is a modification of Figure~2 in Paper I, in that
we have overlaid the host galaxies of the non-cluster radio galaxies and XPSs for comparison.  
The vertical dashed-line identifies L$^*$ galaxies in
the specified filter, corrected for PLE for the mean of that redshift range.  The short-dashed horizontal
lines bound the maximum spread of 0.33 magnitudes from the mean CRS color (see Column 3 of Table~\ref{tab:p2_rtc_optical}).

\subsection{Radio Galaxies and XPS Detections}
\label{subsec:p2_obs_results_sources}

For our $z$$>$0.4 clusters, Table~\ref{tab:p2_radio_gal} lists radio sources 
with flux densities which correspond to \plim\ at the cluster redshift.  (Table~3 in Paper I lists the radio sources for the \lowz\
clusters.)  Table~\ref{tab:p2_radio_gal} lists the number identifier used in subsequent plots, the RA/DEC of the
radio source, the redshift if known, the optical magnitude and color of the host galaxy, the mean CRS color
for the cluster, and the projected distance, or ``radius", from the cluster X-ray emission centroid
(listed in Columns 2-3 of Table~\ref{tab:p2_rtc_xray}).  Column 12 lists the observed 1.4~GHz flux density in mJy.  If the
source redshift is known, then Column 13 lists the 1.4~GHz radio power assuming that F$_\nu\propto\nu^{-\alpha}$ with a
spectral index, $\alpha=0.7$.  For sources with unknown redshifts, Column~13 lists the radio power of the source if
located at the cluster redshift (identified in parentheses).
Assuming a power-law spectrum with $\Gamma$=1.7 (N$_E\propto$E$^{-\Gamma}$), Columns~14-15 of Table~\ref{tab:p2_radio_gal} lists the observed X-ray flux and luminosity for the radio source if at $z_{cluster}$, or a limit on those values for an X-ray non-detection.
The X-ray flux limits are estimated as 3$\sigma$ above the background at the source position.  

Of the 26 detected radio sources in the $z$$>$0.4 cluster sample, 24 radio sources have L$\geq$L$^*$ optical counterparts if at $z$$_{cluster}$.
Eleven are spectroscopically-confirmed cluster members, all with CRS colors, and 3 are spectroscopically-confirmed non-cluster members all with non-CRS colors.
The remaining 10 radio sources (8 with CRS colors, including Source 10 in Table~\ref{tab:p2_radio_gal}, and 
2 with non-CRS color) have unknown redshifts.  Thus, the sample of radio sources listed in Table~\ref{tab:p2_radio_gal} is 
54\% spectroscopically complete in total and 58\% complete for radio sources with CRS host galaxies.

In Figures~\ref{fig:p2_cmd_mid_z}--\ref{fig:p2_cmd_high_z}, these radio sources are identified by red closed (open) triangles for
cluster (non-cluster) sources and by red open stars for radio sources with unknown redshifts.  
These radio sources are labeled
with the number identifiers listed in Column~1 of Table~\ref{tab:p2_radio_gal}, excepting MS1621-R3 and RXJ1716-R3 with
optical non-detections at m$_{lim}$ in the optical images.
Based on the radio source redshift or host galaxy color, Column~3 of Table~\ref{tab:p2_radio_gal} lists a code for 
our best estimate of a radio source's association with the cluster.  (See tablenote (3) of Table~\ref{tab:p2_radio_gal} 
for details.)  Five of the radio sources listed in Table~\ref{tab:p2_radio_gal} have
X-ray counterparts above the required X-ray luminosity limit at the cluster redshift.

For our $z$$>$0.4 clusters Table~\ref{tab:p2_xps_high} only lists XPSs that
are spectroscopically-confirmed cluster members, while Table~\ref{tab:p2_xps_high_all} 
in the Appendix lists our entire sample of XPSs at \lxbb$\geq$\xlimb\ if located at the cluster redshift.
These two tables include the source number (including the letter identifiers used in subsequent plots 
in parentheses), the RA/DEC of the XPS, the source redshift if known, the observed net broadband 
(0.3--8.0 keV) counts and resulting X-ray flux and luminosity.  
The X-ray luminosity at the cluster redshift is tabulated for XPSs with unknown redshifts.  
Also listed is the projected distance of the XPS from the cluster center in Mpc 
(``radius" in Table~\ref{tab:p2_xps_high_all}), the host galaxy's optical magnitude and color, and the 1.4~GHz 
radio power or power limit for each XPS.  

Of the 86 detected XPSs listed in Table~\ref{tab:p2_xps_high_all}, 67 XPSs have optical counterparts on our images.
However, our working sample consists of only 56 XPSs at \lxbb$\geq$\xlimb\ and hosted by L$\geq$L$^*$ galaxies if at $z$$_{cluster}$.
The redshifts for some XPSs come from the literature, 
which includes the optical follow-up campaigns of ChaMP and the Serendipitous Extragalactic X-ray Source Identification 
\citep[SEXSI,][]{2006ApJS..165...19E}.  The SEXSI survey is a 2 deg$^2$ survey of hard X-ray-selected AGN that complements
the \chandra\ Deep Fields to detect XPSs with fluxes between 10$^{-13}$--10$^{-15}$~\flux.

In Figures~\ref{fig:p2_cmd_mid_z}--\ref{fig:p2_cmd_high_z}, XPSs are identified by blue closed (open) circles for
cluster (non-cluster) sources and by blue open stars for the XPSs with unknown redshifts.  
Our working sample of the XPSs listed in Table~\ref{tab:p2_xps_high_all}
is 45\% spectroscopically complete, which increases to 55\% for a subset of XPSs with CRS host galaxies only.
The cluster XPS population is considerably less well-defined than the cluster radio galaxy population in our sample due to
lack of redshifts.
Column~3 of Table~\ref{tab:p2_xps_high_all} lists a classification tag for the XPSs based on spectroscopy and/or host galaxy
color relative to the CRS (see tablenote (3) of Table~\ref{tab:p2_xps_high_all} for more details).

\section{Radio Galaxies in 0.4$<$\lowercase{z}$<$1.1 Clusters}
\label{sec:p2_rg}

\subsection{Statistical Sample}
\label{subsec:p2_rg_statis}

In Paper I, we detected 26 radio galaxies at \rlim\ hosted by L$\geq$L$^*$ galaxies located within 1 Mpc of the
cluster core.  
We found that 87\% of the CRS radio galaxies are cluster members,
whereas, none of the bluer radio galaxies are cluster members.  This result is consistent 
with the expectation that at low-$z$ radio-loud objects are hosted by early-type galaxies \citep{1995AJ....110.1959L}.  Thus, we can 
assume that nearly all high-$z$ radio galaxies with CRS colors are cluster members.
Also, 17 of 20 cluster radio galaxies from Paper~I are detected at \plim\ (the radio limit in
this present work).  The most luminous radio source at low-$z$ (RXJ1758-R6 in Table 3 from Paper I) is detected
at P$_{1.4GHz}$=8.1$\times$10$^{24}$ \radiounits, just below the nominal FR I/FR II power-level boundary 
\citep[P$_{1.4GHz}\sim$10$^{25}$ \radiounits;][]{1996AJ....112....9L}. 

For the ten \combozb\ clusters (excluding RXJ0910 and RXJ0849) Table~\ref{tab:p2_radio_gal} lists the
26 radio sources detected with flux densities sufficient to have \plim\ at the cluster redshift.
Setting aside the 14 sources with spectroscopic redshifts (11 are spectroscopically-confirmed cluster 
members with CRS colors), of the remaining 12 radio
sources, 8 have CRS colors, 2 non-CRS colors, and 2 have no optical counterparts.
While all 8 radio sources with CRS colors should be considered likely cluster members, 
if we extrapolate our low-$z$ results, we conservatively identify only 6 as highly-likely cluster members 
for a total of 17 (of 19 or 87\%) cluster radio galaxies in our \combozb\ sample.

Because there are no obviously distinguishing photometric differences between the 8 CRS radio galaxies
lacking spectroscopy, we have chosen two specific sources as non-cluster members as
follows: (1) we discard the lowest flux source as non-cluster (RXJ1350-R4, Source 14 in Table~\ref{tab:p2_radio_gal}); 
(2) we discard Source 10 (RX J1113-R1) in Table~\ref{tab:p2_radio_gal} due to its unusual nature as follows.
Source 10 is located near the edge of the CRS in Figure~\ref{fig:p2_cmd_mid_z}.  This source is both 0.6 magnitudes 
brighter and 0.3 magnitudes bluer than the other BCGs in the mid-$z$ redshift range.  
At the cluster redshift of $z$$=$0.72, the \oii\ emission line would fall in the Sloan {\it r}
filter and \oiii\ would fall just redward of the Sloan {\it i} filter.  If the flux contribution 
from \oii\ is $\sim$25\% of the total light in this filter, Source 10 
would appear bluer than the CRS at the cluster redshift, similar to the BCG in MS 
1455.0+2232 at $z$$=$0.25 presented in Paper I.  However, this source is
not located at the cluster center, but rather at a projected distance of 500 kpc from the core.  
Source 10 is also one magnitude brighter than the galaxy identified as the RXJ1113 BCG due to its coincidence
with the X-ray centroid.  

For the above reasons, Source 10 is a good candidate for
being a foreground source.  Therefore, we assign Sources 14 and 10 as non-cluster radio sources.  The exclusion of these two sources does not alter
the statistical difference between the low-$z$ and high-$z$ radio luminosity functions presented in \S\ref{sec:p2_rlf}.
Likewise, there is no strong evidence for ``blue" radio galaxies in this sample, and there were none at low-$z$.
We assign the 2 high-$z$ blue radio sources without redshifts as highly-unlikely cluster members.
In summary, we conservatively identify 17 radio galaxies in  Table~\ref{tab:p2_radio_gal}, all on the
CRS, as our high-$z$ radio-loud cluster sample.  However, future spectroscopy is required to verify this assessment and to search
for blue radio galaxies in high-$z$ clusters.

\begin{figure*}
\begin{center}
\plottwo{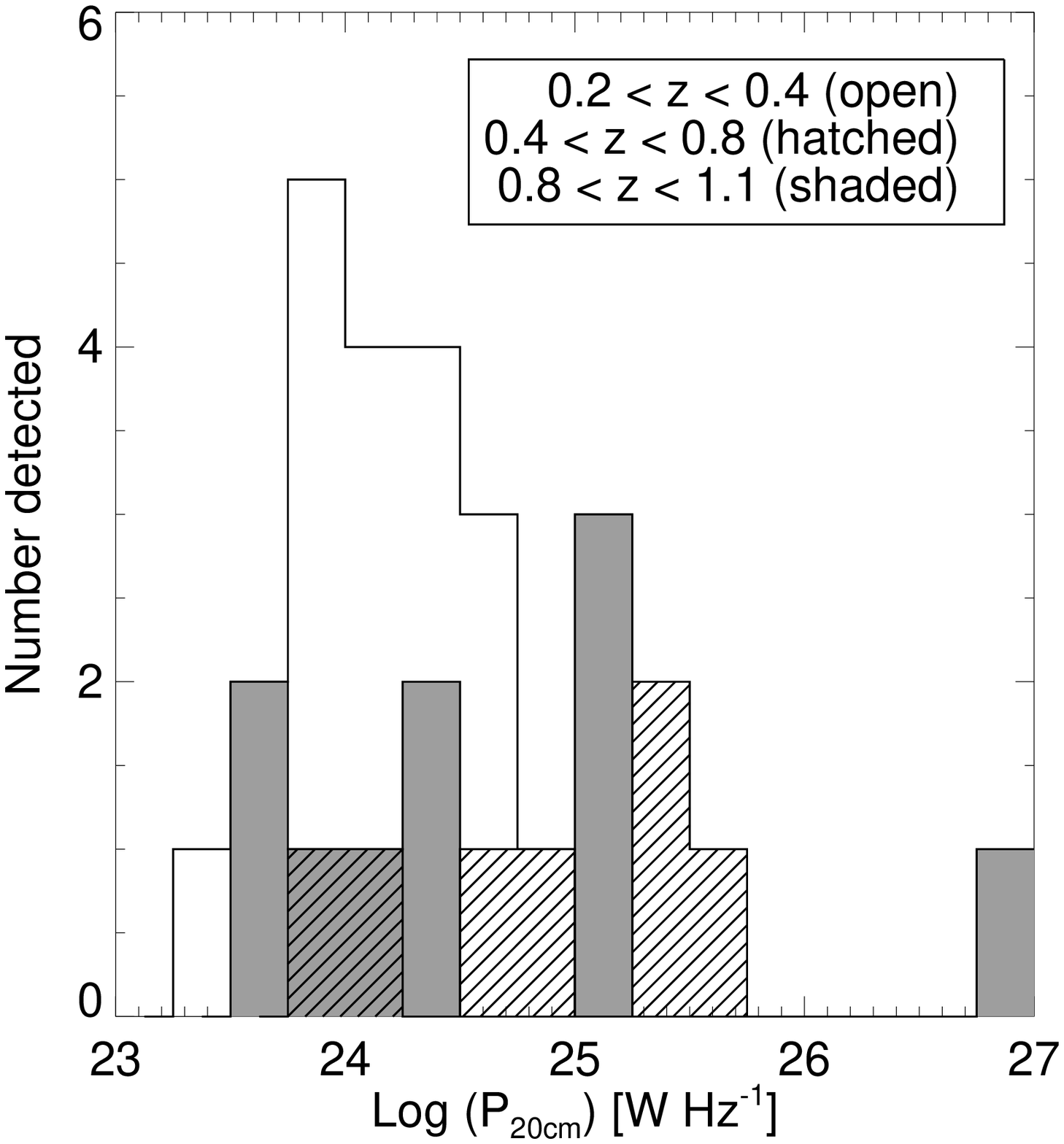}{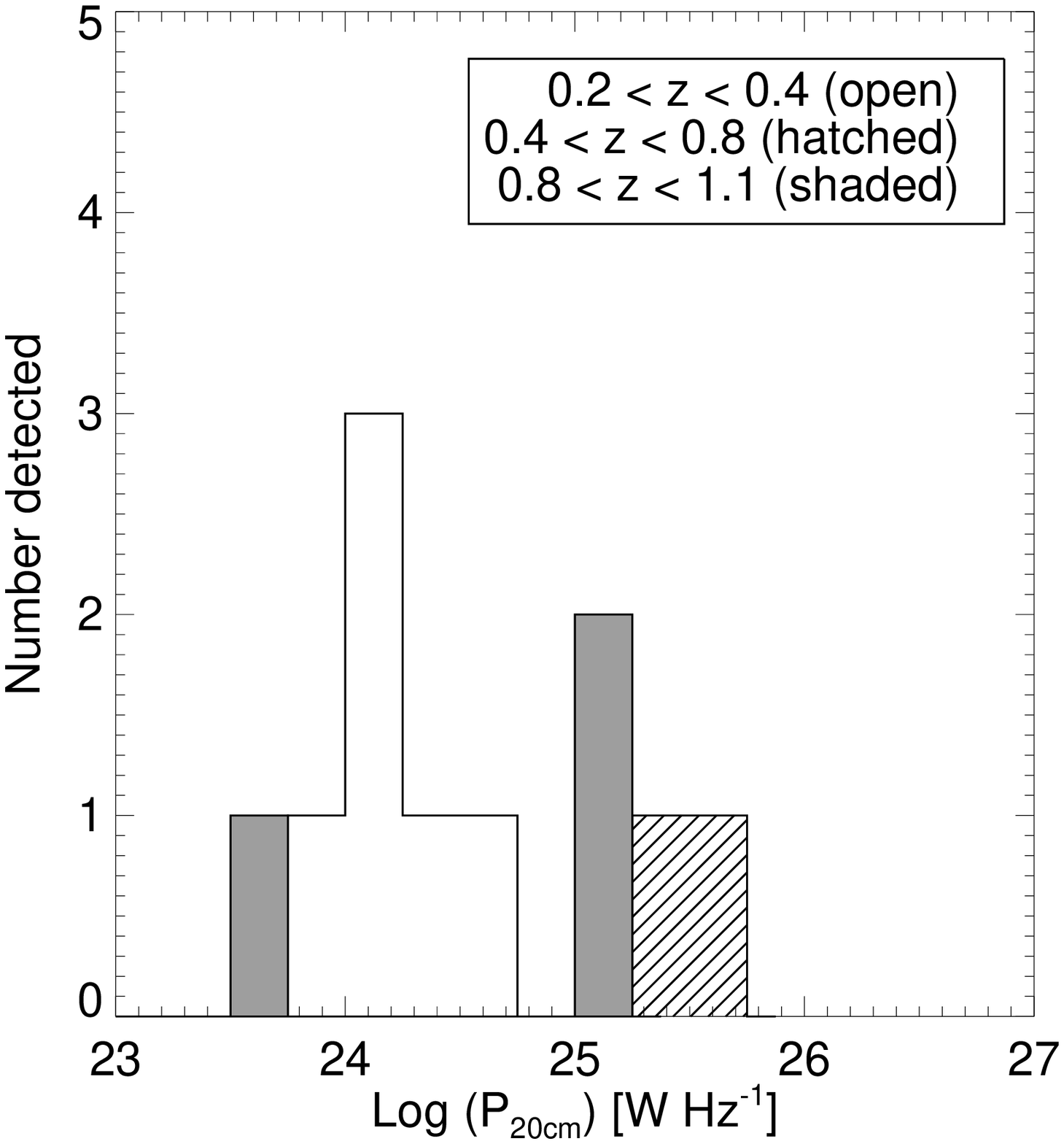}
\caption{
Histogram of radio galaxy powers in 21 clusters at 0.2$<$$z$$<$1.1
(left panel) and of cluster BCGs only (right panel).
The open histogram displays the number of radio sources with \logp$\geq$23.5 \radiounits\
at 0.2$<$$z$$<$0.4 (N=20), while the hatched histogram displays the number of sources at
0.4$<$$z$$<$0.8 with \logp$>$23.5 (N=8).  The
shaded histogram displays the number of radio sources at 0.8$<$$z$$<$1.1 with \logp$>$23.7~\radiounits\ (N=11).
The right panel shows that the decline in radio power with redshift is more dramatic for the BCGs in this cluster sample.
}
\label{fig:p2_rlf_hist}
\end{center}
\end{figure*}

Figure~\ref{fig:p2_rlf_hist} displays the distribution of the cluster radio
galaxy powers with (open, hatched, shaded) regions corresponding to (low-$z$, mid-$z$, high-$z$) 
radio galaxies, respectively.  The right panel of Figure~\ref{fig:p2_rlf_hist} displays the radio-loud
BCGs only (6 at low-$z$, 2 at mid-$z$, 3 at high-$z$).  
The average radio power of cluster radio galaxies is
(2.2$\pm$1.7, 14.5$\pm$13.9, 78.8$\pm$232.0)$\times$10$^{24}$~\radiounits\ in the (low-$z$, mid-$z$, high-$z$) redshift bins, respectively.
Even if we exclude RXJ1716-R1 ($z$$=$0.8), the FR II source at P$_{1.4GHz}$=7.7$\times$10$^{26}$~\radiounits\ (see \S\ref{subsec:p2_obs_results_rxj1716_fr2}), the
average radio power of \combozb\ cluster radio galaxies is 9.4$\pm$10.9$\times$10$^{24}$~\radiounits.
The spread in the average radio powers for $z$$>$0.4 cluster radio galaxies suggest that some are
more luminous than their low-$z$ counterparts presented in Paper~I. 
For BCG radio sources alone, at low-$z$ the average radio power is 1.9$\pm$1.1$\times$10$^{24}$~\radiounits, 
while at \combozb\ it is 1.8$\pm$1.3$\times$10$^{25}$~\radiounits,
evidence that BCG radio sources have faded by about an order of magnitude from \combozb\ to \lowz.
This is consistent with earlier results from samples that were selected differently \citep[e.g.,][]{2001AJ....122.2874H}.

\subsection{Radial Distribution}
\label{subsec:p2_rg_distribution}

\begin{figure*}
        \begin{center}
        \includegraphics[scale=0.75]{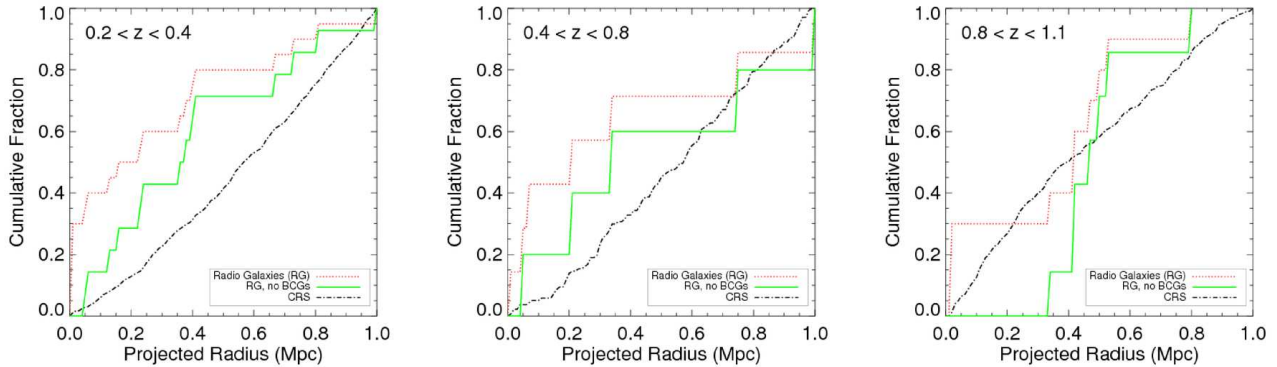}
        \caption{
        Cumulative radial distributions of cluster radio galaxies relative to CRS galaxies in three redshift bins.
        In each panel, the cumulative fraction of cluster radio galaxies (dotted red line), as a function of
        projected distance from the cluster X-ray emission centroid, is compared to the
        CRS galaxy (black dot-dashed line) distribution.  Also displayed is the cumulative fraction of
        cluster radio galaxies excluding the radio-loud BCGs (solid green line).
        Notice that the cluster radio galaxy population at 0.2$<$$z$$<$0.8 (left and middle panels) appears more
        centrally concentrated than the CRS galaxies, while at $z$$>$0.8 (rightmost panel), there appears to be
        an absence of radio-loud galaxies near the cluster centers.  However, the overall impression is that radio galaxies become more
        centrally concentrated with cosmic time (see \S\ref{subsec:p2_rg_distribution} and Table~\ref{tab:p2_kstest} for statistics).
        }
        \label{fig:p2_rg_radial_distr}
        \end{center}
\end{figure*}

Figure~\ref{fig:p2_rg_radial_distr} displays the cumulative radial distribution of (20, 7, 10) cluster 
radio galaxies in the (low-$z$, mid-$z$, high-$z$) cluster samples
as a function of the projected radius from the cluster center (shown as a dotted red line) 
compared to the cumulative radial distribution of CRS galaxies (black dot-dashed line).
In Paper~I, we used a two-sided Kolmogorov-Smirnov (K-S) test to determine the probability
that the cumulative distribution of two samples are drawn 
from the same parent population \citep{press_numerical_1992}. 
We concluded that the radial distributions of the low-$z$ cluster radio galaxy population and the CRS galaxy population as a whole are inconsistent with 
being drawn from the same parent population and that the radio galaxies are more centrally concentrated than the CRS galaxies at the 5.7$\sigma$ level
(see left panel of Figure~\ref{fig:p2_rg_radial_distr}).  

Table~\ref{tab:p2_kstest} summarizes the K-S test results between the cumulative radial distributions of
the radio galaxies in our low-$z$, mid-$z$, and high-$z$ samples compared to their respective CRS galaxy population.
In Table~\ref{tab:p2_kstest} we include the value of the K-S D-statistic, the probability that the two samples are drawn from the same parent
population, and the statistical significance of this probability.
As seen in the middle panel of Figure~\ref{fig:p2_rg_radial_distr}, the mid-$z$
cluster radio galaxy population appears to remain centrally concentrated compared to CRS galaxies at
the 5.9$\sigma$ level and consistent with the results at low-$z$.  However, the radial distribution of
the radio galaxies in the \highzsmall\ clusters (right panel) approximately tracks that of the CRS galaxies.

The probability of a cluster galaxy hosting a radio source increases with optical magnitude 
\citep{1996AJ....112....9L} and the BCG, as the most massive cluster galaxy, is more likely
to host a radio source than other cluster galaxies \citep{2007MNRAS.379..894B}.
Is the special location of the radio-loud BCG at the cluster center skewing the analysis and
only making the population appear more centrally concentrated compared to the CRS galaxies?
To test this, we remove the BCGs from the radial distribution analysis above.  The cumulative radial distribution
of this reduced sample (solid green line in Figure 6) remains statistically different from the underlying CRS galaxy population at the
(4.7$\sigma$, 4.2$\sigma$, 4.0$\sigma$) level for the (low-$z$, mid-$z$, high-$z$) cluster sample.
From Figure~\ref{fig:p2_rg_radial_distr},
the cumulative fraction of radio galaxies within 500 kpc of the cluster core is
(72$\pm$23\%, 60$\pm$35\%, 57$\pm$29\%) in the (low-$z$, mid-$z$, and high-$z$) sample.
Moreover, this fraction decreases to (43$\pm$17\%, 40$\pm$28\%, 0$\pm$14\%) within 300 kpc of the cluster center.
Although the data are insufficient to produce a significant result, we note that our higher redshift 
clusters have few AGN (apart from the BCG) in the cluster cores,
suggesting that radio sources in clusters become more centrally-concentrated over cosmic time.

\subsection{An FR~II Source in a $z$$=$0.8 Cluster}
\label{subsec:p2_obs_results_rxj1716_fr2}

The identification of FR~II radio sources in the high-$z$ cluster sample is somewhat unexpected given that none
were found in the low-$z$ cluster sample from Paper I.  However, it has been known for some time that some luminous quasars 
and FR II radio galaxies are found in clusters at $z$$\geq$0.5 
\citep[e.g.,][]{1991ApJ...371...49E,1991ApJ...367....1H,2001AJ....122.2874H}.
Those detections suggest that FR~IIs located in rich clusters at high redshift rapidly evolve in
luminosity to become FR~Is at $z$$=$0.  Because the fading of cluster FR~IIs occurs over
a similar time-frame ($\sim$1 Gyr) as the development of a deep gravitational potential well and a hotter
ICM in clusters, these results suggest that the AGN evolution is linked to the evolution
of the cluster environment \citep{2001AJ....122.2874H}.  Therefore, a detection of an 
FR~II in our sample at $z$$\approx$0.8 is evidence supporting this idea.

\begin{figure*}
\begin{center}
\includegraphics[scale=0.7]{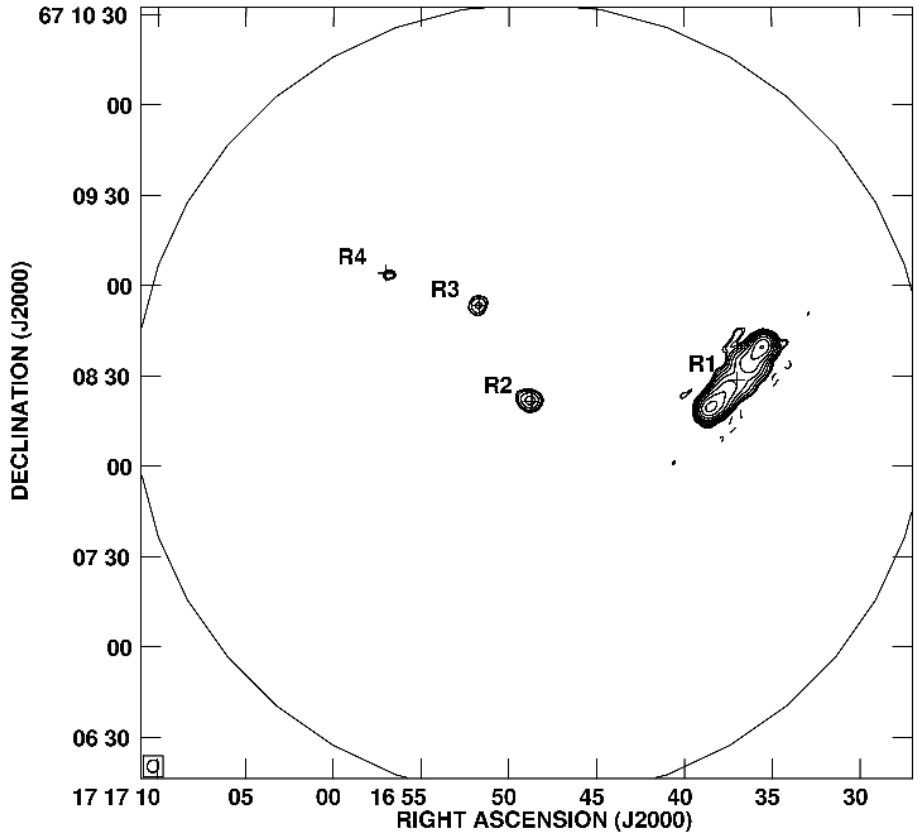}
\caption{
VLA 1.4~GHz radio contour map of RX J1716.6+6708.
Of the four radio sources detected above the 3$\sigma$ map rms of 0.21 mJy/Beam, R1, R2, and R4 are confirmed
cluster members ($z_{cluster}$$=$0.809), while R3 is a confirmed non-cluster source
(see Table~\ref{tab:p2_radio_gal} for source details).
R2, near the map center, is located in the cluster BCG.  However, R1, the double-lobed, FR~II radio source, is
located 1.2$^\prime$ West of the cluster center ($\approx$~530 kpc).  The large outer circle marks a 1 Mpc projected
radius from the cluster center, while the crosses mark the optical counterpart closest to the radio position.
The continuum peak flux of the map is 7.1 mJy/Beam with contour levels of 0.42 mJy $\times$(-1,1,1.18,2,4,8,16,32,64,128,256).
}
\label{fig:p2_rxj1716_fr2}
\end{center}
\end{figure*}

A contour map of the VLA 1.4~GHz continuum image of RXJ1716 ($z$$=$0.809) is displayed in Figure~\ref{fig:p2_rxj1716_fr2} 
with the crosses marking the optical counterpart closest to the radio position.
As noted in \S\ref{subsec:p2_obs_radio},
RXJ1716 was observed for 2 hours with the VLA at 1.4~GHz in its standard B-configuration (VLA program AS873).
Four radio sources, listed in Table~\ref{tab:p2_radio_gal}, are detected above the 3$\sigma$ map rms of 0.21 mJy/beam.
The BCG, source RXJ1716-R2 in Table~\ref{tab:p2_radio_gal}, is located at the center 
of Figure~\ref{fig:p2_rxj1716_fr2}.  However, RXJ1716-R1, which is located 1.2$^\prime$ west of the BCG (or at a
projected radius of 530 kpc), has a clear double-lobed radio morphology (angular extent of $\sim$18\arcsec  or 135 kpc)
and a much higher radio power (P$_{1.4GHz}$=7.7$\times$10$^{26}$~\radiounits).
These properties clearly identify it as an FR~II radio galaxy.

RXJ1716-R1 is also detected at 4.85~GHz and 365 MHz \citep[see][]{1992ApJS...79..331W}, 74 MHz \citep{1998AJ....115.1388C}
and 28.5~GHz \citep{2007AJ....134.1245C}.  Even in the NRAO VLA Sky Survey (NVSS), RXJ1716-R1 appears extended,
as noted by \citet{1998AJ....115.1388C}, although
the lower angular resolution of the D-configuration does not clearly show the double-lobed structure (see
Figure~4 in \citealt{1998AJ....115.1388C}) that is resolved in Figure~\ref{fig:p2_rxj1716_fr2}.
Based on multi-frequency radio observations, the spectral index, $\alpha$, is
1.1 between 6cm and 20cm, and 1.0 between 80cm and 20cm \citep{1992ApJS...79..331W}.
This steep spectral index suggests that the electron population in this source is
older than in typical FR~II radio galaxies.  A higher resolution map might be expected to
show the ``fat double" morphology seen in steep-spectrum FR~IIs in clusters \citep{2002AJ....124.1239H}.

RXJ1716-R1 is also detected as a \chandra\ XPS with \lxbb=46.5$\pm$6.6$\times$10$^{42}$ \ergs.  
The host galaxy ({\it i}=21.23) appears elliptical in our ARC 3.5m optical images (FWHM$\approx$1.3$^{\prime\prime}$), as well as in HST WFPC images
(Proposal ID 7293).  SEXSI identified RXJ1716-R1 
\citep[SEXSI J171636.9+670829; ][]{2003ApJ...596..944H} as an emission-line galaxy \citep{2006ApJS..165...19E} at $z$$=$0.795.
Emission lines of [O II] and [O III], along with a strong \cahkbreak\ are detected (private communication, Megan Eckart).
The radial velocity of this source is 0.9$\sigma$ from the cluster centroid velocity, where $\sigma$ is the observed cluster velocity 
dispersion of 1522~\kms\ \citep{1999AJ....117.2608G}, and so consistent with being a cluster member and not an in-falling galaxy 
seen projected onto the cluster center.  Unlike the radio-loud galaxies situated in nearby clusters found in Paper I, 
RXJ1716-R1 is located 530 kpc from the cluster core, in a region of lower ICM density.  The \chandra\ image
does not reveal any noticeable ``cavities" in the X-ray emitting ICM at the location of the radio lobes, although
the X-ray surface brightness is quite low at that large radial distance.

\section{The Radio Luminosity Function of Cluster Radio Galaxies in Coma Cluster Progenitors}
\label{sec:p2_rlf}

Using the statistical sample of the 17 cluster radio galaxies from the previous section, 
we determine the radio active fraction (RAF) of cluster CRS galaxies and construct the radio luminosity function (RLF)
of radio galaxies in our cluster sample between 0.2$<$$z$$<$1.1.  
In eleven low-$z$ clusters we identified 665 L$\geq$L$^*$ CRS galaxies and also 17 cluster
radio galaxies at \plim\ within 1 Mpc of the cluster core (see Paper~I), 
yielding 1.5$\pm$0.5 radio galaxies {\it per cluster} or a cluster RAF for CRS galaxies of 2.5$\pm$0.8\% \citep[Poisson statistics; ][]{1986ApJ...303..336G}.
Due to the small number of radio sources in the mid-$z$ and high-$z$ bins (7 and 10 sources,
respectively), we combine them together into the ``combined high-$z$ sample".  
In ten \comboz\ clusters (not including RXJ0910 nor RXJ0849),
we identified 443 L$\geq$L$^*$ CRS galaxies and 17 cluster radio galaxies, 
yielding 1.7$\pm$0.5 radio galaxies {\it per cluster} or a cluster RAF of 3.8$\pm$1.2\%
\citep[Poisson statistics; ][]{1986ApJ...303..336G}.
The cumulative binomial probability of detecting $\geq$17 radio galaxies in 443 CRS galaxies in our combined high-$z$ sample based on the low-$z$ results from Paper~I is 6.6\% 
and is therefore only marginal evidence for density evolution of the radio galaxy population
in Coma Cluster progenitors between 0.2$<$$z$$<$1.1. 

\begin{figure*}
\begin{center}
\includegraphics[scale=0.7]{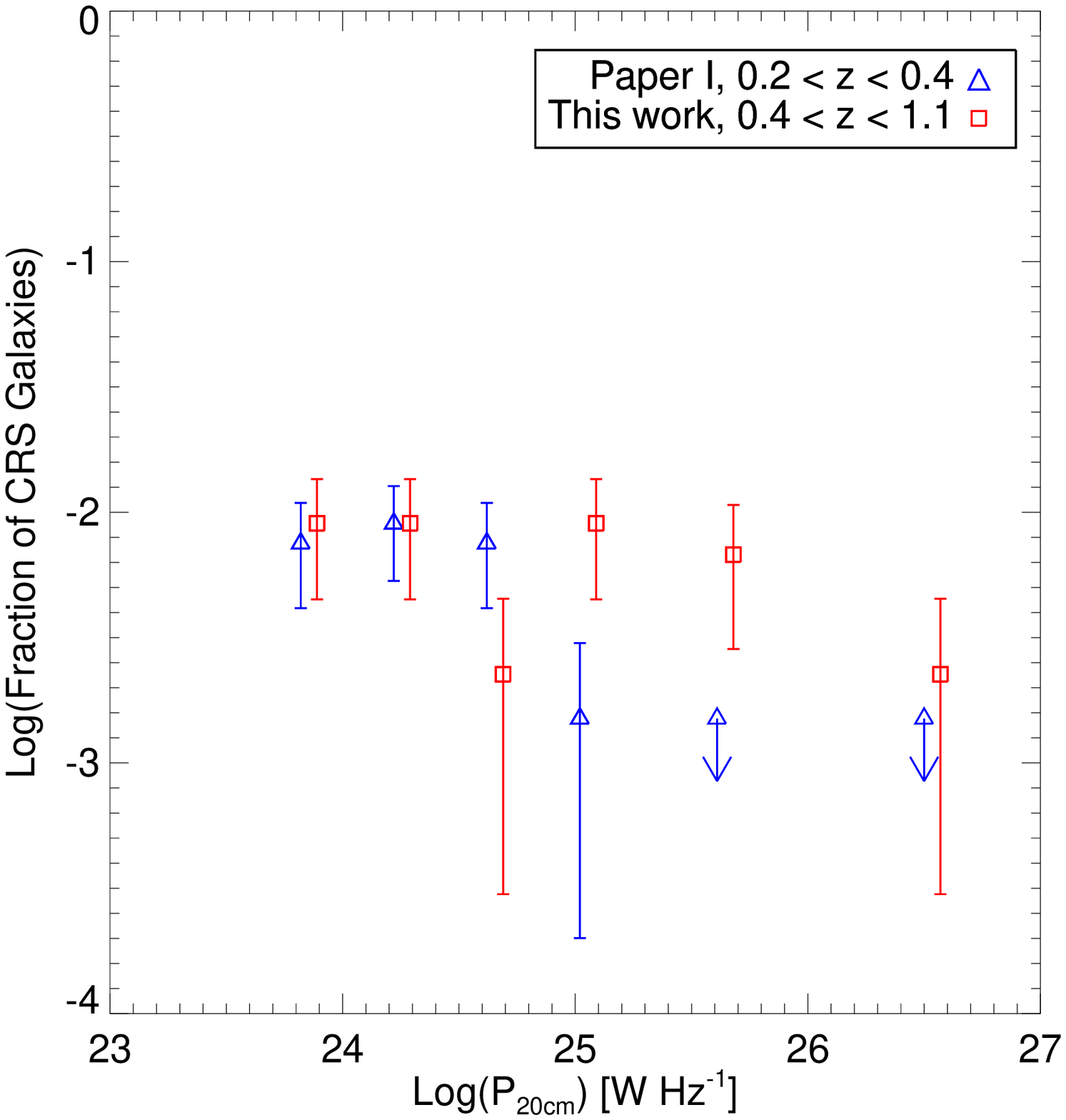}
\caption{
Radio luminosity function of radio galaxies in Coma Cluster progenitors between 0.2$<$$z$$<$1.1.
The y-axis is the fraction of CRS galaxies located within 1 Mpc of the cluster center
that are detected as radio sources within the given radio power bin (refer to Table~\ref{tab:p2_rlf}).
See \S\ref{subsec:p2_obs_results_crs} for details on the calculation of the
total number of CRS galaxies.  In Coma Cluster progenitors the radio galaxy population
at \logp$>$25.0 \radiounits\ evolves from $z$$\sim$1.1 to 0.2 with more powerful radio sources situated
in less-massive clusters at earlier epochs (3.3$\sigma$ significance).
}
\label{fig:p2_rlf_rtc}
\end{center}
\end{figure*}

As noted in \S\ref{subsec:p2_rg_statis}, the spread in the average 1.4~GHz radio power of the \combozb\ cluster
radio galaxies is larger than that at low-$z$.  We quantify this luminosity 
evolution more precisely by constructing a univariate RLF.  Table~\ref{tab:p2_rlf} lists the radio power bins
used to construct the low-$z$ and combined high-$z$ RLF of radio galaxies in this cluster sample,
including upper limits on the bins without radio detections.  The highest radio power bin at \logp=27.0 \radiounits\ includes only RXJ1716-R1,
the cluster FR II at \logp$=$26.8 \radiounits\ described in \S\ref{subsec:p2_obs_results_rxj1716_fr2}.
Figure~\ref{fig:p2_rlf_rtc} is a log-log plot of the RAF fraction with respect to the CRS galaxies
versus \logp.

\subsection{Comparison With Previous Low-$z$ Cluster Results}
\label{subsec:rlf_lowz_compare}

At FR~I power levels between 23.6$<$\logp$<$25.0 \radiounits, we detect 17 cluster radio galaxies in 665 low-$z$ CRS galaxies. (We adopt
a working limit of 10$^{25}$ \radiounits\ for FR~Is based upon the survey of \citet{1996AJ....112....9L} who find few FR~Is above that
power level and more FR IIs above 10$^{26}$ \radiounits.)
Given these statistics, we expect to detect 11 radio galaxies in 443 CRS galaxies at \combozb\ and
we detect 10 sources consistent with the expectation.
This result suggests little density or luminosity evolution in our \allz\ radio galaxy population 
at FR~I radio powers.  However, we detect a substantial increase in the number of cluster radio galaxies at 25.0$<$\logp$<$27.0 \radiounits.
At FR~II radio power levels we detect 7 radio sources in our combined high-$z$ cluster sample 
(of which 4 are located in the BCG), and none were detected at low-$z$ in Paper I.  
Using the 7 detections in our combined high-$z$ clusters to define a detection probability, we expect to detect 10.5 FR~II power-level
radio sources in 665 CRS galaxies at low-$z$ and we detected none.  
The cumulative binomial probability of detecting no radio galaxies at low-$z$ is 
0.003\% (3.3$\sigma$ using binomial statistics).  
In Monte Carlo simulations consisting of 100,000 realization of
our total low-$z$ CRS sample, we find that there is only a 1.1\% probability that our
results would occur randomly (Gaussian significance of 2.6$\sigma$).
Therefore, the radio galaxy population in our cluster samples differs at \logp$>$25.0 \radiounits.
Unlike other high-$z$ cluster radio galaxy surveys \citep[e.g.,][]{1999AJ....117.1967S,2006A&A...446...97B},
this evidence for the evolution of the RLF is more robust because it is entirely internal to our sample,
avoids biases created by different selection and analysis methods, and uses a better sample selection technique.
Table~\ref{tab:p2_rlf_diffs} (rows 1 and 6) summarizes the comparisons between the expected and detected number of radio galaxies in our
low-$z$ and combined high-$z$ cluster samples.

Is this observed excess in the radio galaxy numbers at higher radio powers a result of our sample selection?
To answer this question, we first compare our cluster RLFs to that of local Abell clusters at $z$$<$0.09 from
\citet{1996AJ....112....9L}.  Using 188 cluster radio galaxies, these authors
computed the univariate and bivariate RLF of radio sources hosted by elliptical and S0 galaxies
with M$_{R}$$<$-20.5 and log(P$_{1.4GHz}$)$>$22~\radiounits\ and located within 0.3 Abell radius ($\approx$ 600 kpc) of
the cluster center.  Using Figure~3 from Paper I or Figure~\ref{fig:p2_rg_radial_distr} in this work, we find that 60\%
of our CRS galaxies are located within 600 kpc of the cluster centers.  Therefore, we apply this correction
factor to be consistent with the \citet{1996AJ....112....9L} sample definition and
exclude 2 radio galaxies from our low-$z$ cluster sample (MS1455-R1 and MS1008-R1) and 3 radio galaxies from our 0.4$<$$z$$<$1.1 cluster sample
(MS1621-R4, RXJ1221-R4, and RXJ1350-R1) that are outside that region.

\begin{figure*}
\begin{center}
\plottwo{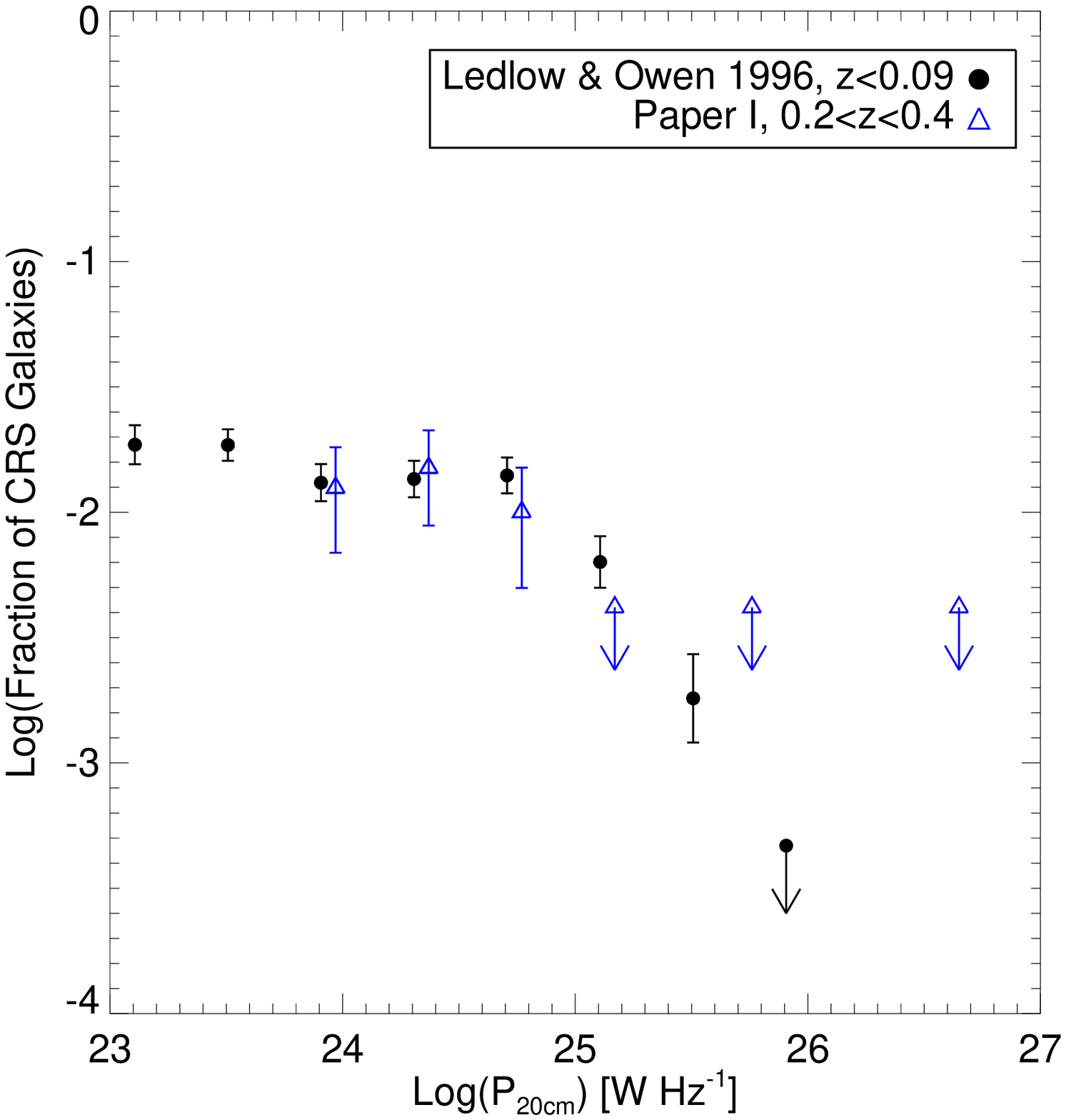}{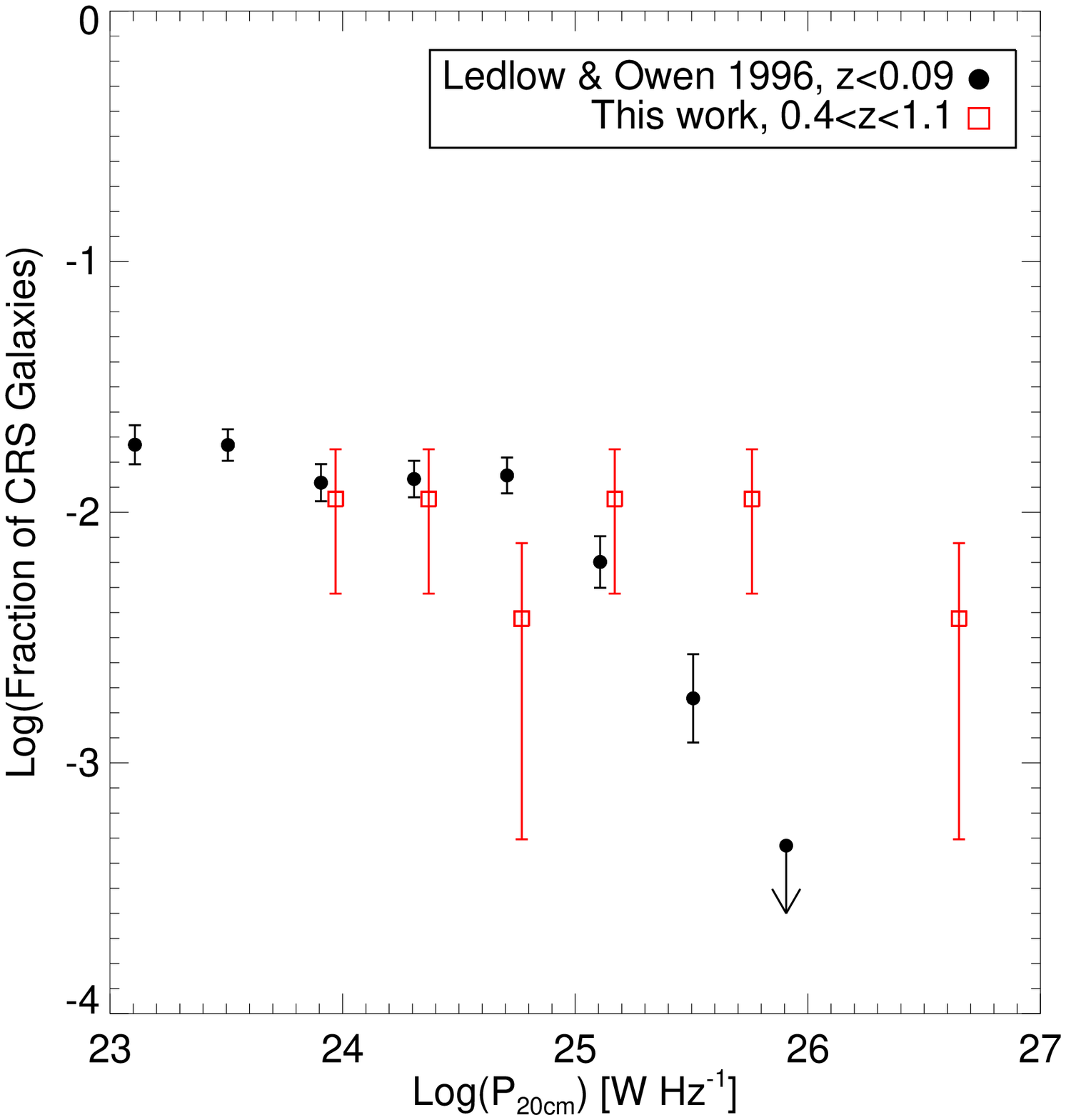}
\caption{
RLF comparison between $z$$<$0.09 Abell cluster radio galaxies and \lowz\ cluster radio galaxies (left)
and \combozb\ cluster radio galaxies (right).
The y-axis is defined as the fraction of CRS galaxies located within 600 kpc
of the cluster center (to match the survey area in \citet{1996AJ....112....9L})
that are detected as radio sources within the given radio power bin.
The RLF for our low-$z$ (\lowz) radio galaxies
is consistent with the radio galaxy populations in nearby Abell clusters (see \citealt{1996AJ....112....9L}).  However,
the RLF of our sources in clusters at $z$$>$0.4 deviates from the nearby Abell clusters at \logp$\geq$25.0 \radiounits.
}
\label{fig:p2_rlf_ledlow}
\end{center}
\end{figure*}

Figure~\ref{fig:p2_rlf_ledlow} displays the RLF of radio galaxies in Abell clusters from the \citet{1996AJ....112....9L} sample compared
to the RLF of our eleven low-$z$ clusters from Paper I (left panel) and our ten combined high-$z$ clusters from this work (right panel).
We adjust the Abell RLF by +0.08 in log(P$_{1.4GHz}$) to account for the different cosmology assumed in their paper.  
At 23.6$<$\logp$<$25.0 \radiounits, the RLFs of our low-$z$ and combined high-$z$ radio galaxies are consistent
with the RLF measured by \citet{1996AJ....112....9L} (cumulative binomial probabilities of 66\% and 85\%, respectively).
However, the RLF of radio galaxies in our combined high-$z$ sample deviates noticeably 
at \logp$>$25.0 \radiounits.  At $z$$>$0.4 we expect to detect 2.2 radio galaxies at \logp$\geq$25.0 \radiounits, but 
we detect 6 radio galaxies above this power level, 5 of which are 
spectroscopically-confirmed cluster members.  The cumulative binomial probability of detecting 
$\geq$6 radio sources at \logp$>$25.0 \radiounits\ in ten \combozb\ clusters is 2\% and thus inconsistent with the expectation.  
Therefore, we conclude that the RLF for cluster radio galaxies does not significantly evolve
from \lowz\ to the present epoch, but does change significantly at \combozb\ entirely at \logp$>$10$^{25}$ \radiounits.

\subsection{Comparison With Previous High-$z$ Cluster Results}
\label{subsec:rlf_highz_compare}

In this sub-section, we compare our RLF results from this work with 
previous work on high-$z$ cluster radio galaxies by \citet{1999AJ....117.1967S}
and \citet{2006A&A...446...97B}. Before the present study, it appeared that 
the results from those two samples were incompatible. \citet{2006A&A...446...97B} claimed probable 
density and luminosity evolution of the RLF of cluster radio galaxies between $z$$\sim$1 and 0, while \citet{1999AJ....117.1967S} 
claimed no evolution over a similar redshift interval. With the 
results of this work before us, we hypothesize that both of these results
(as well as the current results) are correct and their differences are
ascribable to differences in the cluster sample selected for study. Here we compare these
three results to see if they are compatible with this hypothesis.

\begin{figure*}
\begin{center}
\plottwo{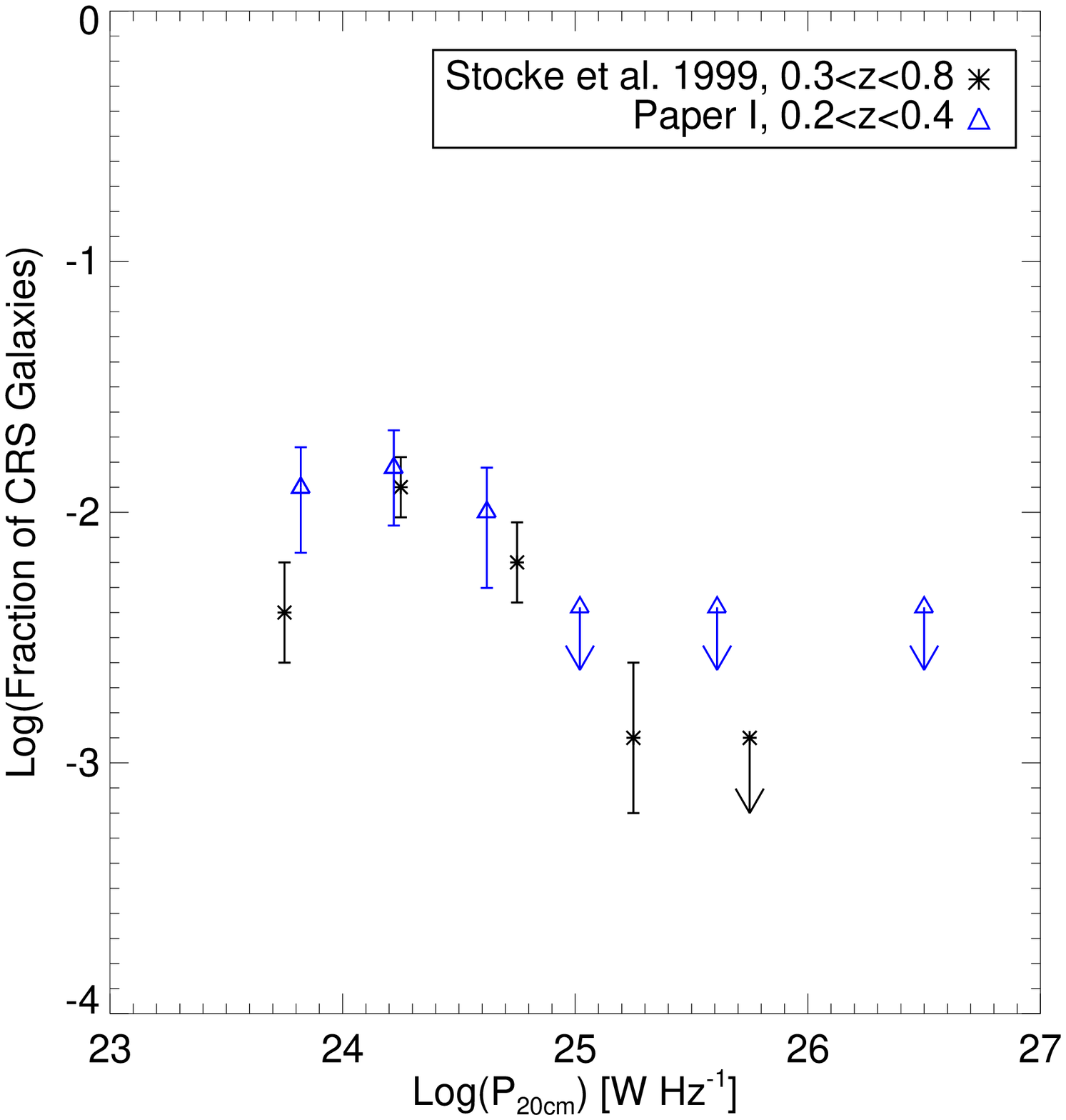}{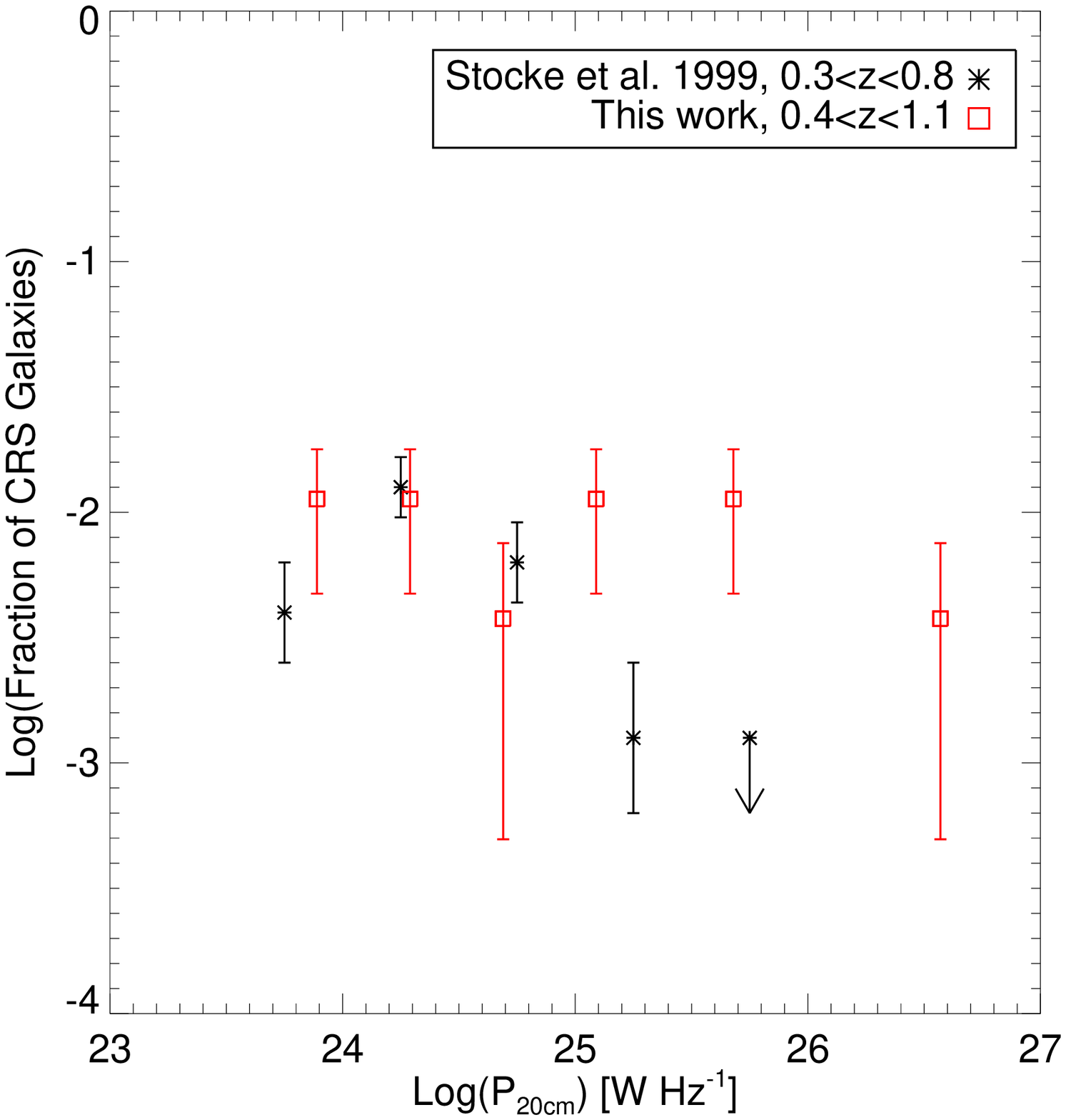}
\caption{
RLF comparison between 0.3$<$$z$$<$0.8 EMSS radio galaxies and \lowz\ cluster radio galaxies (left) and \combozb\ cluster radio galaxies (right).
The y-axis is defined as the fraction of CRS galaxies located within 600 kpc of the cluster center
(to match the survey area in \citet{1999AJ....117.1967S})
that are detected as radio sources within the given radio power bin.
The RLF for our low-$z$ (\lowz) radio galaxies
is consistent with radio galaxies in the flux-limited X-ray sample of massive galaxy clusters
in the EMSS (see \citealt{1999AJ....117.1967S}).  However, the RLF of our cluster radio sources at $z$$>$0.4 deviates from the EMSS RLF
at \logp$>$25.0~\radiounits, similar to \combozb\ radio source
comparisons to the local Abell Cluster radio sources.  Clearly, our unique selection of clusters as Coma
Cluster progenitors is important to understand the evolution of radio galaxies in
cluster environments.  EMSS clusters have X-ray luminosities similar to our \lowz\ clusters
(\lxbb$\sim$5$\times$10$^{44}$ \lum), so it is not surprising that the EMSS RLF is similar to the
RLF of our cluster radio galaxies at \lowz.
}
\label{fig:p2_rlf_emss}
\end{center}
\end{figure*}

\citet{2006A&A...446...97B} detected 32 radio galaxies in eighteen 0.3$<$$z$$<$0.8 clusters from the ROSAT North Ecliptic Pole (NEP) survey
and find potential evidence for evolution in the RLF since $z$$=$0.8.
\citet{1999AJ....117.1967S} find no evolution in the RLF of 21 cluster radio galaxies in 19 EMSS clusters at 0.3$<$$z$$<$0.8 when 
compared to the RLF of $z$$<$0.09 Abell cluster radio sources in \citet{1996AJ....112....9L}.  
However, EMSS clusters were identified in a flux-limited X-ray sample, which preferentially selects very luminous clusters at high-$z$.  
Figure~\ref{fig:p2_rlf_emss} compares the RLF of cluster radio galaxies in this study to the RLF
of EMSS radio galaxies.  We re-calculated the radio powers of
individual EMSS radio galaxies to correct for the different cosmology used in \citet{1999AJ....117.1967S}.
The turn-over in the EMSS RLF at \logp$\leq$24.0 \radiounits\ is due to incompleteness at those power levels \citep{1999AJ....117.1967S}.  
At 24.0$<$\logp$<$25.0 \radiounits, we find that our low-$z$ and combined high-$z$ RLFs 
are well-matched to the RLF from the EMSS survey (refer to Table~\ref{tab:p2_rlf_diffs} and Figure~\ref{fig:p2_rlf_emss}).  
At \logp$>$25.0 \radiounits, our combined high-$z$ RLF values are 
larger than those from the EMSS RLF.  The EMSS RLF predicts a detection of 0.3 radio galaxies for our combined high-$z$ sample, but we
detect 6 radio galaxies (cumulative binomial probability of 7$\times$10$^{-5}$\%).

There are some major differences between the radio galaxy sample drawn from the NEP and EMSS cluster samples
and this current cluster sample.  As noted by \citet{2006A&A...446...97B}, the mean X-ray luminosity of NEP clusters is 
lower than that of a typical EMSS cluster, which is twice as luminous
as a typical \rtc\ progenitor cluster at comparable redshifts.  Thus, the more luminous EMSS clusters at 
high-$z$ have similar masses and X-ray temperatures (see Figure~\ref{fig:cluster_sim}) as our low-$z$ clusters, suggesting that EMSS radio galaxies probably reside 
in similar cluster environments as our low-$z$ radio galaxies. 
Thus, the ``no evolution" conclusion from \citet{1999AJ....117.1967S} is
consistent with our hypothesis that the RLF depends on the cluster sample selection.
In conclusion, the results from our internal and external comparisons of the RLF of cluster radio galaxies
indicate a significant change in the RLF from $z$$=$0--0.8 in an appropriately chosen sample (see \S\ref{sec:p2_rtc}).
The results presented in this section show that the sample selection of distant clusters 
does result in finding cluster AGN populations that differ.

\section{X-ray Point Sources in 0.4$<$$z$$<$1.2 Clusters}
\label{sec:p2_xps}

\subsection{Statistical XPS Excess Over Background?}
\label{subsec:p2_xps_excess}

Several authors have investigated the statistical XPS excess around clusters of galaxies.
\citet{2005AA...430...39C} detected XPS over-densities in several clusters, including RXJ0849, RXJ0910, and MS2053,
but did not detect over-densities in other cluster fields that overlap with our cluster fields,
including RXJ2302, MS1621, and RXJ1113.
\citet{2005AA...430...39C} suggest that the $>$2$\sigma$ excess in XPS counts in cluster fields map the
filamentary structures (R$\sim$3-5 Mpc) around clusters.  These authors note that this XPS excess is statistically
insignificant if larger FOVs are utilized to compute \logns\ plots, suggesting that the
XPS excess is consistent with the increase of the angular correlation signal at small angular scales.
\citet{2005AA...430...39C} detect 20--25\% variations about their XPS non-cluster reference fields.
Due to the small areal coverage in this current work (0.8~deg$^{-2}$), the differences due to cosmic variance
could contribute a similar magnitude of error as Poisson statistics alone. 

In a survey of 51 massive clusters from the Massive Cluster Survey (MACS), \citet{2005ApJ...623L..81R}
find a prominent XPS excess within 0.5 Mpc of the cores of clusters with relaxed X-ray morphology, 
but the XPS excess in MACS clusters with disturbed X-ray morphologies appears more uniformly distributed within a
larger 3.5 Mpc survey region. 
Similarly, within 1 Mpc of 148 galaxy clusters between 0.1$<$$z$$<$0.9, \citet{2009MNRAS.392.1509G} measure a 
3$\sigma$ excess in XPS counts when compared to 44 control fields, resulting in $\sim$~1.5 XPSs per cluster across their sample.
However, these authors find more AGN in the outer 0.5-1.0 Mpc regions than within 0.5 Mpc.
\citet{2007AA...462..449B} find a 2$\sigma$ excess in the total XPS population within
0.5 Mpc in 18 clusters with 0.25$<$$z$$<$1.01. Six of their clusters overlap with our high-$z$ cluster sample,
which includes MS1621, MS2053, RXJ1221, RXJ1350, RXJ1716, WARPSJ1415. 
Similarly, \citet{2007ApJS..168...19M}
find an XPS excess in RXJ0910, RXJ0849, RXJ0152, and MS1054 over most of the soft/hard X-ray bands compared
to the observed \logns\ plots from the \chandra\ Deep Field North and South.
In these large statistical surveys, excess XPSs above the background are assumed to be at
the cluster redshift, but little or no redshift information exists to verify cluster membership.
Therefore, when redshift information is lacking the evidence for a substantial population of XPSs in clusters is quite mixed.

As stated in \S\ref{subsec:p2_obs_results_sources}, only $\sim$50\% of the detected XPSs in the present study
with L$\geq$L$^*$ counterparts have known redshifts.  This incompleteness hampers any thorough
and conclusive analysis of the cluster XPSs at this time.  However, the multiwavelength nature of this
study provides additional information from the optical and/or radio, as well as results from our low-$z$ sample presented in Paper~I,
to assess cluster membership.
Therefore, in this Section we proceed as far as we can in this analysis with only partial redshift
information.

To start, we compare the total number of detected XPSs to the expected background counts to assess the 
potential number of cluster XPSs at $z$$>$0.4 statistically.  Using Figure 4 from \citet{2007ApJ...659...29K}, which
presents the XPS counts and X-ray background results from ChaMP, we
determine the expected number of background XPSs for each cluster's \chandra\ observation within the survey area of
1 Mpc from the cluster center and at X-ray fluxes required to detect cluster XPSs at \xlimb\ at the cluster redshift.  
However, due to the  small survey area of individual cluster observations, we combine the expected numbers into low-$z$,
mid-$z$, and high-$z$ redshift bins as defined in Table~\ref{tab:p2_rtc_sample}.
Based on the cosmic X-ray background observations from the the X-ray Telescope (XRT) onboard {\it Swift} \citep{2009A&A...493..501M} and
the \chandra\ COSMOS survey \citep{2009ApJS..184..158E}, we estimate that cosmic variance contributes 20\% to the
statistical fluctuation in our observed XPS counts and use that amount to compute the total error on the expected number of background XPSs
in Table~\ref{tab:xps_excess}.

Table~\ref{tab:xps_excess} lists the total number of expected background and our detected XPSs at \lxbb$\geq$\xlimb\ if located
at the cluster redshift within 1 Mpc of the cluster core for three different redshift bins.  This table includes the
number of clusters in each redshift bin, the total survey area, and the cut-off X-ray luminosity at each individual cluster redshift.
The statistical significance of the difference between the expected background and our detected XPS counts is listed in Column~7.
We find no strong statistical evidence for an XPS excess 
in our eleven \lowz\ cluster fields nor in our five \midz\ cluster fields.
These results are consistent with results from ChaMP \citep{2004ApJ...600...59K}, who
find no difference in the surface density distribution
of XPSs as a function of flux in \logns\ plots for \chandra\ fields with and without clusters and
no difference in either \logns\ when compared to \chandra\ Deep Field measurements of the cosmic X-ray background.  
In our six \highz\ clusters, we detect a marginal excess of XPSs at \lxbb$\geq$\xlimb\ at the 1.6$\sigma$ significance level.
Therefore, the statistical results from this survey are similar to the ChaMP results and disagree with several earlier studies mentioned previously;
i.e., we find no strong evidence for an excess of XPSs in our cluster fields.

\subsection{X-ray Active Fraction}
\label{subsec:p2_xps_excess_compare}

In Paper I,  we scrutinized the cluster XPS population to determine the nature of those sources.  Here we
summarize the results from that sample that will be used to compare to the 
cluster XPSs in our mid-$z$ and high-$z$ clusters fields.  At low-$z$, we detected 12 XPSs at
\lxbb$\geq$10$^{42}$ \ergs\ in L$\geq$L$^*$ galaxies with CRS colors.
However, only 6 of those (or 50\%) are spectroscopically-confirmed cluster members.
Within the same luminosity and magnitude limits, only 1 of 14 XPSs with colors bluer than
the CRS (or 7\%) is a cluster member (Source A in Figure 2 of Paper I).
This one blue cluster XPS displays the optical properties of a 
typical Seyfert, \lxbb=4.6$\pm$0.1$\times$10$^{43}$ \ergs\ with an X-ray spectrum well-modeled by a power-law 
with spectral index $\Gamma=2.0\pm0.1$, emission from the 6.4 keV Fe K-$\alpha$ line with an equivalent width of 
1.7 keV, and no intrinsic absorption.  

Moreover, we identified two cluster XPSs with slight ``blue excesses" compared to CRS colors
(0.2--0.5 magnitudes bluer than the CRS; See Figure 2 of Paper I).  The bluest of
these two (Source D in Figure 2 of Paper I) is slightly fainter than M$^*$.  These 2 XPSs have passive optical spectra
with no apparent emission lines, and so are very different from the one blue cluster XPS.
We suggested that these two passive blue AGN are most similar to low luminosity, X-ray-detected BL-Lacertae (BL Lac) 
objects that typically have an enhanced blue continuum, which can explain the blue excess \citep{1999ApJ...516..145R}.
Recent spectro-polarimetry at the {\it Subaru} 8m telescope of these two passive blue AGN,
confirm the BL Lac classification by detecting continuum polarization \citep{2011ApJ...732...23T}.  
Because these ``passive blue AGN" are most similar to the CRS XPSs rather than the Seyfert-like AGN, we
included them in the CRS AGN population.

Based on the detection rate of CRS XPSs per cluster from Paper I, we expect to 
detect 1--2 XPSs at \lxbb$\geq$5$\times$10$^{42}$ \ergs\ in our combined high-$z$ cluster sample.
There are a few candidate sources for passive blue AGN in our high-$z$ clusters 
(see Figures~\ref{fig:p2_cmd_mid_z}--\ref{fig:p2_cmd_high_z}), including 2 confirmed
cluster members with passive optical spectra (Object C = RXJ1716-X13 and
Object D = RXJ0910-X12 in Table~\ref{tab:p2_xps_high} and Figure~\ref{fig:p2_cmd_high_z}).  
Thus, while not extremely numerous, passive blue AGN
do appear to be a new type of cluster active galaxy not identified
heretofore. Sources C \& D, as well as those few unconfirmed XPSs slightly
bluer than the CRS, deserve further scrutiny as potential low-luminosity BL Lac Objects.  

At \comboz\, we find 62 XPSs in 6 cluster fields with fluxes that correspond to \lxbb$\geq$\xlimb\ at the cluster redshift.  From
Table~\ref{tab:xps_excess} the expected number of background XPSs is 44.3$\pm$11.1,
so there could be a substantial number of cluster XPSs.  We cannot tell without optical spectroscopy,
but 7 XPSs listed in Table~\ref{tab:p2_xps_high} are spectroscopically-confirmed cluster members (labeled Sources A through G), 
and all 7 lie on or near the CRS.
Thus, at a minimum, these seven CRS XPSs provide a firm {\it lower} limit of 1.4$\pm$0.8\% for the X-ray active fraction (XAF)
in clusters at \comboz\ and at \lxbb$\geq$\xlimb.  This value is an approximate 10-fold increase compared to the XAF of 0.15$\pm$0.15\% 
for CRS galaxies in our low-$z$ clusters.
If, as suggested in Paper I, the CRS XPSs are lower luminosity radio sources, we would expect
a similar evolution for the CRS XPSs and the radio galaxies.
If confirmed, this increase in the XAF is additional evidence that the radio galaxies and CRS XPSs are members of
the same parent population of cluster radio galaxies.

Now we compare our cluster XPS results to \citet{2009ApJ...701...66M}, who
surveyed 32 galaxy clusters at 0.05$<$$z$$<$1.3 (16 clusters at
0.05$<$$z$$<$0.4 and 15 clusters at 0.4$<$$z$$<$1.3) within the virial radius to investigate the
population of luminous \lxbb$>$10$^{43}$ \ergs\ XPSs.  Their study found an eight-fold increase
in the XAF in clusters from 0.05$<$$z$$<$0.3 to 0.4$<$$z$$<$1.3.
Nine clusters overlap with this work (MS1008, MS1455, MS1358 at \lowz;
MS1621, MS2053, RXJ2302 at \midz; RXJ1716, RXJ0910, RXJ0848 at \highz).
Because \citet{2009ApJ...701...66M} lack photometric catalogs for
all of their clusters, they cannot distinguish between red and blue cluster XPSs.
So, while we find comparable evolutionary results, there are significant 
differences in the AGN populations found in these two studies.
\citet{2009ApJ...701...66M} have identified numerous broad and narrow
emission-line (i.e., Seyfert-like) AGN similar to the one Seyfert we found
in our low-$z$ sample mentioned previously.
Approximately 50\% of the X-ray AGN in \citet{2009ApJ...701...66M} are located within 0.35 R$_{200}$, while 45\%
are located within 0.5 and 1.0~R$_{200}$.  These emission-line AGN are located well outside the cluster core at a
large fraction of R$_{200}$, while $>$50\% of our XPSs are CRS absorption-line
galaxies close to the cluster core at a few tenths of R$_{200}$.  And while the \citet{2009ApJ...701...66M} sample does contain
a few CRS AGN, our sample could contain several Seyfert-like AGN if we had spectroscopy available to confirm them.
So, while our evolutionary statistics are broadly consistent in showing a dramatic
increase in X-ray active fractions above and below $z$$\approx$ 0.4, these
two studies have found two different types of cluster AGN. Further study
of both AGN types in high-$z$ clusters clearly are needed.
This can be accomplished by obtaining more spectroscopic redshifts for our sample, while two-color
imaging and VLA mapping would greatly enhance the \citet{2009ApJ...701...66M} study.

\subsection{Cluster Red Sequence XPSs at $z$$>$0.8}
\label{subsec:p2_red_xps}

Seven spectroscopically-confirmed cluster members are detected at $z$$>$0.8.
These sources are marked with letter identifiers (A-G) in Column 1 of Table~\ref{tab:p2_xps_high}, in Table~\ref{tab:p2_xps_high_all}, and
on Figure~\ref{fig:p2_cmd_high_z}.  Sources A-C are located in RXJ1716, Sources D-E are in RXJ0910, and
Sources F-G are in RXJ0849. So while there are other candidate CRS X-ray AGN, spectroscopic confirmation is
limited to these three well-studied cluster fields.  All seven sources have been previously identified by 
SEXSI \citep{2006ApJS..165...19E} and by ChaMP \citep{2004ApJ...600...59K}.
Unless otherwise noted, spectral line identifications are provided by Megan Eckart 
of the SEXSI collaboration.

Sources A-C appear elliptical in our our ARC 3.5m images, as well as in ChaMP images for Source C;
Sources D-F appear elliptical in the HST images.
In the HST image, Source G is hosted by a galaxy that appears slightly irregular with some ``disky" features, 
even though the host galaxy colors are consistent with the CRS.  The host galaxy of Source G has a relative velocity of 2120 \kms, 
or 1.6$\sigma$ from the cluster mean. This XPS may be the BCG of a group of galaxies falling into the main cluster.
The optical spectra of these XPSs are typical, passive elliptical galaxy spectra with \cahk\ absorption and
weak \oii\ emission.  Source A also displays strong \oiii\  emission.  
Source D and Source G are $\sim$1.0 and $\sim$0.5 magnitudes brighter in the rest-frame Sloan {\it i} band than their cluster BCG, 
as might be expected for clusters ``in formation" at high-$z$.

An eighth potential XPS source, RXJ1350-X7, is coincident with the radio counterpart RXJ1350-R2 (Source 12 in Figure~\ref{fig:p2_cmd_high_z}).
In our ARC 3.5m Sloan ({\it r},{\it i}) optical images, two galaxies are located within the 95\% encircled energy
radius for this XPS.  The  best match to the radio AND X-ray source coordinates is the optical source to the SW 
($\alpha$=13:50:50.1 and $\delta$=60:08:03.8), which has a smooth appearance in the ARC 3.5m images.
Although lacking a redshift, we have counted this object as a cluster radio galaxy in \S\ref{sec:p2_rg}. 
Based upon a limited number of examples, the 7 spectroscopically-confirmed CRS XPSs appear 
similar in optical morphology and colors to their low-$z$ counterparts, suggesting that they are a 
similar AGN population to what we found in Paper I.

\section{Implications for Cluster ICM Heating}
\label{sec:p2_implications}

Based on our results from \S\ref{sec:p2_rlf}, we clearly detect
evolution in the radio galaxy population in Coma Cluster progenitors from $z$$=$1.1 to $z$$=$0.2.
In Paper I, we measured an RAF of 2.5$\pm$0.8\% for the cluster radio sources and it is 
3.8$\pm$1.2\% in our combined high-$z$ sample.  Referring to the right panel of
Figure~\ref{fig:p2_rlf_hist}, the radio powers of the BCGs at \combozb\ are also larger
than the BCGs at \lowz.  Cluster radio sources are still hosted by CRS galaxies and
$\sim$75\% of these radio sources are located within 500 kpc of the cluster core.  
At \midz, the cluster radio galaxies remain more centrally concentrated than the CRS galaxies, 
similar to our results from Paper I.  But, at \highzsmall, cluster radio galaxies do not appear as centrally concentrated, even though
their percentage within 500 kpc is similar to those at low-$z$ and mid-$z$.

In Paper I, we used the scaling relationship between AGN jet power and radio luminosity (P$_{jet}$ $\propto$ L$_r^{0.5}$) from
\citet{2008ApJ...686..859B} to estimate that low-luminosity radio
galaxies at \logp$\geq$21.4 \radiounits\ contribute $\sim$55\% of the total heat input to the ICM.  
Recently \citet{2010ApJ...720.1066C} extended the \citet{2008ApJ...686..859B} radio galaxy sample to lower radio power 
(L$_{1.4GHz}$$>$10$^{21}$ \radiounits\=10$^{37}$ \lum) and 
measured a steeper slope of 0.7 for this scaling relationship.  With this steeper scaling
relationship, we estimate that low-luminosity radio galaxies at \logp$\geq$21.4 \radiounits\ contribute 
$\sim$30\% of the total heat input to the ICM, somewhat smaller than our previous estimate.

We now use the updated \citet{2010ApJ...720.1066C} scaling relationship, P$_{jet}$ $\propto$ L$_r^{0.7}$, 
to estimate the increased jet power per cluster from the enhanced population of FR IIs at 
25.0$<$\logp$<$27.0 \radiounits\ in our combined high-$z$ sample.
No radio sources at \logp$\geq$25.0 \radiounits\ were detected in our low-$z$ cluster sample from Paper I.
Using Figure~\ref{fig:p2_rlf_ledlow}, the total integrated radio power per cluster is 3.4$\times$10$^{24}$ 
\radiounits\ at \lowz\ and 8.9$\times$10$^{25}$ \radiounits\ at \combozb, approximately 26$\times$ higher 
than the integrated luminosity in Coma Cluster progenitors at \lowz.
This increase in radio power per cluster corresponds to a 9--10 fold increase in the integrated jet power
and thus more heat is predicted to be available for deposition into 
the surrounding ICM by these radio jets.  Figure~\ref{fig:p2_rxj1716_fr2} shows an excellent example of a
high-$z$ ($z$$=0$.8) cluster from our sample in which three luminous radio
galaxies are found, including in the BCG. However, the most luminous by-far
is well-offset from the cluster center and so can provide significant ICM
heating which is both off-center and mobile with respect to the
intra-cluster gas.  A large majority (75\%) of cluster radio galaxies at $z$$>$0.4 are 
located within 500 kpc, so the additional sources of heating, at higher radio powers,
will be concentrated in the inner ICM regions as well.  

The impact of cluster XPSs on the heating of the ICM is more difficult to assess due to the spectroscopic
incompleteness of our sample, but is likely much less than the radio galaxy population.
Both this paper and \citet{2009ApJ...701...66M} find a marked increase in the X-ray active fraction in
higher redshift clusters, comparable to the increase in the radio galaxy
active fractions. The increase we have found is in the CRS AGN population
close to the cluster core, which we have ascribed to low luminosity BL Lac
Objects and thus to quite weak radio jets. \citet{2009ApJ...701...66M} find this
increase predominantly in Seyfert-like AGN which are much further from the
cluster center, near R$_{200}$ (although these authors do identify a few
absorption line AGN close to the cluster centers as well). The heat input
provided by Seyfert-like (i.e., radio-quiet) AGN is not well-known or
measured \citep[e.g.,][]{2002ApJ...566..699A}, but any heat
they provide would be well outside the dense ICM region. Therefore, the
radio galaxies provide the bulk of the ICM heat input and so the analysis
outlined above provides strong evidence for a greatly increased heat input
to higher-$z$ cluster ICM gas. But, based on the current work and 
\citet{2009ApJ...701...66M}, there is still much work to be done
to better characterize the ICM heat input, including a more complete census
of the absorption and emission-line AGN in high-$z$ clusters and their radial
distribution relative to the ICM. Theoretical work which models the heat
input by non-central, moving AGN sources embedded in an evolving, less dense 
ICM are needed to assess the efficiency of such input as a function of redshift.

\section{Summary and Conclusions}
\label{sec:p2_conclusions}

As a companion paper to \citet[][Paper I]{hart09}, we have presented an extensive investigation of the multiwavelength nature 
of AGN in a sample of 22 clusters of galaxies at 0.2$<$$z$$<$1.2, specifically chosen to be most 
similar to Coma Cluster progenitors at those redshifts.  Using the X-ray temperature of
the ICM as a proxy for cluster mass, we chose massive clusters (or high X-ray temperature clusters) 
at lower redshifts and less massive clusters (or lower X-ray temperature clusters) at higher
redshifts to investigate the X-ray point source (XPS) and radio galaxy populations.
In both wavelength regimes, the limiting luminosities were chosen to sample only the AGN populations in L$\geq$L$^*$ galaxies.

Within 1 Mpc of the cluster center, we detect 19 radio galaxies at \plim\ in our \comboz\ clusters (11 of which are 
spectroscopically-confirmed cluster members and 8 of which are likely cluster members
because they have colors similar to the cluster red sequence (CRS).  
Because 8 CRS AGN do not have confirming spectroscopic redshifts, we conservatively reduce the total number of 
radio galaxies in this sample by 2 to 17 based upon the foreground/background AGN contamination we found within the bounds of the CRS in Paper I.
Similarly, we detect 7 spectroscopically-confirmed cluster XPSs.  Both the X-ray and radio AGN have host galaxies 
with colors consistent with CRS galaxies.  
There is some evidence that the radio galaxies become more centrally concentrated in these clusters at lower redshift.
Among the radio galaxies we have found a clear FR~II morphology source at $z$$=$0.8 with 
P$_{1.4GHz}$$=$7.7$\times$10$^{26}$ \radiounits, while no FR~IIs were detected at P$_{1.4GHz}$$>$10$^{25}$~\radiounits\
in our low-$z$ sample.  In addition to this one luminous FR-II, we detect several other radio galaxies at
P$_{1.4GHz}$$>$10$^{25}$ \radiounits\ in our $z$$>$0.4 clusters, including several BCGs.  These sources
are all more powerful than any we found at low-$z$.

While previous work \citep[e.g.,][]{1991ApJ...371...49E,1991ApJ...367....1H,2002AJ....124.1239H}
has discovered clusters of galaxies around clear
FR~II morphology and power level sources at $z$$>$0.4, here we started
with rich clusters at those redshifts and have discovered an FR~II.
Therefore, there is no doubt that the morphology and power level of some
cluster radio galaxies is changing between $z$$=$1 and 0. Specifically, this
study provides further support for a phenomenological model in which the power
levels of subsequent outbursts of radio galaxies in clusters diminishes
rapidly between these redshifts \citep{1991ApJ...371...49E,2002AJ....124.1239H},
the same range of redshifts over which the ICM density and
temperature dramatically increase (see Figure 1). Evidence presented
herein and in Paper I that radio galaxies are more centrally concentrated
in their clusters than the CRS elliptical galaxy population as a whole 
is further evidence for a link between AGN activity and ICM
density \citep[see also][]{1981ApJ...245..375S}. It is possible that this
AGN/ICM relationship may provide the necessary negative feedback mechanism
to prevent runaway cooling in cluster cores \citep[e.g.,][]{2008MNRAS.386.1309M}.

A comparison between the RLF of our \lowz\ radio sources versus those in our \comboz\
sample reveals a larger number of high-power sources at earlier epochs, evidence for positive evolution of these
sources in Coma Cluster progenitors.  A comparison
between the RLF of our \lowz\ cluster radio sources to local (z$<$0.09) Abell cluster radio 
sources shows that the two populations are similar, suggesting little evolution between \lowz\ and $z$$=$0.  
The clear density and luminosity evolution of radio galaxies found in our
sample of cluster AGN is quite different from the
absence of evolution found by \citet{1999AJ....117.1967S} using a flux-limited
sample of EMSS clusters.  We interpret this lack of agreement as support
for our novel, physically-based method for selecting clusters for study,
the ``Road to Coma'' concept presented in the \S\ref{sec:p2_rtc}. We suggest
that both AGN evolution results are correct and that \citet{1999AJ....117.1967S} failed
to find cluster AGN evolution because their high-$z$ sample has very
similar ICM properties to the low-$z$ sample of \citet{1996AJ....112....9L}, to which
they compared AGN populations. For example, the hot, $z$$=$0.8 cluster, MS
1045-03 (see Figure~\ref{fig:cluster_sim}) is a member of the \citet{1999AJ....117.1967S} high-$z$ sample and
has a similar radio galaxy population to many $z$$=$0 rich Abell clusters.
Our sample was selected to avoid that pitfall and we
conclude that sample selection is as critical for cluster galaxy and AGN
studies as it is for other evolutionary investigations.

We can be far less definitive concerning the evolution of cluster X-ray
AGN due to the large number of XPSs in our high-$z$ cluster fields and to
the absence of redshift information for most of them. However, there are
sufficient redshifts (7) for CRS XPSs to show that the passive,
absorption-line AGN population that was identified in Paper I persists at
high-$z$. Although the statistics are quite sparse, a substantial increase
in either the density and/or the luminosity of this AGN class (at least an
order of magnitude increase in active fraction at \lxbb$\geq$5$\times$10$^{42}$ \ergs\
when compared to low-$z$) is very similar to the rapid increase
recently found in the cluster X-ray AGN population by \citet{2009ApJ...701...66M}.

However, our study is more complementary to this previous
study than confirming because different AGN classes are involved. Optical
spectroscopy, primarily derived from the SEXSI study, and deep
ground-based and HST imaging for our CRS XPSs finds that these passive AGN
are hosted in rather normal, giant elliptical galaxies positioned rather
close to the cluster core, whereas the \citet{2009ApJ...701...66M} AGN are largely
emission-line galaxies similar to Seyferts found further out in the
clusters near R$_{200}$. Since the energy and mass outflow from Seyferts
(i.e., radio-quiet AGN population) is still not well-measured \citep{2002ApJ...566..699A},
it is not known whether these AGN contribute significantly
to ICM heating or not. 

Paper I identified the CRS XPSs as low-luminosity BL Lac Objects, suggesting that they
possess relativistic jets.  But the absence of radio emission from these
sources suggest that these jets are too weak
to inject much heat into the ICM (see Paper I for details). However, these
CRS X-ray AGN are worthy of further study since deep X-ray imaging 
\citep[e.g.,][]{2001AJ....121..662B} finds that a substantial fraction ($\sim$20\%) of the
faint XPS population is due to passive AGN like these cluster sources. So
their detailed nature and cosmological evolution is of considerable
interest.

Using the recent scaling law between AGN jet power and the radio luminosity suggested by \citet{2010ApJ...720.1066C},
P$_{jet}$ $\propto$ L$_r^{0.7}$, we determine that the excess power from
luminous radio galaxies at 23$<$\logp$<$27 \radiounits\ is $\sim$26$\times$ higher
in \combozb\ clusters than in \lowz\ clusters, which corresponds to 9--10 fold increase
in the AGN jet power if FR~II jets have similar kinetic energies to FR~I jets.  
Using the more shallow scaling between AGN jet power and the radio luminosity
from \citet{2008ApJ...686..859B} corresponds to a 5--6 fold increase.  Therefore, a larger amount of heat can be deposited into the ICM
at high-$z$.  The results presented herein and by previous studies of AGN in high-$z$
clusters \citep[e.g.,][]{1991ApJ...371...49E,1991ApJ...367....1H,2002AJ....124.1239H,2009ApJ...701...66M}
invites new theoretical work on the co-evolution of AGN and the cluster ICM and its effect on ICM
heating and cooling.

\appendix
\section{Detected XPS\lowercase{s} in \comboz\ Clusters}
This section contains Table~\ref{tab:p2_xps_high_all}, a complete list of detected XPSs in our \comboz\ clusters
listed in Table~\ref{tab:p2_rtc_sample}.  These XPSs, if located at the cluster redshift, are at
\lxbb$\geq$\xlimb\ and are located at a projected distance less than 1 Mpc from
the cluster X-ray emission centroid (see Columns 2-3 of Table~\ref{tab:p2_rtc_xray}).  For completeness we include the sources from
Table~\ref{tab:p2_xps_high}, which lists only the spectroscopically-confirmed cluster XPSs within the above limits.

\acknowledgments
QNH and JTS acknowledge the support from the {\it Chandra} Theory/Archive Grant AR8-9013X.
AEE acknowledges support from NSF AST-0708150 and NASA NNX10AF61G.
We thank Brian Keeney for assistance with our optical observations,
Yasuhiro Hashimoto for providing use of the Subaru images of RXJ1053,
Eric Perlman and Harald Ebeling of the WARPS collaboration for providing 
1.4 GHz maps and spectroscopic information,
and Megan Eckart for providing us with detailed spectral line identifications of
X-ray AGN in clusters that overlap with the SEXSI study.
QNH acknowledges useful conversation with Jack Burns and Eric Hallman.

We would like to acknowledge Paul Green and the {\it Chandra} Multi-wavelength Project
\citep[ChaMP; ][]{2004ApJS..150...43G} collaboration for providing use of the their photometric datasets on
some cluster fields.  ChaMP is supported by NASA. Optical data for ChaMP are
obtained in part through NOAO, operated by the Association of Universities
for Research in Astronomy, Inc. (AURA), under cooperative agreement with the National Science Foundation.

Funding for the SDSS and SDSS-II has been provided by the Alfred P. Sloan Foundation, the Participating Institutions, the National Science Foundation, the U.S. Department of Energy, the National Aeronautics and Space Administration, the Japanese Monbukagakusho, the Max Planck Society, and the Higher Education Funding Council for England. The SDSS Web Site is http://www.sdss.org/.
This research has made use of data obtained from the {\it Chandra} Data Archive and software provided by the 
{\it Chandra} X-ray Center (CXC) in the application packages CIAO, ChIPS, and Sherpa.

Some of the data presented in this paper were obtained from the Multimission Archive at the Space Telescope 
Science Institute (MAST). STScI is operated by the Association of Universities for Research in Astronomy, Inc., 
under NASA contract NAS5-26555. Support for MAST for non-HST data is provided by the NASA Office of Space Science 
via grant NNX09AF08G and by other grants and contracts. 

This research has made use of the NASA/IPAC Extragalactic Database (NED) which is operated by the 
Jet Propulsion Laboratory, California Institute of Technology, under contract with the National Aeronautics and Space Administration,
as well as NASA's Astrophysics Data System.

{\it Facilities:} \facility{CXO}, \facility{ARC (3.5m)}, \facility{VLA}, \facility{Sloan}, \facility{HST}

\bibliographystyle{apj}

\clearpage
\setlength{\tabcolsep}{12pt}
\begin{deluxetable}{lclcc}
\tabletypesize{\scriptsize}
\tablecaption{The ``Road to Coma'' Sample: 23 Galaxy Clusters between 0.2$<$$z$$<$1.2 \label{tab:p2_rtc_sample}}
\tablewidth{0pt}
\tablehead{
\colhead{Redshift} & \colhead{No. of} & \colhead{Cluster} & \colhead{{\it z}} & \colhead{kT} \\
\colhead{Bin} & \colhead{Clusters} &  \colhead{Name} &  \colhead{} & \colhead{(keV)}\\
\colhead{(1)} &  \colhead{(2)} & \colhead{(3)} & \colhead{(4)} & \colhead{(5)}
}
\startdata
\lowz\ (Low-$z$) & 11 & MS 0440.5+0204  & 0.197 & 6.1$^{+0.6}_{-0.5}$  \\
 & & Abell 963       & 0.206 & 7.1$^{+0.2}_{-0.2}$  \\
 & & RX J0952.8+5153 & 0.214 & 5.2$^{+0.2}_{-0.2}$  \\
 & & Abell 2111      & 0.229 & 7.2$^{+0.6}_{-0.5}$  \\
 & & MS 1455.0+2232  & 0.260 & 4.5$^{+0.1}_{-0.1}$  \\
 & & Abell 1758      & 0.280 & 7.5$^{+0.5}_{-0.3}$  \\
 & & MS1008.1-1224   & 0.306 & 6.1$^{+0.4}_{-0.5}$  \\
 & & MS2137.3-2353   & 0.313 & 4.6$^{+0.1}_{-0.1}$  \\
 & & Abell 1995      & 0.319 & 8.8$^{+0.5}_{-0.5}$  \\
 & & MS1358.4+6245   & 0.330 & 8.2$^{+0.7}_{-0.7}$  \\
 & & Abell 370       & 0.373 & 8.5$^{+0.5}_{-0.5}$  \\
\midz\ (Mid-$z$) & 5 & MS 1621.5+2640  & 0.426 & 6.2$^{+0.7}_{-0.6}$ \\
 & & MS 2053.7-0449  & 0.587 & 4.5$^{+1.0}_{-0.7}$ \\
 & & RX J1221.4+4918 & 0.700 & 6.9$^{+0.7}_{-0.7}$ \\
 & & RX J2302.8+0844 & 0.720 & 7.4$^{+2.4}_{-1.0}$  \\
 & & RX J1113.1-2615 & 0.725 & 5.6$^{+1.7}_{-1.0}$  \\
\highz\ (High-$z$) & 7 & RX J1350.0+6007 & 0.796 & 4.5$^{+0.9}_{-0.7}$  \\
 & & RX J1716.6+6708 & 0.809 & 6.7$^{+1.0}_{-0.9}$  \\
 & & RX J0152.7-1357$^a$ & 0.835  & 5.0$^{+0.7}_{-0.7}$  \\
 & & WARPS J1415.1+3611 & 1.013 & 5.1$^{+0.6}_{-0.5}$  \\
 & & RX J0910.7+5422 & 1.106  & 5.3$^{+3.7}_{-1.7}$  \\
 & & RX J1053.7+5735 & 1.134  & 3.3$^{+0.7}_{-0.6}$ \\
 & & RX J0848.9+4452 & 1.260  & 3.4$^{+2.1}_{-0.8}$ 
\enddata
\tablecomments{
Columns: 
(1) Redshift bin 
(2) Number of clusters in this redshift bin
(3) Cluster name 
(4) Redshift 
(5) Bulk X-ray temperature of the intracluster medium. See \S~\ref{subsec:p2_obs_xray} for details.}
\tablenotetext{a}{The X-ray emission of RXJ0152 is double-peaked. The quoted bulk ICM temperature is extracted from a 
region that includes a 500 kpc radius around EACH X-ray clump. The resulting extraction radius is 900 kpc.  
The individual X-ray temperature within 500 kpc of the clump centers is 5.6$^{+1.0}_{-0.7}$ for the North clump and 
5.2$^{+0.7}_{-0.7}$ for the South clump.}
\end{deluxetable}

\setlength{\tabcolsep}{6pt}	

\begin{deluxetable}{lllccccc}
\tabletypesize{\scriptsize}
\tablecaption{\chandra\ X-ray Observations of Twelve 0.4$<$$z$$<$1.2 Clusters \label{tab:p2_rtc_xray}}
\tablewidth{0pt}
\tablehead{
\colhead{Cluster} & \colhead{RA} & \colhead{DEC} &  \colhead{{\it Chandra}} & \colhead{ACIS} & \colhead{Exp. Time} & \colhead{F$^{limit}_{X}$} & \colhead{L$^{limit}_{X}$} \\
\colhead{} & \colhead{} & \colhead{} & \colhead{ObsID} & \colhead{Chip} & \colhead{(ksec)} &\colhead{$\times$10$^{-15}$} & \colhead{$\times$10$^{41}$} \\
\colhead{(1)} &  \colhead{(2)} & \colhead{(3)} & \colhead{(4)} & \colhead{(5)} & \colhead{(6)} & \colhead{(7)} & \colhead{(8)}
}
\startdata
MS 1621.5+2640  	& 16:23:35.3 & +26:34:20.6 & 546   & I3 & ~29.9 & 3.2 & 2.1 \\
MS 2053.7-0449  	& 20:56:21.4 & -04:37:49.1 & 1667  & I3 & ~43.9 & 0.8 & 2.3 \\
RX J1221.4+4918 	& 12:21:26.4 & +49:18:26.8 & 1662  & I3 & ~78.5 & 1.3 & 2.8 \\
RX J2302.8+0844 	& 23:02:48.2 & +08:43:51.2 & 918   & I3 & 107.6 & 0.8 & 1.8 \\
RX J1113.1-2615$^{a}$ 	& 11:13:05.3 & -26:15:39.8 & 915   & I3 & 104.6 &  1.0 & 2.5 \\
RX J1350.0+6007$^{b}$ 	& 13:50:48.3 & +60:07:10.8 & 2229  & I3 & ~57.9 &  1.4 & 4.2 \\
RX J1716.6+6708$^{b}$ 	& 17:16:48.9 & +67:08:23.0 & 548   & I3 & ~51.5  & 1.6 & 5.0 \\
RX J0152.7-1357$^{c}$ 	& 01:52:42.0 & -13:58:01.0 & 913   & I3 & ~36.3  & 2.5 & 7.5$^*$  \\
WARPS J1415.1+3611$^{b}$ & 14:15:11.2 & +36:12:01.5 & 4163 & I3 &  ~88.8 &  0.9 & 4.9 \\
RX J0910.7+5422$^{b,d}$  & 09:10:45.0 & +54:22:03.7 & 2227 & I3 & 105.1 & 0.8 & 5.1 \\
RX J1053.7+5735$^{b,d}$  & 10:53:39.9 & +57:35:16.9 & 4936 & S3 & ~88.4 &  0.7 & 4.6 \\
RX J0848.9+4452$^{b,d}$  & 08:48:58.7 & +44:51:52.5 & 927,1708  & I2 & 185.1 &  0.5 & 4.8
\enddata
\tablecomments{
Columns: 
(1) Cluster name 
(2-3) RA and DEC (J2000) of the ICM emission centroid based on 2D elliptical Gaussian profile fits.
(4)\chandra\ observation ID 
(5) ACIS aimpoint 
(6) Exposure time of the observation after filtering for flaring events
(7) X-ray flux limit (0.3-8.0 keV) in \flux\ for a point source near the edge of our survey region (R$=$1~Mpc).  See
\S~\ref{subsec:p2_obs_xray} for more details.  
(8) K-corrected X-ray luminosity limit (0.3-8.0 keV) in \ergs\ for the flux limits in Column~7.
}
\tablenotetext{*}{Although the X-ray luminosity limit for this cluster is \lxbb$\geq$5.0$\times$10$^{42}$ \ergs, RXJ0152 is included in radio-related statistics in the later sections because the 1.4~GHz radio map detects sources at P$_{1.4GHz}$$>$1.3$\times$10$^{23}$~\radiounits.}
\tablenotetext{a}{The ICM spectrum was extracted between 0.5--6.0 keV.}
\tablenotetext{b}{The ICM spectrum was extracted within a 500 kpc radius from the cluster core.}
\tablenotetext{c}{The X-ray emission of RXJ0152 is double-peaked and this coordinate is approximately between the
peaks.}
\tablenotetext{d}{The ICM spectrum was binned to a minimum of 40 counts per bin prior to XSPEC model fitting.}
\end{deluxetable}

\begin{deluxetable}{llccc}
\tabletypesize{\scriptsize}
\tablecaption{1.4~GHz Observations of Twelve 0.4$<$$z$$<$1.2 Clusters \label{tab:p2_rtc_radio}}
\tablewidth{0pt}
\tablehead{
\multicolumn{1}{l}{Cluster} &
\multicolumn{1}{l}{VLA} & 
\colhead{S$_{1.4,lim}$} & 
\colhead{P$_{1.4,lim}$} & 
\colhead{L$_{1.4,lim}$} \\
\multicolumn{1}{l}{Name} & 
\multicolumn{1}{l}{1.4~GHz Map} & 
\colhead{(mJy/Beam)} & 
\colhead{($\times 10^{23}$)} & 
\colhead{($\times$10$^{39}$)} \\
\multicolumn{1}{l}{(1)} &  
\multicolumn{1}{l}{(2)} & 
\colhead{(3)} & 
\colhead{(4)} & 
\colhead{(5)}
}
\startdata
MS 1621.5+2640$^a$  	& AP245 (A)  & 0.14 & ~0.8 & ~1.1 \\
MS 2053.7-0449  	& AP219 (B)  & 0.18 & ~2.2 & ~3.1 \\
RX J1221.4+4918 	& AS873 (B)  & 0.11 & ~2.0 & ~2.8 \\ 
RX J2302.8+0844 	& AP422 (B)  & 0.18 & ~3.5 & ~5.0 \\   
RX J1113.1-2615 	& AS873 (AB) & 0.11 & ~2.2 & ~3.1 \\
RX J1350.0+6007 	& AS873 (B)  & 0.14 & ~3.6 & ~5.0 \\
RX J1716.6+6708 	& AS873 (B)  & 0.16 & ~4.2 & ~5.8 \\
RX J0152.7-1357 	& AC757 (A)  & 0.15 & ~4.2 & ~5.9 \\
		    	& AL713 (B)  & 0.45 & 12.6 & 17.6 \\
WARPS J1415.1+3611 	& AP439 (B)  & 0.03 & ~1.3 & ~1.8 \\
RX J0910.7+5422$^{b}$  	& AS873 (B)  & 0.21 & 11.3 & 15.7 \\
RX J1053.7+5735 	& AC587 (B)  & 0.03 & ~1.7 & ~2.4 \\
RX J0848.9+4452$^{b}$ 	& AB1036 (AB)& 0.18 & 13.0 & 18.3
\enddata
\tablecomments{
Columns: 
(1) Cluster name
(2) VLA program ID with the array configuration in parentheses 
(3) 1.4~GHz radio flux density limit in mJy/Beam and 
(4) radio power limit in W~Hz$^{-1}$ and
(5) radio luminosity limit at 1.4~GHz in \ergs\ for a 3$\sigma$ detection at the edge of our survey region.
}
\tablenotetext{a}{FIRST detects sources at P$_{1.4GHz}$$>$2.3$\times$10$^{23}$~\radiounits\ or L$_{1.4GHz}$$>$3.2$\times$10$^{39}$~\ergs\ for this cluster.}
\tablenotetext{b}{This cluster is excluded from radio-related statistics because the radio power limit is P$_{1.4GHz}$$\geq$5$\times$10$^{23}$~\radiounits.}
\end{deluxetable}

\clearpage
\begin{turnpage}
\begin{deluxetable}{lccccccl}   
\tabletypesize{\scriptsize}
\tablecaption{Optical Images of Twelve 0.4$<$$z$$<$1.2 Clusters \label{tab:p2_rtc_optical}} 
\tablewidth{0pt}
\tablehead{
\colhead{Cluster} & \colhead{Color} &	\multicolumn{2}{c}{CRS Color} & \colhead{m$_{lim}$} & \colhead{m$^*$} & \colhead{N$_{CRS}$}  & \colhead{Survey Images} \\ 
\colhead{} 	  & \colhead{}      & \colhead{Observed}   & \colhead{Expected}	& \colhead{} & \colhead{} & \colhead{}    & \colhead{} \\
\colhead{(1)} 	  & \colhead{(2)}   & \colhead{(3)} 	  & \colhead{(4)}   & \colhead{(5)} & \colhead{(6)} & \colhead{(7)}  & \colhead{(8)}   }
\startdata
MS 1621.5+2640 & Sloan (r-i) 				& 0.66$\pm$0.10	& 0.89		& ({\it r,i})=(22.2, 21.3) & r=21.4 & 36  & SDSS DR6  \\
MS 2053.7-0449 & Sloan (r-i) 				& 1.27$\pm$0.15	& 1.14		& ({\it r,i})=(24.8, 24.2) & r=22.5 & 29  & ChaMP \\
RX J1221.4+4918 & Sloan (r-i) 				& 1.27$\pm$0.27	& 1.32		& ({\it r,i})=(24.0, 23.2) & r=23.8 & 52  & ARC 3.5m, 2006 \\
RX J2302.8+0844 & Sloan (r-i) 				& 1.31$\pm$0.27	& 1.35		& ({\it r,i})=(25.5, 24.6) & r=24.0 & 27  & ChaMP \\
RX J1113.1-2615 & Sloan (r-i) 				& 1.41$\pm$0.24	& 1.35		& ({\it r,i})=(25.8, 25.1) & r=24.0 & 31  & ChaMP  \\
RX J1350.0+6007 & Sloan (r-i) 				& 1.12$\pm$0.21	& 1.46		& ({\it r,i})=(24.7, 24.0) & r=24.2 & 37  & ARC 3.5m, 2006  \\
RX J1716.6+6708 & Sloan (r-i) 				& 1.30$\pm$0.15	& 1.47		& ({\it r,i})=(23.8, 23.5) & r=23.5 & 56  & ARC 3.5m, 2006  \\
RX J0152.7-1357 & (F625W-F775W) 			& 1.20$\pm$0.30	& 1.47		& ({\it i,z})=(27.1, 27.0) & i=22.9 & 49  & \hst\ ACS WFC, Program 9290 \\
WARPS J1415.1+3611 & (F775W-F850LP)			& 0.90$\pm$0.31	& $\sim$1.0$^a$ & ({\it i,z})=(26.8, 27.0) & i=23.9 & 52  & \hst\ ACS WFC, Program 10496 \\
RX J0910.7+5422 & (F775W-F850LP)		 	& 0.93$\pm$0.32 & $\sim$1.0$^a$ & ({\it i,z})=(26.6, 26.5) & i=24.4 & 31  & \hst\ ACS WFC, Program 9919  \\
RX J1053.7+5735 & Sloan (i-z) 				& 1.00$\pm$0.33	& $\sim$1.0$^a$ & ({\it i,z})=(26.7, 24.8) & i=24.5 & 74  & Subaru Suprime-Cam  \\
RX J0848.9+4452 & (F775W-F850LP) 			& 0.91$\pm$0.33 & $\sim$1.0$^a$ & ({\it i,z})=(27.4, 27.3) & i=25.1 & 60  & \hst\ ACS WFC, Program 9919 
\enddata
\tablecomments{
Columns:
(1) Cluster name
(2) Optical colors used to create the color-magnitude diagram for the cluster
(3) Mean color and the 3$\sigma$ spread of cluster red sequence (CRS) galaxies determined by fitting a single component Gaussian profile to the color
distribution of cluster red galaxies.  See \S~\ref{subsec:p2_obs_results_crs} for details.
(4) The expected observed color of CRS galaxies, as estimated from \citet{fukugita95} for E0 galaxies
(5) The 3$\sigma$ limiting optical magnitude of the images in the specified bandpass
(6) Observed magnitude of an L$^*$ galaxy at the cluster redshift.  See \S~\ref{subsec:p2_obs_results_crs} for details on L$^*$.
(7) Number of CRS galaxies brighter than L$^*$
(8) Origin of the survey images, which includes the Sloan Digital Sky Survey Data Release 6 \citep[SDSS DR6;][]{2008ApJS..175..297A},
the {\it Chandra} Multi-wavelength Project \citep[ChaMP;][]{2004ApJS..150...43G}, the Astrophysical Research Consortium (ARC) 3.5m Telescope
at Apache Point Observatory, the Hubble Space Telescope Advanced Camera for Surveys (\hst\ ACS) with Program ID, and the National Astronomical Observatory of Japan's 8.2m Subaru Telescope at Mauna Kea.}
\tablenotetext{a}{The expected (r-i) color of a passive galaxy at $z$$\sim$1.0 is estimated from Figure~2 in \citet{2006ApJ...644..759M}.  
Using the stellar population models of \citet{2003MNRAS.344.1000B}, these authors estimated the (F775W-F850LP) colors
of passive ellipticals at $z$$\sim$1.0, assuming solar metallicities and a stellar population age of 4 Gyr.}
\end{deluxetable}
\clearpage
\end{turnpage}

\clearpage
\begin{turnpage}
\begin{deluxetable}{clcllccccccccccp{.8in}}
\tablewidth{0pt}
\setlength{\tabcolsep}{0.05in}
\tablecaption{Radio Sources within 1 Mpc of Clusters at $z$$>$0.4 \label{tab:p2_radio_gal}}
\tablehead{
\colhead{No.} & 
\colhead{Object} & 
\colhead{In?} & 
\colhead{$\alpha$} &
\colhead{$\delta$} &
\colhead{$z$} &
\colhead{mag} &
\colhead{(r-i)} &
\colhead{(i-z)} &
\colhead{CRS} &
\colhead{Radius} &
\multicolumn{1}{c}{S$_{1.4GHz}$} &
\multicolumn{1}{c}{P$_{1.4GHz}$} & 
\multicolumn{1}{c}{F$_X$} &
\multicolumn{1}{c}{L$_X$} &
\multicolumn{1}{l}{Comments} \\
\colhead{(1)} & 
\colhead{(2)} & 
\colhead{(3)} &
\colhead{(4)} & 
\colhead{(5)} &
\colhead{(6)} & 
\colhead{(7)} &
\colhead{(8)} &
\colhead{(9)} & 
\colhead{(10)} & 
\colhead{(11)} & 
\multicolumn{1}{c}{(12)} & 
\multicolumn{1}{c}{(13)} & 
\multicolumn{1}{c}{(14)} &
\multicolumn{1}{c}{(15)} &
\multicolumn{1}{l}{(16)} 
}
\startdata
1 & MS1621-R1  & S	& 16:23:35.1 & 26:34:27.2 & 0.427$^a$ & r=19.36 & 0.66 & \nodata & 0.66$\pm$0.10 		& 0.042 & 14.0$\pm$0.2 &  82.5$\pm$1.2 & $<$5.6 & $<$3.7 & \\
2 & MS1621-R2  & S	& 16:23:32.2 & 26:33:44.0 & 0.431$^a$ & r=20.48 & 0.65 & \nodata & 0.66$\pm$0.10 		& 0.337 &  8.3$\pm$0.2 &  47.7$\pm$1.2 & $<$3.3 & $<$2.2 & \\
3 & MS1621-R3  & NP	& 16:23:31.0 & 26:34:07.0 & \nodata & r$>$22.2 & \nodata & \nodata & 0.66$\pm$0.10      	& 0.298 &  3.5$\pm$0.2 &       \nodata & $<$3.6 & \nodata & \\
4 & MS1621-R4  & S 	& 16:23:47.3 & 26:34:33.0 & 0.426$^a$ & r=18.91 & 0.77 & \nodata & 0.66$\pm$0.10 		& 1.012 &  2.2$\pm$0.2 &  12.9$\pm$1.2 & $<$8.9 & $<$5.8 & \\
5 & RXJ1221-R1$^{*}$ & P   & 12:21:25.7 &49:18:34.2  & \nodata 	   & r=21.94 & 1.27 & \nodata & 1.27$\pm$0.27 		& 0.055 & 13.70$\pm$0.09 & (253.7$\pm$1.6) & $<$2.6 		& ($<$5.7) & \\
6 & RXJ1221-R2       & S   & 12:21:29.2 &49:18:16.1  & 0.6998$^b$  & r=21.26 & 1.32 & \nodata & 1.27$\pm$0.27 		& 0.224 & 13.41$\pm$0.07 & 248.1$\pm$1.6   & $<$1.9 		& $<$4.1 & \\
7 & RXJ1221-R3       & NP  & 12:21:20.2 &49:18:46.0  & \nodata     & r=21.10 & 0.15 & \nodata & 1.27$\pm$0.27 		& 0.451 &   1.1$\pm$0.06 & \nodata         & 94.7$\pm$4.2 	& \nodata & RXJ1221-X1 (Source 7) in Table~\ref{tab:p2_xps_high_all} \\
8 & RXJ1221-R4       & P   & 12:21:31.1 &49:16:53.9  & \nodata     & r=22.13 & 1.32 & \nodata & 1.27$\pm$0.27 		& 0.768 &   0.5$\pm$0.07 & (9.1$\pm$1.4)   & $<$1.3  		& ($<$2.8) & \\
9 & RXJ2302-R1$^{*}$ & S   & 23:02:48.1 &08:43:50.8  & 0.722$^c$   & r=21.29 & 1.41 & \nodata & 1.31$\pm$0.27		& 0.027 & 17.88$\pm$0.15 & 355.8$\pm$2.9   & $<$2.7             & $<$6.3 & \\
10 & RXJ1113-R1      & NP & 11:13:09.7 &-26:16:10.1 & \nodata & r=20.56 & 1.05 & \nodata & 1.41$\pm$0.24 & 0.488 & 0.50$\pm$0.03 &  \nodata  & $<$1.2 &  \nodata &  \\
11 & RXJ1350-R1      & P & 13:50:59.6 &60:06:09.6  & \nodata & i=20.40 & 0.98 & \nodata & 1.12$\pm$0.21 & 0.784 & 5.56$\pm$0.05 & (138.8$\pm$1.3) & $<$1.4 		& ($<$4.2) & \\
12 & RXJ1350-R2      & P & 13:50:50.1 &60:08:03.2  & \nodata & i=21.49 & 1.01 & \nodata & 1.12$\pm$0.21 & 0.409 & 0.96$\pm$0.05 & (23.9$\pm$1.3)  & 3.1$\pm$1.2 	& (7.7$\pm$3.9) & RXJ1350-X7 (Source 31) in Table~\ref{tab:p2_xps_high_all} \\
13 & RXJ1350-R3      & P & 13:50:41.9 &60:07:15.2  & \nodata & i=21.13 & 1.10 & \nodata & 1.12$\pm$0.21 & 0.357 & 0.60$\pm$0.05 & (14.9$\pm$1.3)  & $<$1.2 		& ($<$3.8) & \\
14 & RXJ1350-R4      & NP & 13:50:45.7 &60:05:53.9  & \nodata & i=23.16 & 1.10 & \nodata & 1.12$\pm$0.21 & 0.596 & 0.24$\pm$0.05 & (6.0$\pm$1.3)   & $<$1.7 		& ($<$5.1) & \\
15 & RXJ1716-R1      	& S  & 17:16:36.9 & 67:08:30.4 & 0.7947$^d$ & i=21.15 & 1.22 & \nodata & 1.30$\pm$0.27 & 0.533 		& 297$\pm$12    & 7.7$\pm$0.3$\times$10$^3$ 	& 17.7$\pm$2.5 	& 46.0$\pm$6.6 & RXJ1716-X2 (Source 35) in Table~\ref{tab:p2_xps_high_all} \\
16 & RXJ1716-R2$^{*}$ 	& S  & 17:16:48.8 & 67:08:21.9 & 0.8256$^e$ & i=19.36 & 1.36 & \nodata & 1.30$\pm$0.27 & 0.011 		& 5.02$\pm$0.06 & 130.4$\pm$1.5 		& $<$7.1 		& $<$22.3 & \\
17 & RXJ1716-R3		& NS & 17:16:51.7 & 67:08:53.5 & 2.067$^d$  & r$>$24.3 & \nodata & \nodata & 1.30$\pm$0.27 & 0.260 	& 2.08$\pm$0.06 & 465.5$\pm$13.4		& 27.1$\pm$2.8	 	& 606.5$\pm$62.7 & RXJ1716-X1 (Source 34) in Table~\ref{tab:p2_xps_high_all} \\
18 & RXJ1716-R4      	& S  & 17:16:56.8 & 67:09:03.7 & 0.8009$^e$ & i=22.55 & 1.34 & \nodata & 1.30$\pm$0.27 & 0.461 		& 0.75$\pm$0.06 & 19.0$\pm$1.5 			& $<$0.6 		& $<$1.9 & \\
19 & RXJ1415-R1$^{*} $ & S 	& 14:15:11.2 & 36:12:04.2  & 1.01$^c$ & i=21.98 & \nodata 	 & 1.01 & 0.94$\pm$0.15 & 0.012 & 3.15$\pm$0.14 & 137.7$\pm$5.9 & $<$10.0 & $<$54.1 & \\
20 & RXJ1415-R2    & P 	& 14:15:08.6 & 36:14:44.0  & \nodata & i=23.78 & \nodata & 0.74 & 0.94$\pm$0.15 & 0.425 & 0.15$\pm$0.02 & (6.5$\pm$0.8) & $<$0.5 & ($<$2.6) & \\
21 & RXJ1053-R1		& NS & 10:53:51.1 & 57:35:27.9 & 0.391$^g$ & i=21.10 & \nodata & 0.43 & 1.00$\pm$0.33 & 0.756 & 0.22$\pm$0.02 & 1.1$\pm$0.1 & $<$0.8 & $<$0.4 & \\
22 & RXJ1053-R2      	& NS & 10:53:30.8 & 57:33:49.0 & 0.420$^g$ & i=24.56 & \nodata & 0.65 & 1.00$\pm$0.33 & 0.938 & 0.09$\pm$0.02 & 0.5$\pm$0.1 & $<$1.6 & $<$0.9 & \\
23 & RXJ1053-R3  	& NP & 10:53:27.1 & 57:34:28.7 & \nodata   & i=23.56 & \nodata & 0.33 & 1.00$\pm$0.33 & 0.933 & 0.09$\pm$0.02 & \nodata & 4.6$\pm$0.8 & \nodata & RXJ1053-X4 (Source 68) in Table~\ref{tab:p2_xps_high_all} \\
24 & RXJ1053-R4$^{*}$ 	& S  & 10:53:40.0 & 57:35:17.5 & 1.135$^f$ & i=21.64 & \nodata & 1.11 & 1.00$\pm$0.33 & 0.015 & 0.09$\pm$0.02 & 5.1$\pm$1.2 & $<$3.2 & $<$23.3 & \\
25 & RXJ1053-R5 	& NS & 10:53:28.9 & 57:35:35.6 & 0.741$^h$ & i=20.09 & \nodata & 0.57 & 1.00$\pm$0.33 & 0.751 & 0.08$\pm$0.02 & 1.7$\pm$0.04 & $<$0.9 & $<$1.9 & \\
26 & RXJ1053-R6 	& S  & 10:53:46.9 & 57:35:10.5 & 1.130$^f$ & i=21.89 & \nodata & 1.13 & 1.00$\pm$0.33 & 0.462 & 0.08$\pm$0.02 & 4.5$\pm$1.1 & $<$1.2 & $<$8.6 &
\enddata
\tablecomments{
Only radio sources at P$_{1.4GHz}$$\geq$\rlimb\ at the cluster redshift are included in this table.
Columns:   
(1) Object number 
(2) Radio galaxy name with brightest cluster galaxies (BCGs) identified with an asterisk.  No radio galaxy was detected at \rlim\ in MS2053 nor RXJ0152.
(3) Cluster membership is identified as follows: S $=$ spectroscopically-confirmed cluster member; NS $=$ non-cluster
member based on spectroscopy; P $=$ probable cluster member based on photometry, which places the host galaxy on the CRS; NP $=$
probable non-cluster based on either being a blank field, host galaxy at L$\geq$L$^*$, and/or bluer than the CRS. 
See \S\ref{subsec:p2_obs_results_sources} for details.
(4-5) RA/DEC of the radio source
(6) Object $z$ with references listed below
(7) Observed Sloan {\it r}-band or {\it i}-band magnitude
(8) Observed \rminusi\ or (9) \iminusz\ colors
(10) Mean color of L$\geq$L$^{*}$ CRS galaxy and the corresponding 3$\sigma$ spread in color.  See \S\ref{subsec:p2_obs_results_crs} 
for details.
(11) Projected radial distance in Mpc from the cluster X-ray emission centroid
(12) Observed 1.4~GHz radio flux density in mJy
(13) 1.4~GHz radio power in units of 10$^{23}$~\radiounits\ at the known redshift, 
or at the cluster redshift if the source redshift is unknown (enclosed in parentheses)
(14) Observed X-ray flux (0.3-8.0 keV) in units of 10$^{-15}$~\flux, assuming a power-law spectrum with $\Gamma$=1.7 (N$_E\propto$E$^{-\Gamma}$).
See \S~\ref{subsec:p2_obs_xray} for details.
(15) Rest-frame X-ray luminosity (0.3-8.0 keV) or limit in units of 10$^{42}$~\lum\ at the source redshift, or at the cluster
redshift if the source redshift is unknown (enclosed in parentheses).
(16) Comments.
References for source redshifts listed in Column 5: (a) \citet{1999AJ....117.1967S} (b) \citet{2003ApJ...594..154M} 
(c) \citet{2002ApJS..140..265P} (d) \citet{2006ApJS..165...19E} (e) \citet{1997AJ....114.1293H} (f) \citet{2005AA...439...29H} (g) \citet{2005AA...434..801Z} (h) \citet{1998AA...340L..27H} }
\end{deluxetable}
\clearpage
\end{turnpage}

\clearpage
\begin{turnpage}
\begin{deluxetable}{lllcccccccccc}
\tabletypesize{\Tinytwo}
\tablewidth{0pt}
\tablecaption{Cluster XPSs within 1 Mpc of Clusters at $z$$>$0.4 \label{tab:p2_xps_high}}
\tablehead{
	\multicolumn{1}{l}{} &
	\multicolumn{1}{l}{} &
        \multicolumn{1}{l}{} &
	\multicolumn{1}{c}{} & 
	\multicolumn{1}{c}{} & 
	\multicolumn{1}{c}{} & 
	\multicolumn{1}{c}{Net} & 
	\multicolumn{1}{c}{} & 
	\multicolumn{1}{c}{} & 
	\multicolumn{1}{c}{} & 
	\multicolumn{1}{c}{} & 
	\multicolumn{1}{c}{} & 
	\multicolumn{1}{c}{} \\
        \multicolumn{1}{l}{No.} &
        \multicolumn{1}{l}{Object} &
        \multicolumn{1}{l}{In?} &
        \multicolumn{1}{c}{$\alpha$} &
        \multicolumn{1}{c}{$\delta$} &
        \multicolumn{1}{c}{$z$} &
        \multicolumn{1}{c}{Counts} &
        \multicolumn{1}{c}{F$_X$} &
        \multicolumn{1}{c}{L$_X$} &
        \multicolumn{1}{c}{R} &
        \multicolumn{1}{c}{m} &
        \multicolumn{1}{c}{Color} &
        \multicolumn{1}{c}{P$_{1.4GHz}$} \\
        \multicolumn{1}{l}{(1)} & 
        \multicolumn{1}{l}{(2)} &
        \multicolumn{1}{l}{(3)} & 
        \multicolumn{1}{c}{(4)} & 
        \multicolumn{1}{c}{(5)} & 
        \multicolumn{1}{c}{(6)} & 
        \multicolumn{1}{c}{(7)} & 
        \multicolumn{1}{c}{(8)} & 
        \multicolumn{1}{c}{(9)} & 
        \multicolumn{1}{c}{(10)} &
        \multicolumn{1}{c}{(11)} &
        \multicolumn{1}{c}{(12)} &
        \multicolumn{1}{c}{(13)} \\
}
\startdata
35 (A) &   RXJ1716-X2 & XC   & 17:16:37.0 &    +67:08:29.3 & 0.795$^d$  &  69.8$\pm$10.0 &    17.7$\pm$2.5 &    44.1$\pm$6.2 &   0.527 & r=22.36  & (r-i)=1.22  & 7.7e3$\pm$0.3e3 \\
\multicolumn{13}{l}{Comments: CXOSEXSI/CXOMP~J171636.9+670829; RXJ1716-R1; [O II]; strong Ca HK; D4000; very strong [O III]} \\ 
      &                &      &            &                &            &                &                 &                 &         &          &             &        \\
40 (B) &   RXJ1716-X9 & XC   & 17:16:37.8 &    +67:07:30.8 & 0.8044$^d$ &   16.1$\pm$5.5 &     4.4$\pm$1.5 &    11.4$\pm$3.9 &   0.630 & r=22.60  & (r-i)=1.31 & $<$4.2 \\
\multicolumn{13}{l}{Comments: CXOSEXSI/CXOMP~J171637.6+670730; Source 33 from Table 1 of \citet{1999AJ....117.2608G}; only Ca H\&K detected} \\
      &                &      &            &                &            &                &                 &                 &         &          &             &        \\
44 (C) &   RXJ1716-X13 & XC   & 17:17:03.9 &    +67:08:59.7 & 0.813$^a$ &  8.3$\pm$4.4 &     2.0$\pm$1.0 &     5.2$\pm$2.6 &   0.714 & r=21.81  & (r-i)=1.12 & $<$4.2  \\
\multicolumn{13}{l}{Comments: CXOSEXSI~J171703.8+670900, very weak [O II], Ca HK; CXOMP J171703.8+670859; Object 9 from Table 1 of \citet{1999AJ....117.2608G}} \\
      &                &      &            &                &            &                &                 &                 &         &          &             &        \\
62 (D) &   RXJ0910-X11 & XB   & 09:10:42.7 &    +54:20:36.4 & 1.108$^a$   &   14.1$\pm$5.1 &     1.6$\pm$0.6 &     8.6$\pm$3.2 &   0.732 & i=22.04  & (i-z)=0.49  & $<$11.3 \\
\multicolumn{13}{l}{Comments: CXOMP J091042.7+542036; CXOSEXSI J091042.7+542034, weak [O II], elliptical host; Object 66 from Table 6 of \citet{2007ApJS..168...19M}}  \\
      &                &      &            &                &            &                &                 &                 &         &          &             &        \\
66 (E) &   RXJ0910-X16 & XC   & 09:10:48.3 &    +54:22:29.6 & 1.1108$^g$  &    8.7$\pm$4.7 &     1.0$\pm$0.5 &     5.4$\pm$2.7 &   0.315 & i=23.88  & (i-z)=1.00  & $<$11.3 \\
\multicolumn{13}{l}{Comments: CXOMP J091048.3+542229; CXOSEXSI J091048.3+542228; Object 023 from Table 1 of \citet{2002AJ....123..619S}}\\
\multicolumn{13}{l}{Object 51 from Table 6 of \citet{2007ApJS..168...19M}} \\
      &                &      &            &                &            &                &                 &                 &         &          &             &        \\
80 (F) &   RXJ0849-X4 & XC   & 08:49:05.3 &    +44:52:04.0 & 1.266$^a$ &  65.0$\pm$9.6 &     4.5$\pm$0.7 &    32.5$\pm$4.8 &   0.597 & i=24.00 & (i-z)=0.91     & $<$13.0  \\
\multicolumn{13}{l}{Comments: CXOMP J084905.2+445202; CXOSEXSI J084905.3+445203 with weak [O II], weak Ca HK; Object 60 from Table 5 of \citet{2007ApJS..168...19M}} \\
      &                &      &            &                &            &                &                 &                 &         &          &             &        \\
83 (G) &   RXJ0849-X8 & XC   & 08:49:04.0 &    +44:50:24.8 & 1.276$^a$ &  16.9$\pm$6.2 &     1.1$\pm$0.4 &     8.2$\pm$2.9 &   0.872 & i=23.22 & (i-z)=1.02     & $<$13.0  \\
\multicolumn{13}{l}{Comments: CXOMP J084904.0+445024; CXOSEXSI J084903.9+445023 with [O II] emission;} \\
\multicolumn{13}{l}{Object 156 from Tables 1-2 of \citet{2002AJ....123.2223S}; Object 67 from Table 5 of \citet{2007ApJS..168...19M}}
\enddata
\tabletypesize{\footnotesize}
\tablecomments{
Comments for each source, including alternative names of the XPS from other X-ray surveys and optical spectral line identifications, follow each source's entry.\\
Columns: 
(1) Source number (and letter identifier in parentheses) 
(2) Cluster X-ray point source ID 
(3) Classification tag as follows: 
XC = spectroscopically-confirmed cluster member on the CRS; 
XB = spectroscopically-confirmed cluster member bluer than the CRS; 
(4-5) J2000 RA and DEC from CIAO {\it wavdetect} routine;
(6) Source redshift with references listed below;
(7) Net X-ray (0.3-8.0 keV) counts;
(8-9) X-ray (0.3-8.0 keV) flux in units of 10$^{-15}$~\fluxunits\ and corresponding luminosity in units of 10$^{42}$~\ergs\ 
at the source redshift or at the cluster redshift, if the source redshift is unknown;
(10) Projected distance, R, from the cluster X-ray emission centroid in Mpc;
(11-12) Apparent optical magnitude and color;
(13) 1.4 GHz Radio power or limit of the X-ray source in units of 10$^{23}$~\radiounits;
The following notation is used in this column: BLAGN~=~Broadline AGN, ELG~=~Emission-line galaxy.  References for source redshifts:
(a) \citet{2006ApJS..165...19E}(b) \citet{2005ApJ...618..123S} (c) \citet{2006ApJ...644..829K} (d) \citet{1999AJ....117.2608G} (e) \citet{2005AA...432..381D}
(f) \citet{2007ApJS..168...19M} (g) \citet{2002AJ....123..619S} (h) \citet{1998AA...340L..27H} (i) \citet{2005AA...444...79M} (j) \citet{2002AJ....123.2223S}
}
\end{deluxetable}
\clearpage
\end{turnpage}

\setlength{\tabcolsep}{18pt}
\begin{deluxetable}{ccccc}
\tabletypesize{\scriptsize}
\tablecaption{Cumulative Radial Distribution of Cluster Radio Galaxies Compared to CRS Galaxies \label{tab:p2_kstest}}
\tablewidth{0pt}
\tablehead{
\colhead{Redshift Bin}  & 
\colhead{With BCGs?} & 
\colhead{K-S D-statistic} & 
\colhead{Probability} &
\colhead{Significance} \\
\colhead{(1)}  & 
\colhead{(2)} &  
\colhead{(3)}   &  
\colhead{(4)} &
\colhead{(5)} 
}
\startdata
\lowz\  &       Y       &  0.43 & 1.1$\times$10$^{-8}$  &  5.7$\sigma$  \\
        &       N       &  0.36 & 3.3$\times$10$^{-6}$  &  4.7$\sigma$  \\
\midz\ 	& 	Y	&  0.44	& 4.5$\times$10$^{-9}$ 	&  5.9$\sigma$ 	\\
 	& 	N	&  0.33 & 2.7$\times$10$^{-5}$	&  4.2$\sigma$ 	\\
\highzsmall\ & 	Y	&  0.31	& 1.0$\times$10$^{-4}$ 	&  3.9$\sigma$ 	\\
        & 	N	&  0.32	& 5.4$\times$10$^{-5}$  &  4.0$\sigma$ 	
\enddata
\tablecomments{
Columns: 
(1) Cluster redshift bin 
(2) Are the BCGs included in the cumulative radial distribution function?
(3) Two-sided Komolgorov-Smirnov (K-S) test D-statistic, which is a measure of the maximum difference between the radio galaxy
cumulative radial distribution function compared to the CRS galaxy distribution.
(4) Two-sided K-S test probability that the distributions are drawn from the same parent distribution.
(5) Significance level of the probability.
}
\end{deluxetable}

\setlength{\tabcolsep}{6pt}

\begin{deluxetable}{ccccc}
\tabletypesize{\scriptsize}
\tablecaption{``Road to Coma" Cluster Radio Luminosity Function \label{tab:p2_rlf}}
\tablewidth{0pt}
\tablehead{
\colhead{Redshift}  & \colhead{N$_{CRS}$} & \colhead{Log(P$_{1.4GHz}$) Bin} & \colhead{N$_{Radio}$} & \colhead{Log(N$_{Radio}$/N$_{CRS}$)} \\
\colhead{(1)}  & \colhead{(2)} &  \colhead{(3)}   &  \colhead{(4)} & \colhead{(5)}
}
\startdata
\lowz\ & 665 
&    23.6--24.0 	&  5 & -2.12 $\pm$ 0.16 \\
& &  24.0--24.4 	&  6 & -2.04 $\pm$ 0.15 \\
& &  24.4--24.8 	&  5 & -2.12 $\pm$ 0.16 \\
& &  24.8--25.2 	&  1 & -2.82 $\pm$ 0.30 \\
& &  25.2--26.0 	&  0 & $<$-2.8 \\
& &  26.0--27.0 	&  0 & $<$-2.8 \\
0.4$<$$z$$<$1.1 & 443 
   & 23.6--24.0 	&  4 & -2.04 $\pm$ 0.18 \\
&  & 24.0--24.4 	&  4 & -2.04 $\pm$ 0.18 \\
&  & 24.4--24.8 	&  1 & -2.65 $\pm$ 0.30 \\
&  & 24.8--25.2 	&  4 & -2.04 $\pm$ 0.18 \\
&  & 25.2--26.0 	&  3 & -2.17 $\pm$ 0.20 \\
&  & 26.0--27.0 	&  1 & -2.65 $\pm$ 0.30 
\enddata
\tablecomments{
Columns: (1) Cluster redshift bin (2) Number of L$\geq$L$^*$ CRS galaxies within 1 Mpc of the cluster center
(3) Log(P$_{1.4GHz}$) bin in units of \radiounits\  
(4) Number of radio sources detected in the specified log(P$_{1.4GHz}$) bin
(5) Log of the fraction of CRS galaxies that host radio sources in the specified log(P$_{1.4GHz}$) bin.
See \S\ref{subsec:p2_obs_results_crs} for the details on the CRS galaxies number calculation.  Quoted errors
are one-sided Poisson errors, using \citet{1986ApJ...303..336G} approximation.}
\end{deluxetable}

\clearpage
\setlength{\tabcolsep}{10pt}
\begin{deluxetable}{lp{2.2in}ccc}
\tabletypesize{\scriptsize}
\tablecaption{Differences in the Expected vs. Detected Cluster Radio Galaxy Counts \label{tab:p2_rlf_diffs}}
\tablewidth{0pt}
\tablehead{
\colhead{Log(P)} &
\colhead{Surveys} &
\colhead{N$_{exp}$} &
\colhead{N$_{det}$} &
\colhead{\% Probability} \\
\colhead{(1)} &
\colhead{(2)} &
\colhead{(3)} &
\colhead{(4)} &
\colhead{(5)}
}
\startdata
23.6-25.0     & Paper I vs. This work	     			& 11.3 &  10 &   70    \\	
23.6-25.0     & \citet{1996AJ....112....9L} vs. Paper I		& 16.2 &  15 &   66    \\	
23.6-25.0     & \citet{1996AJ....112....9L} vs. This work	& 10.8 &  8  &   85    \\	
24.0-25.0$^*$ & \citet{1999AJ....117.1967S} vs. Paper I		&  7.6 &  10 &   23    \\	
24.0-25.0$^*$ & \citet{1999AJ....117.1967S} vs. This work	&  5.0 &  5  &   56    \\	
$>$25.0       & Paper I vs. This work				&    0  &  7  &   0$^{**}$   \\	
$>$25.0       & \citet{1996AJ....112....9L} vs. This work	&   2.2 &  6  &   2    \\	
$>$25.0       & \citet{1999AJ....117.1967S} vs. This work       &   0.3 &  6  &   7$\times$10$^{-5}$  
\enddata
\tablecomments{
Columns:
(1) log(P$_{1.4GHz}$) in \radiounits\
(2) Surveys that are being compared
(3) Number of expected radio galaxies based on the first survey listed in Column~2
(4) Number of detected radio galaxies in the second survey in Column~2
(5) Cumulative binomial probability (\%) of detecting $\geq$N$_{det}$ radio galaxies in the second survey
}
\tablenotetext{*}{The sample of cluster radio galaxies from \citet{1999AJ....117.1967S} is incomplete at log(P$_{1.4GHz}$)$<$24 \radiounits.}
\tablenotetext{**}{Based on our detected number of \logp$>$25 \radiounits\ radio galaxies at \comboz, we expect to detect 10.5 radio sources at \logp$>$25 \radiounits\
in our \lowz\ cluster sample, but we detected none.  The binomial probability of detecting no source is 0.003\%).}
\end{deluxetable}

\setlength{\tabcolsep}{6pt}	

\begin{deluxetable}{ccccccccc}
\tabletypesize{\scriptsize}
\tablecaption{Number of Expected vs. Detected X-ray Point Sources \label{tab:xps_excess}}
\tablewidth{0pt}
\tablehead{
\colhead{Redshift Bin} & \colhead{N$_{clusters}$} & \colhead{Area} & \colhead{Cut-Off Limit} & \colhead{N$_{detected}$} & \colhead{N$_{bgd}$} & \colhead{Excess} \\
\colhead{(1)}  & \colhead{(2)} &  \colhead{(3)} & \colhead{(4)} &  \colhead{(5)} & \colhead{(6)} &  \colhead{(7)} 
}
\startdata
0.2--0.4   &  11      &   0.165       	& \lxbb$\geq$5$\times$10$^{42}$ & ~22 &  	~19.9$\pm$6.8   &      	0.3$\sigma$ \\
\nodata   &  \nodata  &   \nodata      	& \lxbb$\geq$1$\times$10$^{42}$ & 105 &  	103.8$\pm$23.1  &      	0.05$\sigma$ \\
0.4--0.8  &  ~5      &   0.028       & \lxbb$\geq$5$\times$10$^{42}$ & ~24 & 	~23.6$\pm$6.8   &      	0.06$\sigma$ \\
0.8--1.26 &  ~6      &   0.022       & \lxbb$\geq$5$\times$10$^{42}$ & ~62 &       ~44.3$\pm$11.1  &      	1.6$\sigma$ 
\enddata
\tablecomments{
Columns: (1) Redshift range (2) Number of clusters in the given redshift bin (3) Total area surveyed in square degrees
(4) Limiting luminosity for an an individual source (0.3--8.0 keV) in \ergs\ at the cluster redshift.
(5) Total number of detected XPSs in the cluster observations within 1 Mpc of the cluster center
(6) Number of expected background XPSs above the specific limit listed in Column 4.
The quoted error is the contribution from statistical fluctuations 
and $\approx$20\% due to cosmic variance in the estimated background counts added in quadrature.
(7) Statistical excess of detected XPSs above the expected XPS background from Column 6 in the units of $\sigma$.
Note that RXJ0152 is not included in this specific analysis.
}
\end{deluxetable}

\clearpage
\begin{turnpage}
\begin{deluxetable}{llp{0.15in}p{0.55in}cccccccp{0.1in}p{1.6in}}
\tabletypesize{\tiny}
\tablewidth{0pt}
\setlength{\tabcolsep}{0pt}
\tablecaption{XPSs within 1 Mpc of Clusters at $z$$>$0.4 \label{tab:p2_xps_high_all}}
\tablehead{
	\multicolumn{1}{l}{} &
	\multicolumn{1}{l}{} &
        \multicolumn{1}{l}{Cl} &
	\multicolumn{1}{c}{RA} & 
	\multicolumn{1}{c}{} & 
	\multicolumn{1}{c}{Net} & 
	\multicolumn{1}{c}{} & 
	\multicolumn{1}{c}{} & 
	\multicolumn{1}{c}{} & 
	\multicolumn{1}{c}{} & 
	\multicolumn{1}{c}{} & 
	\multicolumn{1}{c}{} &
	\multicolumn{1}{l}{}  \\
        \multicolumn{1}{l}{No.} &
        \multicolumn{1}{l}{Object} &
        \multicolumn{1}{l}{Src?} &
        \multicolumn{1}{c}{DEC} &
        \multicolumn{1}{c}{$z$} &
        \multicolumn{1}{c}{Counts} &
        \multicolumn{1}{c}{F$_X$} &
        \multicolumn{1}{c}{L$_X$} &
        \multicolumn{1}{c}{R} &
        \multicolumn{1}{c}{m} &
        \multicolumn{1}{c}{Color} &
        \multicolumn{1}{c}{P$_{1.4GHz}$} &
        \multicolumn{1}{l}{Comments}  \\
        \multicolumn{1}{l}{(1)} & 
        \multicolumn{1}{l}{(2)} &
        \multicolumn{1}{l}{(3)} & 
        \multicolumn{1}{c}{(4)} & 
        \multicolumn{1}{c}{(5)} & 
        \multicolumn{1}{c}{(6)} & 
        \multicolumn{1}{c}{(7)} & 
        \multicolumn{1}{c}{(8)} & 
        \multicolumn{1}{c}{(9)} & 
        \multicolumn{1}{c}{(10)} &
        \multicolumn{1}{c}{(11)} &
        \multicolumn{1}{c}{(12)} &
        \multicolumn{1}{l}{(13)} 
}
\startdata
1 &    MS1621-X1 & BG & 16:23:43.8 +26:32:44.7 & 0.659$^a$  &  430.3$\pm$22.5 &   171.5$\pm$9.0 &  276.1$\pm$14.5 &   0.826 & r=19.52  & (r-i)=0.00  & $<$0.8 & CXOSEXSI J162343.7+263244, BLAGN \\
2 &    MS1621-X2 & BG & ~~16:23:27.3 +26:32:07.3 & 1.438$^a$  &   59.5$\pm$9.1 &    23.6$\pm$3.6 &  231.8$\pm$35.4 &   0.959 & r=20.86  & (r-i)=0.55  &  $<$0.8 & CXOSEXSI J162327.2+263207, BLAGN \\ 
3 &    MS1621-X3 & CRS2 & ~16:23:36.1 +26:36:52.6 & \nodata    &   40.5$\pm$7.7 &    15.3$\pm$2.9 &     9.0$\pm$1.7 &   0.849 & r=22.8 & (r-i)=0.54     &  $<$0.8 & CXOSEXSI J162335.9+263652 \\
4 &    MS1621-X4 & BG & 16:23:45.9 +26:33:35.0 & 1.283$^a$ &    22.7$\pm$6.4 &     9.0$\pm$2.6 &   68.0$\pm$19.6 &   0.826 & r$>$22.2 & \nodata     &  $<$0.8 & CXOSEXSI J162345.7+263335, ELG \\ 
5 &    MS2053-X1 & BG & 20:56:22.3 -04:40:07.3 & 0.641$^a$ & 114.2$\pm$12.1 &  37.3$\pm$3.9 &   56.3$\pm$5.9 &   0.918 & r=21.21 & (r-i)=0.51   & $<$2.2 & CXOSEXSI J205622.2-044005, ELG \\
6 &    MS2053-X2 & FG & 20:56:24.8 -04:35:35.4 & 0.260$^a$ &  28.0$\pm$6.9 &    9.9$\pm$2.5 &    1.9$\pm$0.5 &   0.945 & r=19.48  & (r-i)=0.55   & $<$2.2 & CXOSEXSI J205624.7-043533, ELG \\
7 &   RXJ1221-X1 & BLU  & 12:21:20.1 +49:18:45.8 & \nodata & 598.5$\pm$26.4 &    94.7$\pm$4.2 &   175.4$\pm$7.7 &   0.462 & r=21.10  & (r-i)=0.15    & 20.7$\pm$1.2 & CXOMP J122120.1+491846, unlikely cluster member based on radio \& optical color; Also RXJ1221-R3 (Source 7 in Table~\ref{tab:p2_radio_gal}) \\
8 &   RXJ1221-X2 & BLU2 & 12:21:29.0 +49:16:45.4 & \nodata &   19.7$\pm$5.9 &     3.5$\pm$1.0 &     6.5$\pm$1.9 &   0.747 & r=24.19  & (r-i)=0.94    & $<$2.0 & CXOMP J122129.0+491645 \\ 
9 &   RXJ1221-X3 & CRS  & 12:21:30.8 +49:17:58.8 & \nodata &   16.8$\pm$5.9 &     2.9$\pm$1.0 &     5.3$\pm$1.9 &   0.365 & r=22.87  & (r-i)=1.20    & $<$2.0 & V 1221+4918, Source 5 from Table 2 of 
\citet{2007AA...462..449B} \\
10 &   RXJ2302-X1 & BG & 23:02:46.1 +08:45:22.8 & 1.944$^b$  &  441.6$\pm$22.7 &   156.1$\pm$8.0 &  3043.6$\pm$156.0 &   0.684 & r=19.08  & (r-i)=0.52  & $<$3.5 & WARP J2302.7+0845; CXOMP~J230246.0+084523, BLAGN \\ 
\enddata
\tabletypesize{\footnotesize}
\tablecomments{
Table~\ref{tab:p2_xps_high_all} is published in its entirety in the electronic edition of the Astrophysical Journal. A portion is shown here for guidance regarding its form and content.\\
Columns: 
(1) Source number (and letter identifier in parentheses)
(2) Cluster X-ray point source ID 
(3) Classification tag as follows: 
XC = spectroscopically-confirmed cluster member on the CRS; 
XB = spectroscopically-confirmed cluster member bluer than the CRS; 
FG = spectroscopically-confirmed foreground non-cluster source;
BG = spectroscopically-confirmed background non-cluster source; 
CRS = source host galaxy with CRS colors at an unknown redshift and with L$\geq$L$^*$ if located at the cluster redshift; 
CRS2 = similar to CRS but with L$<$L$^*$ host galaxy;
BLU = XPS hosted in L$\geq$L$^*$ host galaxy with colors bluer than typical CRS galaxies and with unknown redshift;
BLU2 = similar to BLU, but with L$<$L$^*$ host galaxy at the cluster redshift;
BLU3 = similar to BLU, but with L$\geq$L$_{BCG}$ host galaxy;
NO = No optical counterpart is detected in one or both optical images;
NP = No photometric data in one or both optical images;
(4) J2000 RA and DEC from CIAO {\it wavdetect} routine;
(5) Source redshift with references listed below;
(6) Net X-ray (0.3-8.0 keV) counts;
(7-8) X-ray (0.3-8.0 keV) flux in units of 10$^{-15}$~\fluxunits\ and X-ray (0.3-8.0 keV) luminosity in units of 10$^{42}$~\ergs\ 
at the source redshift or at the cluster redshift, if the source redshift is unknown;
(9) Projected distance, R, from the cluster X-ray emission centroid in Mpc;
(10-11) Optical magnitude and color;
(12) 1.4 GHz radio power or limit of the X-ray source in units of 10$^{23}$~\radiounits;
(13) Comments, including alternative names of the XPS from other X-ray surveys and optical spectral line identifications.
The following notation is used in this column: BLAGN~=~Broadline AGN, ELG~=~Emission-line galaxy per
SEXSI identification by \citet{2006ApJS..165...19E}.  References for source redshifts:
(a) \citet{2006ApJS..165...19E}(b) \citet{2005ApJ...618..123S} (c) \citet{2006ApJ...644..829K} (d) \citet{1999AJ....117.2608G} (e) \citet{2005AA...432..381D}
(f) \citet{2007ApJS..168...19M} (g) \citet{2002AJ....123..619S} (h) \citet{1998AA...340L..27H} (i) \citet{2005AA...444...79M} (j) \citet{2002AJ....123.2223S}
}
\end{deluxetable}
\clearpage
\end{turnpage}

\end{document}